\documentclass[a4paper,fleqn]{cas-dc}
\usepackage[authoryear,round]{natbib}
\usepackage{lineno}
\usepackage{setspace}
\usepackage{appendix}
\usepackage{blindtext}
\usepackage[T1]{fontenc}
\usepackage{graphicx}
\usepackage{float}
\usepackage{rotating}
\usepackage[section]{placeins} %keep things from going past into the references
\usepackage[version=4]{mhchem}
\usepackage{array}
\usepackage{booktabs}
\usepackage{pdflscape}
\usepackage{longtable}
\usepackage{tablefootnote}
\usepackage{threeparttablex}

\newcolumntype{P}[1]{>{\raggedright}p{#1}}
\newcolumntype{s}{>{\scriptsize\raggedright\arraybackslash}p}
\newlength{\ninetabcol}
\newlength{\eighttabcol}
\newlength{\sixtabcol}
\newlength{\fivetabcol}
\newcommand{\tn}{\tabularnewline}

\usepackage{xcolor}

\usepackage{graphicx}
\usepackage{subcaption}
\newcommand{\GIN}[1]{\textcolor{black}{#1}} % CHANGED FROM RED TO BLACK

\usepackage{caption}
\begin{document}
\let\WriteBookmarks\relax
\def\floatpagepagefraction{1}
\def\textpagefraction{.001}

% \linenumbers %turn on line numbers

\shorttitle{Earth accreted increasingly oxidized material}

\shortauthors{Nathan et~al.}

% \title [mode = title]{Varying Solar System formation scenarios successfully build Earth if initial accreted material is reduced and later accreted material is oxidized} %(old title)    
\title[mode = title]{Earth accreted increasingly oxidized material as it grew constraining Theia to be small and relatively oxidized}

%\tnotetext[1]{This project was funded by the NASA FINESST program. NASA award number: 80NSSC23K1376.}

\author[1]{Gabriel I. Nathan}[orcid=0000-0002-4261-6677]
\cormark[1]
\cortext[cor1]{Corresponding author}
\ead{gabriel.i.nathan@jpl.nasa.gov}
\credit{Conceptualization, Methodology, Software, Validation, Formal analysis, Investigation, Data Curation, Writing - Original Draft, Writing - Review \& Editing, Visualization, Supervision, Funding acquisition}

\author[1]{Hung D. Nguyen}
\credit{Validation, Formal analysis, Investigation, Data Curation, Writing - Review \& Editing}

\author[1]{Seth A. Jacobson}[orcid=0000-0002-4952-9007]
\ead{seth@msu.edu}
\credit{Conceptualization, Methodology, Software, Resources, Writing - Review \& Editing, Supervision, Project administration, Funding acquisition}

% \affiliation[2]{organization={NASA Jet Propulsion Laboratory},
%                 addressline={4800 Oak Grove Dr}, 
%                 city={Pasadena},
%                 state={CA},
%                 postcode={91109},             
%                 country={USA}} 

\affiliation[1]{organization={Department of Earth \& Environmental Sciences, Michigan State University},
                city={East Lansing},
                state={MI},
                postcode={48824},             
                country={USA}}

\begin{abstract}
Many different Solar System formation scenarios successfully reproduce the main astrophysical properties of the inner Solar System. 
Here, we use the chemistry of Earth's mantle to determine whether proposed Solar System formation scenarios can produce Earth and to constrain the characteristics of both the Proto-Earth and the Moon-forming impactor, Theia.
To do this, we use a coupled \textit{N}-body accretion and core-mantle differentiation model \citep{rubie2011heterogeneous, rubie2015accretion} applied to a range of proposed formation scenarios (\GIN{Classical, Annulus, and Grand Tack scenarios}).
This model calculates the mantle chemistry of Earth-like planets in Solar System formation simulations and adjusts the initial redox state of the building blocks in a synthetic protoplanetary disk until the mantle of the simulated Earth closely matches the bulk silicate Earth. 
\GIN{We find that a strong oxidizing gradient among the initial solids of the protoplanetary disk is favored to reproduce the chemistry of Earth's mantle, in agreement with \citet{rubie2015accretion}.}
The Solar System formation scenarios examined here successfully reproduce the composition of Earth if two related conditions are met: 1) there is a strong positive oxidation gradient with distance in the inner protoplanetary disk and 2) Earth accretes increasingly more oxidized material (on average) as it grows.
Furthermore, in simulations that successfully reproduce Earth's composition, Proto-Earth and Theia may possess a wide range of mantle compositions and mass ratios. 
However, because late accretion is oxidizing across all proposed Solar System formation scenarios, Theia and Proto-Earth cannot both possess mantles that are more oxidized than the present-day Earth's mantle. 
Similarly, Theia's mantle may be significantly more oxidized than present-day Earth's mantle, but only if it is relatively small compared to the Proto-Earth.
\end{abstract}

\begin{highlights}
\item The Earth mantle can be reproduced in a variety of dynamical models
\item Late accretion to the Earth is more oxidized than earlier stages
\item A small Theia could have had a range of redox states and is favored in our models
\item If Theia was large it must have possessed a similar redox state to present-day Earth
\end{highlights}

% Keywords

\maketitle

\section{Introduction} \label{sec:intro}

Many proposed dynamical scenarios of terrestrial planet formation reproduce the architecture of the inner Solar System and have been explored with numerical astrophysical \textit{N}-body algorithms that simulate the dynamical evolution and growth of planetary bodies \citep[for review, see][]{morbidelli2012building}.
These computational experiments have revealed persistent trends in the way initial populations of planetary materials collide and grow \citep[e.g.,][]{kokubo1996runaway,kokubo1998oligarchic} and how Solar System architecture evolved to its current state from different possible starting disk configurations \citep[e.g.,][]{chambers2001making,raymond2009building,hansen2009formation}.
However, the scientific achievement of matching Solar System astrophysical constraints (e.g., the masses and orbits of the terrestrial planets) has now been met by a variety of mutually incompatible terrestrial planet formation scenarios \citep[e.g.,][]{hansen2009formation, walsh2011low, jacobson2014lunar, clement2019early, nesvorny2021role, izidoro2022planetesimal, lykawka2023terrestrial, Nesvorny2025terrestrial}.
Thus, additional constraints must be used to differentiate these planet formation scenarios, and we turn to geologic, meteoritic, and planetary sample records to distinguish between these hypothesized histories.

Models of planetary evolution have demonstrated that formation conditions strongly affect the composition of terrestrial planets \citep{wade2005core,rubie2011heterogeneous}. 
Because of this, a detailed understanding of the terrestrial geologic record is a potentially powerful set of constraints to discriminate between proposed planetary formation scenarios.
For instance, using a variety of Solar System formation scenarios, \citet{gu2023comparisons} found that the mantle composition of Earth analogs depends most strongly on the fraction of equilibrating embryo cores and the size of initial embryo masses.
It is not clear if all Solar System formation scenarios are compatible with producing Earth analogs with the same mantle composition as the bulk silicate Earth.
Prior studies have demonstrated that the Grand Tack scenario is able to reproduce Earth analogs with accurate composition \citep[e.g.,][]{rubie2015accretion} as well as Mars, though not both of these planets simultaneously \citep{nathan2023constraining}.
Furthermore, the mantle composition of Earth can be reproduced from so-called ring models \citep{dale2025compositional,Dale2026}, but comprehensive examinations between different proposed terrestrial planet formation scenarios are limited.

The formation history of the Earth likely shaped its compositional evolution \citep[e.g.,][]{dauphas2017isotopic} and is inextricable from the Moon-forming impact \citep[e.g.,][]{canup2001origin}, which was the most significant accretion event during Earth's formation. 
In the canonical Moon formation hypothesis, the Moon formed from the debris disk produced by a collision between the Proto-Earth and a Mars-sized impactor, often called Theia \citep{hartmann1975satellite,Benz1986,canup2001origin,Canup2004,canup2008lunar}. 
However, each of many proposed Moon-forming impact scenarios \citep[e.g.,][]{canup2001origin, cuk2012making, lock2018origin} can seemingly reproduce the astrophysical properties of the Earth-Moon system if a specific dynamical history is invoked.
Because the astrophysical properties of the Moon's size and orbit can be reproduced by a range of dynamical pathways, its origin and formation history remain undetermined. 

Here, the goals are twofold; first, we test whether the geochemistry of the Earth's mantle is compatible with each of the major Solar System formation scenarios: Classical, Grand Tack/Instability, and Annulus, and second, we determine what constraints these scenarios place on the mass ratio and composition of Proto-Earth and the Moon-forming impactor, Theia.
To calculate the evolving mantle and core compositions of the bodies in these simulations, we \GIN{couple} accretion histories from astrophysical \textit{N}-body simulations to a planetary differentiation model, which simulates geochemical evolution due to multiple stages of impact-induced melting and core-mantle equilibration events.

Our model is built on a legacy of prior work \citep[e.g.,][]{rubie2011heterogeneous, rubie2015accretion, rubie2016highly, jennings2021metal, blanchard2022metal, nathan2023constraining, Dale2023} and combines dynamics of Solar System accretion, mixing of impacting planetary bodies, and laboratory data describing elemental partitioning between metal and silicate.
With this combined planetary accretion and differentiation model, we compare and contrast the three end-member classes of formation scenarios and determine if it is possible to build Earth analogs with accurate bulk silicate compositions in each of these scenarios. 
We use many astrophysical \textit{N}-body simulations from or inspired by the scientific literature that successfully reproduce the dynamical properties of the Earth \citep[e.g.,][]{raymond2009building,hansen2009formation, walsh2011low,jacobson2014lunar,O'Brien2014,izidoro2015terrestrial,nesvorny2021role}.
By doing this for each simulation, we determine what the redox state of the protoplanetary disk must have been in order to recreate the composition of the Earth and how the mantle composition of Earth evolved with time.
Finally, we determine which combinations of Theia and Proto-Earth are compatible with the composition of the bulk silicate Earth. 

\subsection{Terrestrial planet formation} \label{sec:background}

Numerical models of the formation of the Solar System with astrophysical \textit{N}-body simulations have become generally successful at reproducing the orbits and masses of the major planets and stable minor planet populations. 
In particular, there now exist dynamical formation scenarios that generally reproduce the astrophysical characteristics of the terrestrial planets in the Solar System \citep[e.g.,][]{hansen2009formation, walsh2011low, jacobson2014lunar, clement2019early, nesvorny2021role, izidoro2022planetesimal, lykawka2023terrestrial, Nesvorny2025terrestrial}. 
These different proposed planet formation scenarios are able to match the major astrophysical constraints of the Solar System such as the number of terrestrial planets as well as their masses, semi-major axes, and net angular momentum deficit, which statistically accounts for deviations of their orbits from in-plane and circular geometries.
These scenarios also generally reproduce the mass and orbital distribution of the main asteroid belt including a two-component inner and outer compositional belt associated with the S- and C-complex asteroids, respectively, and water delivery from the outer component to Earth.
Some of the most successful proposed scenarios can robustly reproduce these properties despite the stochasticity that ensues from the chaotic dynamics of the era of planet formation.

These modern scenarios are in contrast to what is sometimes called the Classical or Canonical scenario, which, despite some distinct problems, reproduces many features of the inner Solar System, including the number of terrestrial planets and their orbits, while requiring minimal assumptions regarding the initial mass distribution in the protoplanetary disk \citep{chambers2001making,obrien2006terrestrial,raymond2009building,fischer2014dynamics,jacobson2015earth}.
This scenario asserts that the terrestrial planets grew from a protoplanetary disk with a structure derived from the minimum mass solar nebula model \citep{weidenschilling1977aerodynamics,hayashi1981structure}.
In essence, this disk spreads the solid mass of the eight planets quasi-smoothly between the semi-major axes of Mercury and Neptune in small protoplanets according to a surface density inspired by the distribution of mass in the final planets \citep{Wetherill1978, wetherill1989accumulation}.
Consistent with models of giant planet growth and orbital evolution of large planets in the presence of a gas disk, the giant planets start on circular orbits in their present-day locations and at their present sizes, having grown faster due to the presence of ices in the outer Solar System \citep[e.g.,][]{morbidelli2015great}.
This scenario is referred to as the Circular Jupiter-Saturn (CJS) scenario because of the circular orbits assigned to Jupiter and Saturn, which are important to the dynamical evolution of the inner Solar System \citep{raymond2009building}.
This scenario often results in planets that match the masses and orbits of Venus and Earth. 
In particular, the accretion histories of Earth-like planets in this scenario tend to be consistent with aspects of Earth's growth history as inferred from geochemistry, such as late giant impacts like the Moon-forming impact \citep{hartmann1975satellite,cameron1976origin,Benz1986,canup2001origin}, a post-Moon-forming impact late veneer of planetesimal accretion consistent with the highly siderophile element abundances in Earth's mantle \citep{jones1986geochemical,day2016highly}, and a terrestrial growth timescale in agreement with geochemical evidence from the short-lived Hf-W radiometric system \citep{halliday2004mixing,Jacobsen2005hf,kleine2009hf,nimmo2010tungsten}.

However, the Classical scenario consistently fails to reproduce Mars and properties of the main asteroid belt \citep{chambers2001making,obrien2006terrestrial,raymond2009building,fischer2014dynamics,jacobson2015earth}.
For both Mars and the asteroid belt, a primary challenge is to reproduce their small masses relative to Earth---this theoretical challenge is the so-called Small Mars problem. 
A too large Mars and a too massive asteroid belt also affects the orbital evolution and growth of Earth, so the problem cannot be ignored even if Earth is one's primary concern \citep[e.g.,][]{wetherill1991isn, chambers2001making, raymond2006high}.
The Small Mars problem is a direct consequence of the assumed initial distribution of materials throughout the protoplanetary disk in the Classical scenario. 
The continuous distribution of solid mass spread among a sea of protoplanets from the innermost regions of the disk ($\le 1$ AU) to the area just interior of the giant planets ($\lesssim 5$ AU) creates a Mars feeding zone that is too large. 
In fact, the zone centered at Mars' current location ($\sim 1.5$ AU) has an amount of material roughly as massive as that in the area of Earth's growth ($\sim 1$ AU) in the Minimum Mass Solar Nebula \citep{weidenschilling1977aerodynamics,hayashi1981structure}.
In such a case, there is roughly the same amount of building materials for Mars as for Earth and, as a consequence, Mars grows to a similar size as Earth in the Classical scenario \citep{raymond2009building,Raymond2022}.
The asteroid belt is generally too dynamically stirred to accrete a planetary-sized body but it is also not stirred enough to reduce the mass in the main belt region from that inferred by the Minimum Mass Solar Nebula to an amount comparable to the modern asteroid belt.

A slightly modified version of the Classical scenario asserts that the orbits of the giant planets, particularly Jupiter and Saturn, possessed eccentric orbits; this scenario is typically referred to as the eccentric Jupiter-Saturn (EJS) scenario. 
If Jupiter and Saturn begin on excited orbits, for instance with eccentricities consistent with their modern values, then the giant planets scatter mass out of the asteroid belt region and, to a lesser extent, from the Mars region \citep{raymond2009building,fischer2014dynamics,jacobson2015earth}.
This results in a reduced Mars analog mass, but Mars-like planets typically still exceed the mass of Mars by factors of a few, rather than ten as in the circular case \citep{raymond2009building,fischer2014dynamics,jacobson2015earth}.
In rare cases, small Mars-like planets form \citep{fischer2014dynamics}, but these are often in planetary systems that otherwise do not resemble the inner Solar System \citep{jacobson2015earth}.
Starting with an over-excited giant planet system, e.g., Jupiter and Saturn possessing eccentricities far in excess of their modern values, further reduces the mass of simulated Mars analogs to Mars-like values, but this strong excitation often destabilizes much of the terrestrial planet region, for instance reducing the similarities of Earth analogs to Earth \citep{raymond2009building}.
Furthermore, there is no reason to expect the giant planets to be on excited orbits at all, since growth within the solar nebula is expected to fully dynamically dampen their orbits---an observation consistent with exoplanet systems \citep[e.g.,][]{kley2019planet}.

A straightforward solution to the Small Mars problem is forming the terrestrial planets from an initially truncated protoplanetary disk of material.
If the region of Mars and the main asteroid belt begins depleted, then they cannot grow too large. 
Indeed, the first annulus scenario did not place any mass within the Mars or main belt regions, and instead mass had to be scattered by the growing Earth and Venus into these regions \citep{hansen2009formation}.
This initial proposal did not motivate why the inner disk would be a sharp narrow annulus instead of a smooth wide disk.
However, such a truncation of the protoplanetary disk may have been due to dynamical interactions between the giant planets (and the nebular gas disk) or due to an initial shortage of solid material reflecting protoplanetary disk processes.

The most dramatic sculpting mechanism proposed is a `Grand Tack' of Jupiter and Saturn as they migrated through the gas disk \citep{walsh2011low}. 
In the Grand Tack scenario, as Jupiter and Saturn interact separately with a viscous proto-planetary disk, they migrate initially inward; Saturn migrates faster and catches Jupiter into a mean motion resonance, which reverses the direction of their migration \citep{masset2001reversing}.
The inward movement of the giant planets scatters protoplanets out of the main belt and Mars regions, compressing rocky planetesimals and embryos from a wide disk into a narrower annulus.
The later outward migration of the giant planets then scatters outer icy planetesimals into the inner disk region, and the inner growing planets are also able to scatter rocky protoplanets back into the Mars and main belt regions \citep{walsh2012populating}. 
This scenario was the first to reproduce not just the small masses of Mars and the main belt, but also the two-component composition of the asteroid belt \citep{gradie1989distribution,mothediniz2003distribution,demeo2013taxonomic} and the rapid formation timescale of Mars as inferred from the short-lived radiometric Hf-W system \citep{nimmo2007mars, dauphas2011hf,jacobson2014lunar}, while retaining all of the other positive features of the Classical scenario \citep{walsh2011low,jacobson2014lunar,brasser2016analysis,deienno2016grand}.
The Grand Tack scenario invokes a very specific migration history for the giant planets and hence a very specific evolution of the Solar System's protoplanetary disk \citep{brasser2016analysis}.
Furthermore, like the self-consistent circular Classical scenario, a late giant planet instability is required to explain the orbits of the outer giant planets \citep{tsiganis2005origin} as well as small body populations like the Jupiter Trojans \citep{morbidelli2005chaotic}.

A late giant instability, though, risks destabilizing the fully grown inner Solar System \citep{kaib2016fragility}---a problem present for both the Classical and Grand Tack scenarios.
Note that the Solar System could be a statistically rare outcome, since it is possible for four inner Solar System planets to survive.
However, an earlier giant planet instability occurring during the main stage of terrestrial planet accretion alleviates this problem.
This is because, although an early dynamical instability still scatters a fraction of the existing protoplanets into the Sun, enough can survive to build the terrestrial planets \citep{kaib2016fragility}.
Felicitously, an early giant planet instability would preferentially remove material from the outer parts of the terrestrial planet forming region, so the instability could also have dynamically excited and emptied the Mars and main asteroid belt regions, solving the Small Mars Problem too \citep{clement2018mars}. 
Numerical experiments exploring the timing of the giant planet instability indicate that the best matches to the Solar System date the instability to the first 10 million years \citep{clement2018mars,clement2019early}, possibly caused by the dissipation of the gas disk itself \citep{Liu2022}.

Alternatively, giant planet evolution and the inner disk may have evolved relatively separately \citep[e.g.,][]{nesvorny2021role}, and the low mass in the outer parts of the inner disk was inherited from evolution of the protoplanetary disk itself \citep[e.g.,][]{hansen2009formation,izidoro2014terrestrial,raymond2017empty,izidoro2021effect,nesvorny2021role,lykawka2023terrestrial,Nesvorny2025terrestrial}.
The exact rationale for a narrow annulus may be only qualitatively motivated by a specific hypothesis \citep[e.g.,][]{izidoro2014terrestrial,raymond2017empty,lykawka2023terrestrial} but often quantitative models invoke nebular condensation fronts associated with water, organics, and/or silicates \citep[e.g.,][]{lichtenberg2021bifurcation,izidoro2021effect,morbidelli2022contemporary}. 
Indeed, distant protoplanetary disks show substructures indicating strong radial solid mass surface density concentrations and depletions, sometimes invoking condensation fronts and sometimes invoking nearby giant planets \citep[e.g.,][]{Andrews2018}. 
If invoked, the consequent annuli can be tuned to accurately produce the terrestrial planets, including a solution to the Small Mars Problem \citep[e.g.,][]{hansen2009formation,izidoro2014terrestrial,raymond2017empty,izidoro2021effect,nesvorny2021role,lykawka2023terrestrial,Nesvorny2025terrestrial}.
Some Annulus scenarios begin with an asteroid belt that has no material at all \citep[e.g.,][]{raymond2017empty}, and the asteroid belt is built entirely from scattered material.
Other Annulus scenarios assert a low-mass disk of planetesimals in addition to a more massive annulus of protoplanets \citep[e.g.,][]{Nesvorny2025terrestrial}.

\GIN{Yet another proposed planet formation scenario is pebble accretion \citep[e.g.,][]{Lambrechts2019,Bizzarro2025}.
This formation mechanism involves accretion of pebble-sized objects to form the cores of planets on a timescale orders of magnitude more rapid than that of traditional planetesimal accretion.
If the terrestrial planets were formed solely from pebble accretion \citep[e.g.,][]{johansen2021pebble}, then one would not need to consider multistage accretion. 
However, forming the terrestrial planets without multistage accretion seems very unlikely given the Moon-forming impact and other geochemical evidence that the terrestrial planets did grow via collisions between planetary embryos, which indicate longer timescales of accretion than are possible with pebble accretion \citep[e.g.,][]{zube2019constraints,morbidelli2025did,Yap2025}.
Thus, we do not consider pebble accretion scenarios here and, instead, focus on planetesimal accretion scenarios.
The simulations we use in this study reproduce the general features of planet formation scenarios that contain a giant impact epoch, regardless of whether the initial planetary embryos are created due to runaway growth by planetesimal accretion \citep[e.g.,][]{morbidelli2012building,jacobson2014highly} or pebble accretion \citep[e.g.,][]{Lambrechts2019,Bizzarro2025}.}

We hypothesize that these different \GIN{planetesimal accretion} terrestrial planet formation scenarios have very different implications for geochemical gradients across the inner disk.
The Grand Tack and Early Instability scenarios share a characteristic similar to the Classical scenario that the primary inner disk building blocks originate from a wide multi-AU region, extending typically from about 0.5 to 3 AU.
However, unlike the Classical Scenario, the giant planets dynamically stir the inner disk much more vigorously, enhancing radial mixing across any radial geochemical gradient.
All Annulus scenarios share the characteristic that the primary inner disk building blocks originate from a sub-AU region, often posited to be just a few tenths of an AU.
This would seem to imply a homogenized geochemical reservoir since dynamical radial mixing of planetesimals is a rapid process \GIN{\citep[e.g.,][]{obrien2006terrestrial,Kaib2015}}, and so it is unclear if a geochemical gradient could establish itself across such a narrow annulus.
Perhaps, a possible solution is to invoke two or more narrow annuli each associated with different condensation fronts \citep[e.g.,][]{morbidelli2022contemporary,Nesvorny2025terrestrial}.
For now, we focus on whether it is possible to reproduce the Earth's mantle in Earth-like planets from each terrestrial planet scenario and, if so, determine what geochemical gradient is required to reproduce the Earth's mantle composition.

\begin{figure}[pos=bt!]
    \centering
    \includegraphics[width=0.95\columnwidth]{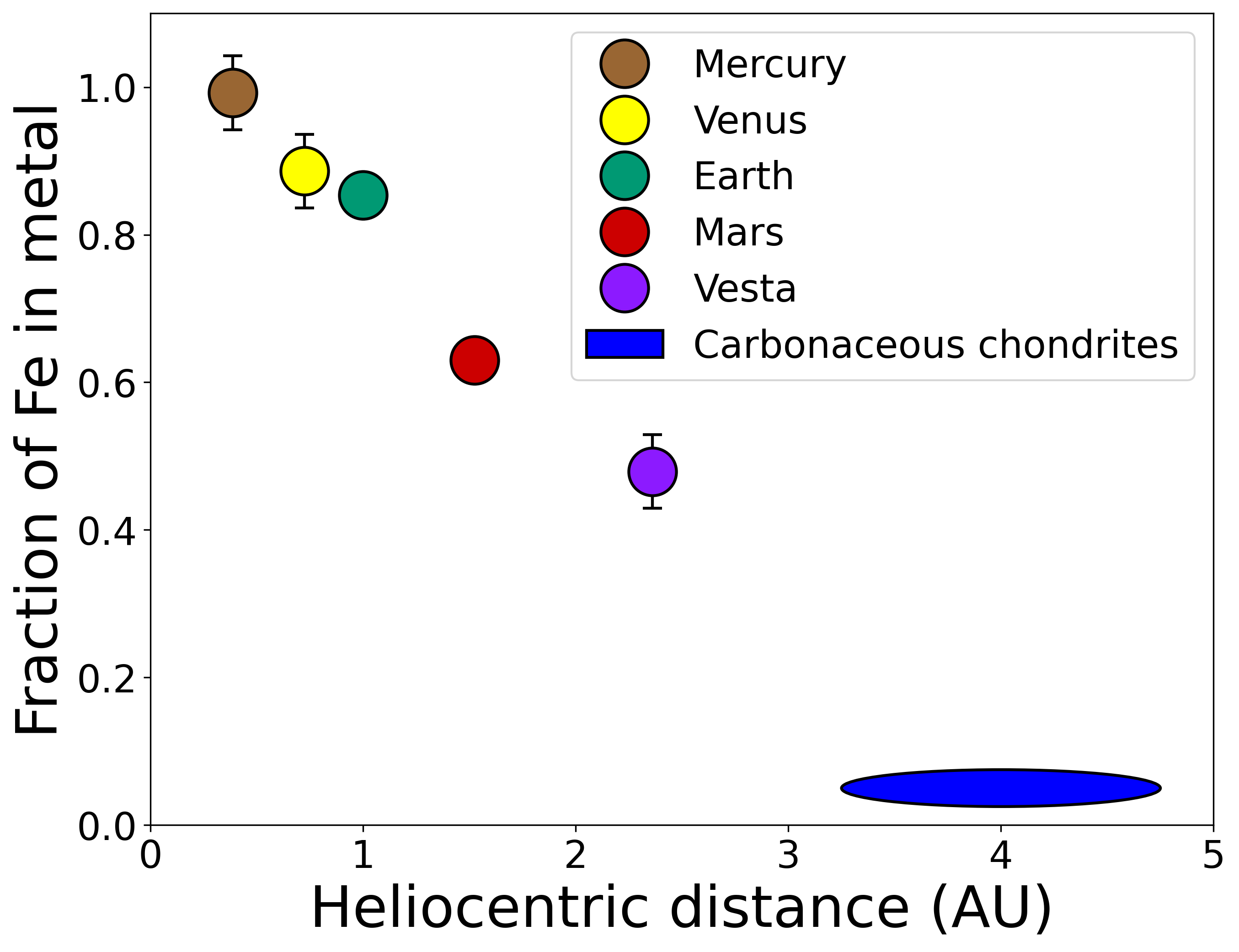}
    \caption{Representative estimates of observed Fe-in-metal fraction for a range of planetary bodies show a trend of increasingly oxidized bulk composition at greater heliocentric distances.
    Mercury, for example, has a large iron core ($\sim$70\% of Mercury's total mass) and a small amount of FeO in its mantle ($\sim$2 wt\% of mantle), meaning nearly all of its Fe is reduced \GIN{\citep{Hauck2013,Nittler2011}}.
    \GIN{While the core and mantle of these bodies are not in complete equilibrium, they are indicative of average and evolving redox states of the accreted material, and the observed trend suggests the presence of a redox gradient throughout the early protoplanetary disk.
    The fraction of a body's total Fe inventory residing in metal is estimated as $f_{\mathrm{Fe,metal}} = 0.85\,f_{\mathrm{core}}/\left[0.85\,f_{\mathrm{core}} + 0.78\,X_{\mathrm{FeO}}^{\mathrm{mantle}}\,(1-f_{\mathrm{core}})\right]$, where $f_{\mathrm{core}}$ is the body's core mass fraction, $X_{\mathrm{FeO}}^{\mathrm{mantle}}$ is its mantle FeO mass fraction, the factor 0.85 assumes Fe constitutes 85 wt\% of the core (the remainder being light elements), and the factor 0.78 converts FeO mass to Fe mass.
    This figure is adapted from Figure 1B in \citet{righter2020redox}, which shows trends in mantle FeO wt\% and core mass fraction among inner Solar System planetary bodies.}}
    \label{fig:fe-vs-au}
\end{figure}

\subsection{Redox states in the Solar System} \label{sec:solarsystemredoxstates}

The redox states of Solar System materials vary greatly, ranging from IW-8 to IW+2 \citep{righter2016redox}. 
The meteoritic record indicates there were heterogeneous formation environments in the proto Solar System \citep{jones2012petrographic, zhang2024compositions}.
One of the major planetary-scale geochemical trends in the Solar System is the apparent oxidation trend between the inner terrestrial planets. 
Mercury, the planetary body nearest the Sun, is highly reduced, with nearly all of its total Fe in metallic form \GIN{\citep{Hauck2013,Nittler2011}}. 
The meteoritic record and the variety of planetary materials suggests existence of a compositional redox gradient in the early protoplanetary disk. 
Evidence of varied redox state \GIN{is} found in the meteoritic record and in the estimated bulk iron composition of major Solar System bodies, shown in Figure~\ref{fig:fe-vs-au}. 
Thus, the source of variation in oxidation states amongst Solar System material is likely a combination of inheritance from heterogeneous building blocks sourced from different environments including the incorporation of differing amounts of water \citep{Ciesla2006,Grewal2024} and possible condensation processes \citep{Charnoz2026} as well as modification by fractionating planetary processes such as multi-stage core formation \citep{galimov2005redox} as well as contemporaneous Fe disproportionation within the mantle \GIN{\citep{wordsworth2018redox, armstrong2019deep}}.
The combination of the varied initial redox states of material in the protoplanetary disk and the subsequent dynamical evolution of those bodies has wide-ranging implications for the geochemical evolution of the Earth, including for the subset of contributors to the Earth such as Proto-Earth, Theia, and any late accreted material to the Earth. 

\subsection{Theia and Proto-Earth} \label{sec:theiaintroduction}

\GIN{The formation of the Earth-Moon system is frequently hypothesized to be the result of a giant impact \citep{hartmann1975lunar,cameron1976origin}, but both the physical and chemical properties of the impactor are debated. 
The giant impact hypothesis is compelling because it directly explains the angular momentum state of the Earth-Moon system without invoking subsequent dynamics to drain excess angular momentum from the Earth-Moon system; these constraints prompted the canonical theory of lunar formation involving a roughly Mars-sized body, Theia, impacting Proto-Earth \citep{Benz1986,canup2001origin}.}
However, dynamical interactions in the Sun-Earth-Moon system, such as the evection resonance, can transfer angular momentum from the Earth-Moon orbit to the Sun-Earth orbit \citep{cuk2012making,wisdom2015early,cuk2016tidal,tian2017coupled,tian2020vertical,cuk2021tidal}.
This opens up the possibility of alternative high-energy and angular momentum scenarios such as a `hit-and-run' \citep{reufer2012hit}, an impactor on a rapidly spinning Earth \citep{cuk2012making}, a collision between a Proto-Earth and Theia of similar size \citep{canup2012forming}, a series of impacts \citep{rufu2017multiple}, and a very high energy impact forming a planetary `synestia' \citep{lock2018origin}.
All of these various formation scenarios leave the question of the size of the Moon-forming impactor, Theia, unanswered. 
However, in all proposed lunar formation scenarios, a non-zero portion of the Earth is derived from Theia, providing a potential constraint on the viable size and chemistry of Theia.
 
The composition of the Moon provides potential clues to its provenance.
For instance, the Moon has an elevated mantle FeO abundance \citep[9.24 wt.\%;][]{warren2005new} compared to the Earth \citep[8.05 wt.\%;][]{mcdonough1995composition}, and a very small metallic core \GIN{\citep[$\lesssim2$ wt.\%;][]{Hood2000, gaffney2023magmatic}}.
In the canonical Moon-forming impact hypothesis much of the mass of the Moon is derived from Theia \citep{Benz1986,canup2001origin}, so a straightforward solution to the elevated lunar FeO composition is a scenario where Theia possessed an elevated FeO mantle abundance \citep[$\sim$20 wt.\%;][]{Meier2014}.
However, not all Earth-Moon formation models propose an oxidized Theia; for example \citet{Wade2016} propose a very reduced Theia ($\sim$0.3 wt.\% FeO mantle) impacting a moderately oxidized Proto-Earth ($\sim$10.7  wt.\% FeO mantle).

The Earth and Moon are isotopically identical across a wide range of isotopic systems, further complicating the potential origin of this system. 
Apollo samples show that lunar and terrestrial oxygen isotopic compositions (ratios of $^{16}$O, $^{17}$O, $^{18}$O) are identical within measurement error \citep{wiechert2001oxygen,spicuzza2007oxygen,hallis2010oxygen,young2016oxygen}. 
This isotopic similarity extends to many other isotopic systems, even among more refractory elements, such as Cr \citep{lugmair1998early}, and Zr \citep{schonbachler2005nb,akram2016zirconium}, W \citep{touboul2007late}, Si \citep{georg2007silicon,fitoussi2012silicon,armytage2011silicon,armytage2012silicon}, and Ti \citep{zhang2012proto}, as well as to nucleosynthetic $\mu^{54}\text{Fe}$ isotopic anomalies \citep{hopp2025moon}.
More precise measurements of lunar W isotopes indicate a slight excess of $^{182}\text{W}$ in lunar material compared to terrestrial samples, but this may be due to variable late accretion on the Earth and Moon, rather than a compositional difference between the initial Proto-Earth and Theia \citep{touboul2015tungsten,kruijer2015lunar}. 
These geochemical observations have strong repercussions for lunar origin hypotheses.

Typically, simulations of lunar impacts show that most of the Moon originates from Theia material \citep{canup2001origin,Canup2004,canup2008lunar}, so this means that the Moon would obtain an isotopic signature that \GIN{reflects} Theia's composition.
Theia's composition is typically presumed to be distinct from Earth's because all other differentiated bodies in the Solar System have isotopic compositions distinct from Earth's, in particular, inner Solar System bodies like Mars and Vesta possess resolvable differences with Earth \citep[for review, see][]{dauphas2016mass}.

To reconcile the isotopic similarity of the Earth and Moon with proposed Moon-forming impact scenarios, three solutions have been proposed.
One approach is to assert that there was already a high degree of initial isotopic similarity between Proto-Earth and Theia \citep{dauphas2017isotopic}. 
A Theia with Earth-like isotopic composition and elevated mantle FeO content is the most straightforward way to reconcile the Earth-Moon isotopic compositions \citep{Meier2014}. 
The isotopic composition of Earth is similar to enstatite chondrites across many isotopic systems \citep{javoy1995integral,javoy2010chemical,dauphas2016mass,dauphas2017isotopic}, although not all \citep{fitoussi2012silicon,hopp2025moon}.
However, given the wide variety of isotopic compositions amongst Solar System materials, the odds that the Earth and Theia are isotopically identical to one another appear low \citep{pahlevan2007equilibration}.
\citet{Kaib2015} estimates the likelihood that Proto-Earth and Theia had the same stable isotopic composition to be between 5-8\%, while others have argued that this possibility is as high as a $\sim$20\% chance \citep{Mastrobuono-Battisti2015}.
Alternatively, in a pebble accretion scenario, Venus and Theia would obtain quite similar compositions to that of the Earth as they would grow from the same pebble flux source \citep[e.g.,][]{Johansen2021}. 
However, the radiometric W system would also require that the Proto-Earth and Theia have the same core formation history prior to impact---a challenging proposition unless the two bodies have very similar sizes and accretion histories \citep{kruijer2015lunar,Johansen2021}.
In this approach, the Proto-Earth and Theia originate from isotopically similar initial materials.

Alternatively, a second solution appeals to physical mixing of Proto-Earth and Theia material during the impact itself to create a circumplanetary disk with the same mix of Proto-Earth and Theia material as the Earth itself.  
Since Theia must contribute as much to Earth as the Proto-Earth, this requires the collision and physical mixing of nearly equal-sized Theia and Proto-Earth \citep{canup2012forming}. 
A complete physical mixing has the consequence of erasing any isotopic compositional differences post-impact regardless of any initial difference between Theia and the Proto-Earth. 
Thus, in this approach, the Proto-Earth and Theia can originate from very isotopically distinct initial materials.

Lastly, the third solution to this problem appeals to thermodynamic equilibration between the isotopically distinct Earth and the post-impact circumplanetary disk, removing isotopic differences by the time the Moon accretes out of the disk. 
This supposes isotopic equilibration occurred in a molten proto-lunar disk due to high energy conditions following the Moon-forming impact \citep[e.g.,][]{pahlevan2007equilibration}.
Alternatively, higher energy impacts could create the formation of a planetary synestia, which results in efficient chemical equilibration \citep{lock2018origin}.
These post-impact processes could provide the appropriate degree of chemical and isotopic equilibration between Earth and post-impact circumplanetary material prior to the accretion of the Moon.
Thus, in this approach, the Proto-Earth and Theia can originate from moderately isotopically distinct initial materials, depending on the precise details of the proposed equilibration scheme.

Using coupled \textit{N}-body formation models and geochemical differentiation modeling, we simultaneously study both Solar System formation and lunar formation, with the chemistry of the bulk silicate Earth serving as the primary constraint, the composition of which is well characterized compared to that of other planetary bodies. 
Because the present-day Earth's mantle composition is a combined mixture of Proto-Earth, the Earth just before the Moon-forming impact, and Theia, the Moon-forming impactor, the mantle of the Earth must contain a chemical record of the Moon-forming impact. 
If certain combinations of Theia and Proto-Earth are or are not compatible with the present-day Earth's mantle composition, this can constrain the qualities of the Moon-forming impact. 
Given reasonable starting conditions for the composition of the Solar System, we determine both the viable oxidation state of the early Solar System, and the characteristics of the Moon-forming impactor, Theia. 

\section{Methods}
\label{sec:methods}

\begin{table*}[tb!]
\centering
\begin{ThreePartTable}
\caption{Summary of the initial conditions of the simulations used in this study.
Disk size refer to the region in which bodies are initially populated in each simulation suite.
Disk $\alpha$ describes the surface-density power-law slope of the protoplanetary disk.
The initial population of objects is characterized by the total mass (in Earth masses, $M_\oplus$) and number of embryos, which gravitationally self-interact with all bodies in the simulation, and the total mass and number of planetesimals, which do not gravitationally self-interact with other planetesimals.
Not all simulation runs produced Earth analogs. These simulations listed here were published previously \citep{O'Brien2014, walsh2011low,jacobson2014lunar,izidoro2015terrestrial,nesvorny2021role} or are reproductions of prior published work \citep{raymond2009building,hansen2009formation}, as indicated in the table. \label{table_of_simulations}}
% \begin{longtable}{@{}P{1.8cm}P{1.0cm}P{0.8cm}P{1.4cm}P{2.2cm}P{0.8cm}P{1.4cm}P{2.1cm}P{2.1cm}P{2.4cm}@{}}
\begin{tabular*}{\textwidth}{@{}s{1.6cm}s{0.8cm}s{0.6cm}s{0.9cm}s{3.3cm}s{0.7cm}s{1.0cm}s{2.75cm}s{1.6cm}s{2.75cm}@{}}
\toprule
Scenario & Name & Runs & Earth(s) & Disk size (AU) & Disk $\alpha$ & \GIN{Number of embryos} & \GIN{Individual embryo masses} & \GIN{Number of planetesimals} & \GIN{Individual planetesimal masses} \tn
\midrule
%\endfirsthead
% ---------- HEADER FOR PAGE 2+ ----------
%\multicolumn{10}{c}{
%  \textbf{Table \thetable\ (continued)}: Summary of types of simulations and their characteristics
%} \tn
%\toprule
%Scenario Type & name & \# of runs & \# of Earth-like & Disk Limits (AU) & Disk $\alpha$ & \# Embryos & Mass in Embryos & \# Planetesimals & Mass in Planetesimals \tn
%\midrule
%\endhead
Circular Jupiter-Saturn  & CJS\tnote{$\parallel$}  & 50 & 20 & 0.5--4.0 & 3/2 & 100 & 0.005--0.1 $M_\oplus$\tnote{$\dagger\dagger$} & 1000 &  0.0025 $M_\oplus$ \tn  % CJS-*  
Eccentric Jupiter-Saturn & EJS\tnote{$\parallel$}  & 50 & 11  & 0.5--4.0 & 3/2 & 100 & 0.005--0.1 $M_\oplus$\tnote{$\dagger\dagger$} & 1000 &  0.0025 $M_\oplus$ \tn  % EJS-*
Truncated Disk           & TD\tnote{$\Delta$}      & 50 & 25  & 0.7--1.0 & 0 & 400 & 0.005 $M_\oplus$\tnote{$\dagger\dagger$} & 0 &  -- \tn  % TD-*
Steep Profile            & SP\tnote{$\circ$}       & 15 & 6  & 0.7--4.0 & 5.5 & 60 & 0.00075 $M_\oplus$ -- 0.4 $M_\oplus$  \tnote{$\dagger\ast$} & 1282 &  0.00075 $M_\oplus$ \tn  % SP-*
Ring Model               & Ring\tnote{$\ast\ast$}  & 22 & 24  & 0.7--1.0 (embryos), 0.3--4.0 (planetesimals) & 1 & 100 & 0.02 $M_\oplus$   & 1000 & 0.002 \tn  % ring-* get from Nesvorny output (and send to seth)
Grand Tack & GT\_1                                 & 10 & 5  & 0.4\tnote{$\ast$}, 0.7--3.0 (inner), outer disk\tnote{$\dagger$} & 3/2 & 56 & 0.055 $M_\oplus$, 0.08 $M_\oplus$ & 2782, 500\tnote{$\dagger$} &  0.0004 $M_\oplus$ (inner), 0.0007 $M_\oplus$ (outer) \tn % intGT5.3-4to1-0.8 
           & GT\_2                                 & 10 & 4  & 0.4\tnote{$\ast$}, 0.7--3.0 (inner), outer disk\tnote{$\dagger$} & 3/2 & 171 & 0.055 $M_\oplus$, 0.025 $M_\oplus$ & 2782, 500\tnote{$\dagger$} &  0.0004 $M_\oplus$, 0.0007 $M_\oplus$ \tn % intGT5.3-4to1-0.25 
           & GT\_3a\tnote{$\ddagger$}              & 4  & 1   & 0.7--3.0 & 3/2 & 37 & 0.05 $M_\oplus$ & 727 &  0.003 $M_\oplus$ \tn % newGT4.3-1to1-0.5 runs 1 through 4
           & GT\_3b\tnote{$\S$}                    & 8  & 5  & 0.7--3.0 (inner), outer disk\tnote{$\dagger$} & 3/2 & 37 & 0.05 $M_\oplus$ & 727, 767\tnote{$\dagger$} &  0.003 $M_\oplus$ \tn % newGT4.3-1to1-0.5 runs 5 through 12 (O'Brien)
           & GT\_4\tnote{$\#$}                     & 10 & 2  & 0.7--3.0 (inner), outer disk\tnote{$\dagger$} & 3/2 & 27 & 0.08 $M_\oplus$ & 5252, 1563 &  0.0004 $M_\oplus$, 0.00007 $M_\oplus$ \tn % newGT4.3-1to1-0.8
           & GT\_5a\tnote{$\ddagger$}              & 4  & 2   & 0.7--3.0 & 3/2 & 74 & 0.025 $M_\oplus$ & 727 &  0.003 $M_\oplus$ \tn % newGT4.3-1to1-0.25 runs 1 through 4
           & GT\_5b\tnote{$\S$}                    & 8  & 2  & 0.7--3.0 (inner), outer disk\tnote{$\dagger$} & 3/2 & 74 & 0.025 $M_\oplus$ & 727, 766 &  0.003 $M_\oplus$, 0.0002 $M_\oplus$ \tn  % newGT4.3-1to1-0.25 runs 5 through 12  (O'Brien)          
           & GT\_6\tnote{$\#$}                     & 10 & 2  & 0.7--3.0 (inner), outer disk\tnote{$\dagger$} & 3/2 & 66 & 0.05 $M_\oplus$ & 4289, 1563 &  0.0004 $M_\oplus$, 0.00007 $M_\oplus$ \tn  % newGT4.9-2to1-0.5
           & GT\_7\tnote{$\#$}                     & 10 & 3  & 0.7--3.0 (inner), outer disk\tnote{$\dagger$} & 3/2 & 42 & 0.08 $M_\oplus$ & 4289, 1563 &  0.0004 $M_\oplus$, 0.00007 $M_\oplus$ \tn  % newGT4.9-2to1-0.8
           & GT\_8\tnote{$\#$}                     & 10 & 3  & 0.7--3.0 (inner), outer disk\tnote{$\dagger$} & 3/2 & 123 & 0.025 $M_\oplus$ & 4026, 1563 &  0.0004 $M_\oplus$, 0.00006 $M_\oplus$ \tn  % newGT4.9-2to1-0.25
           & GT\_9\tnote{$\#$}                     & 10 & 8  & 0.7--3.0 (inner), outer disk\tnote{$\dagger$} & 3/2 & 87 & 0.05 $M_\oplus$ & 2836, 1563 &  0.0004 $M_\oplus$, 0.00005 $M_\oplus$ \tn  % newGT5.3-4to1-0.5
           & GT\_10\tnote{$\#$}                    & 10 & 2  & 0.7--3.0 (inner), outer disk\tnote{$\dagger$} & 3/2 & 55 & 0.08 $M_\oplus$ & 2783, 500 &  0.0004 $M_\oplus$, 0.0005 $M_\oplus$ \tn  % newGT5.3-4to1-0.8
           & GT\_11\tnote{$\#$}                    & 10 & 3  & 0.7--3.0 (inner), outer disk\tnote{$\dagger$} & 3/2 & 170 & 0.025 $M_\oplus$ & 2783, 1563 &  0.0004 $M_\oplus$, 0.00005 $M_\oplus$ \tn  % newGT5.3-4to1-0.25
           & GT\_12\tnote{$\#$}                    & 20 & 9  & 0.7--3.0 (inner), outer disk\tnote{$\dagger$} & 3/2 & 107 & 0.05 $M_\oplus$ & 1750, 500 &  0.0004 $M_\oplus$, 0.0004 $M_\oplus$ \tn  % newGT6.0-8to1-0.5
           & GT\_13\tnote{$\#$}                    & 20 & 5  & 0.7--3.0 (inner), outer disk\tnote{$\dagger$} & 3/2 & 68 & 0.08 $M_\oplus$ & 1750, 500 &  0.0004 $M_\oplus$, 0.0004 $M_\oplus$ \tn  % newGT6.0-8to1-0.8
           & GT\_14\tnote{$\#$}                    & 10 & 3  & 0.7--3.0 (inner), outer disk\tnote{$\dagger$} & 3/2 & 213 & 0.025 $M_\oplus$ & 1750, 500 &  0.0004 $M_\oplus$, 0.0003 $M_\oplus$ \tn  % newGT6.0-8to1-0.25
           & GT\_15                                & 10 & 1  & 0.3--3.0 (inner), outer disk\tnote{$\dagger$} & 3/2 & 59 & 0.02--0.08 $M_\oplus$\tnote{$\dagger\dagger$} & 2714, 500\tnote{$\dagger$} &  0.0004 $M_\oplus$, 0.0005 $M_\oplus$ \tn % merGT3.1-2to1-0.8 (variations on things from jacobson2014)
           & GT\_16                                & 10 & 2  & 0.3--3.0 (inner), outer disk\tnote{$\dagger$} & 3/2 & 80 & 0.02--0.08 $M_\oplus$\tnote{$\dagger\dagger$} & 1839, 500\tnote{$\dagger$} &  0.0004 $M_\oplus$, 0.0002 $M_\oplus$ \tn % merGT3.5-4to1-0.8 (variations on things from jacobson2014)
           & GT\_17\tnote{$\#$}                    & 10 & 2  & 0.5--3.0 (inner), outer disk\tnote{$\dagger$} & 3/2 & 56 & 0.02--0.08 $M_\oplus$\tnote{$\dagger\dagger$} & 3065, 500 &  0.0004 $M_\oplus$, 0.0005 $M_\oplus$ \tn  % upGT3.5-2to1-0.8
           & GT\_18\tnote{$\#$}                    & 10 & 2  & 0.5--3.0 (inner), outer disk\tnote{$\dagger$} & 3/2 & 82 & 0.02--0.08 $M_\oplus$\tnote{$\dagger\dagger$} & 2260, 500 &  0.0004 $M_\oplus$, 0.0003 $M_\oplus$ \tn  % upGT4.3-4to1-0.8 (up means still alpha 3/2 but embryo mass is increased)
           & GT\_19\tnote{$\#$}                    & 10 & 2  & 0.7--3.0 (inner), outer disk\tnote{$\dagger$} & 3/2 & 63 & 0.02--0.08 $M_\oplus$\tnote{$\dagger\dagger$} & 3853, 500 &  0.0004 $M_\oplus$, 0.001 $M_\oplus$ \tn  % venGT4.4-2to1-0.8          
           & GT\_20\tnote{$\#$}                    & 10 & 3  & 0.7--3.0 (inner), outer disk\tnote{$\dagger$} & 3/2 & 84 & 0.02--0.08 $M_\oplus$\tnote{$\dagger\dagger$} & 2575, 500 &  0.0004 $M_\oplus$, 0.0004 $M_\oplus$ \tn  % venGT4.9-4to1-0.8  
           & GT\_21\tnote{$\#$}                    & 10 & 7  & 0.7--3.0 (inner), outer disk\tnote{$\dagger$} & 3/2 & 104 & 0.02--0.08 $M_\oplus$\tnote{$\dagger\dagger$} & 1606, 500 &  0.0004 $M_\oplus$, 0.0004 $M_\oplus$ \tn  % venGT5.5-8to1-0.8  
\bottomrule
\end{tabular*}
\begin{tablenotes}
\item[$\ast$] For these runs there is an interior embryo     %\item[a] 
\item[$\dagger$] The outer disk objects are from \citet{O'Brien2014}   %\item[b]  
\item[$\ddagger$] Simulations from \citet{walsh2011low}   %\item[c]  
\item[$\S$] Simulations from \citet{O'Brien2014}   %\item[d]  
\item[$\#$] Simulations from \citet{jacobson2014lunar}   %\item[e]  
\item[$\dagger\dagger$] Embryo mass increases with heliocentric distance   %\item[f]  
\item[$\parallel$] Identical to \citet{raymond2009building}   %\item[g] %CJS EJS 
\item[$\Delta$] Based on \citet{hansen2009formation}   %\item[h]  % Truncated Disk
\item[$\circ$] From \citet{izidoro2015terrestrial}   %\item[i]   %Steep profile
\item[$\dagger\ast$] Embryo mass decreases with heliocentric distance   %\item[j]  
\item[$\ast\ast$] From \citet{nesvorny2021role}   %\item[k]  %Ring Model
\end{tablenotes}
\end{ThreePartTable}
\end{table*}

In order to assess the geochemistry of Earth-like planets in many different terrestrial planet formation scenarios, the growth histories of Earth-like planets from astrophysical \textit{N}-body simulations are post-processed with a planetary accretion and differentiation model.
The model has a significant legacy reproducing aspects of Solar System history \citep[e.g.,][]{rubie2011heterogeneous, rubie2015accretion,jennings2021metal,nathan2023constraining}.
It assigns a starting composition to each protoplanet from the \textit{N}-body simulation based on the body's location and a modified chondritic composition. 
Planetary bodies collide and grow as prescribed by a given \textit{N}-body accretion history.
Compositions of the growing planets evolve due to core formation events resulting from impact-generated melt; core-forming liquids from projectile bodies equilibrate with a fraction of the silicate mantle of the target body and silicate material from the projectile mixes with the unmelted portion of the target body mantle.
After the final accretion event, the composition of the Earth-like planet's mantle is compared to that of the bulk silicate Earth using a goodness-of-fit metric.
Critically, the planetary accretion and differentiation model produces not only a final Earth-like planet but it also reports the compositions of the Proto-Earth, Theia, and late accreting impactors. 

\subsection{Astrophysical \textit{N}-body simulations} \label{sec:nbody}

All the \textit{N}-body simulations used in this work are organized into suites of simulations which have similar initial conditions, and are summarized in Table~\ref{table_of_simulations}.
They are either from the published literature \citep{raymond2009building,hansen2009formation,O'Brien2014, walsh2011low,jacobson2014lunar,izidoro2015terrestrial,nesvorny2021role} or, in the case of some simulations, very similar to published simulations from \citet{raymond2009building,hansen2009formation,jacobson2014lunar}.

These suites of simulations span the three main classes of formation scenarios, as briefly reviewed in Section~\ref{sec:background}: the Classical scenario simulations include the Circular Jupiter-Saturn (CJS) and Eccentric Jupiter-Saturn (EJS) \citep{raymond2009building}; the Annulus scenario simulations include Truncated Disk (TD) \citep{hansen2009formation}, Steep Profile (SP) \citep{izidoro2015terrestrial}, and Ring (ring) \citep{nesvorny2021role}; and the Grand Tack (GT) scenario \citep{walsh2011low, O'Brien2014, jacobson2014lunar}, which includes some unpublished Grand Tack simulations with similar initial conditions to published studies.
All simulation suites are described in Table~\ref{table_of_simulations}, which lists the initial conditions of literature simulations in the study and the number of Earth analogs from those simulations. 
Scenario types differ based on the number of initial embryos and planetesimals, the ratio of total mass in embryos and planetesimals, the disk limits, and the disk surface density $\alpha$. 
Disk limits are the semi-major axis range of the initial planetesimal and embryo population in a simulation. 
In most simulations, initial bodies are not distributed uniformly; the disk $\alpha$ prescribes the surface-density power-law distribution ($\Sigma(r)\propto r^{-\alpha}$) of planetesimals and embryos in the disk, where $r$ is heliocentric distance.

The initial numbers and total mass of planetesimals and embryos for each simulation suite are also shown in Table~\ref{table_of_simulations}.
In \textit{N}-body simulations, planetary embryos are larger than planetesimals and gravitationally self-interact with all other bodies in the simulation.
Each individual planetesimal, in contrast, represents a swarm of smaller bodies which interact with embryos but do not gravitationally self-interact with other planetesimals.
Though simulated planetesimals represent a small population of bodies, nearly all of the mass is concentrated in the largest of the swarm. 
Thus, we reasonably assert both (1) that planetesimals contain cores if metal is present and (2) that melting occurs during planetesimal impacts.

From these astrophysical \textit{N}-body simulations, 164 viable Earth analogs \GIN{are} selected from a larger set of Solar System formation simulations from either the existing literature or reproduced based on the existing literature.
We \GIN{define} an Earth-like planet as a body with a final mass between 0.75 and 1.55 $M_\oplus$ that accretes on a timescale of 50 to 200 million years, where $M_{\oplus}$ is one Earth mass. 
We consistently \GIN{apply} this definition across all scenarios, which has two consequences: (1) the Earth-like planets identified in this paper are not necessarily the same ones identified in the original studies and (2) different scenarios have different yields of Earth-like planets.
It is even the case that some simulations \GIN{do} not produce any Earth-like planets, and so no planet from that simulation is included in this study.
\GIN{In contrast, certain simulations may produce more than one Earth-like planet.
To focus on the role of feeding zones and accretion history, the mass of each Earth-like planet is normalized to an Earth mass, and that same normalization factor is applied to each of its building blocks.
The range of Earth analog masses we consider is between $0.75 M_\oplus$ and  $1.55M_\oplus$. 
As such, the normalization factors range between $1.33$ and $0.65$. 
In this way, we are then comparing Earth-like planets, each of an Earth mass.}

\begin{figure*}[pos=bt!]
    \centering
    \includegraphics[width=0.95\textwidth]{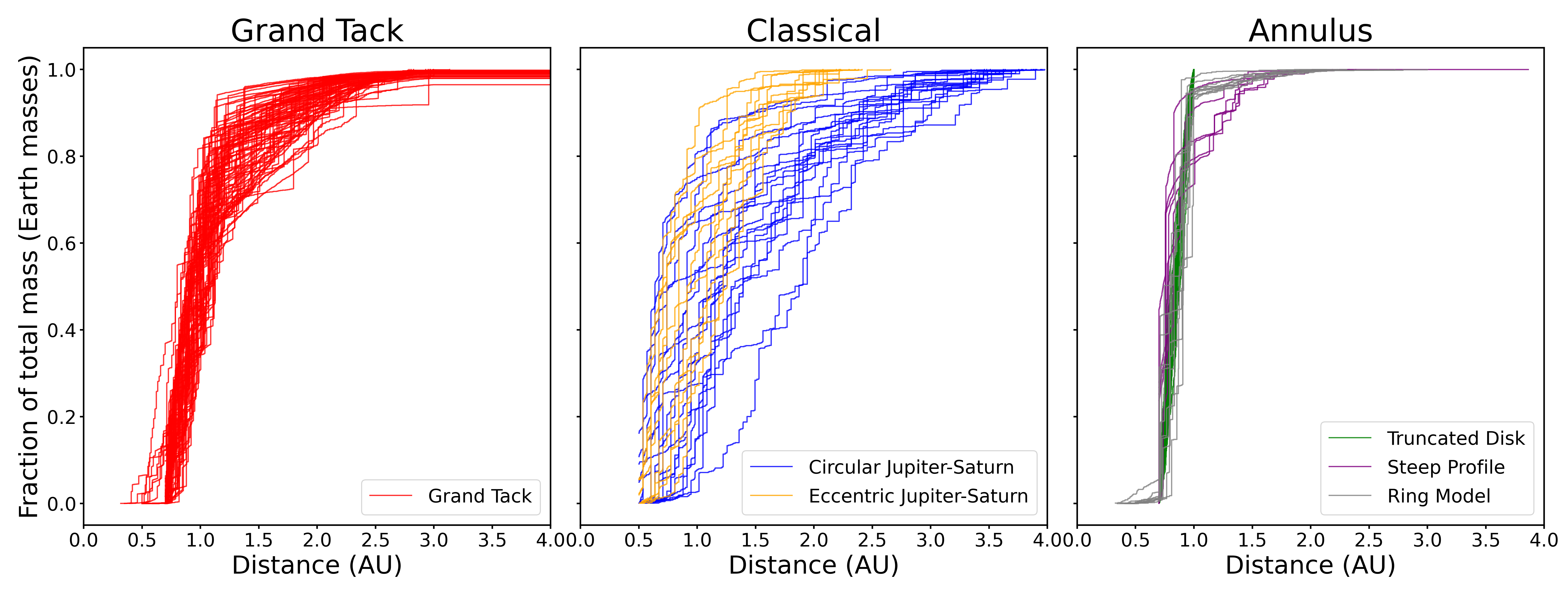}
    \caption{The feeding zone of each Earth-like planet shown as a cumulative distribution function (CDF) of the normalized planetary mass as a function of the initial semi-major axis location of each building blocks.
    The initial location distribution of planetary building blocks is the primary difference between the terrestrial planet formation scenarios.
    Colors correspond to the class of formation scenario to which a simulation belongs: Classical model (Circular Jupiter Saturn (CJS) and Eccentric Jupiter Saturn (EJS)), Annulus model (Truncated Disk (TD), Steep Profile (SP), and Ring), and Grand Tack model (GT).}
    \label{fig:3subplots_Earth_Mass_CDF_Grouped}   
\end{figure*}

The Earth-like planets in this study are built from material that samples portions of initial disk of the Solar System in the \textit{N}-body simulations used here, corresponding to how the material in the disk is initially distributed.
The cumulative fraction of the material that contributes to the Earth analog and originates interior to a given heliocentric distance varies both on the formation scenario and among individual simulations within a scenario type. 
A wide distribution of material (seen in Grand Tack (GT) and Classical (CJS, EJS) Types) indicates that the Earth analog forms from material from throughout the inner Solar System.
Narrower distributions (seen in the Annulus Types (TD, SP, ring)) indicate that the Earth analog is built from material that initially formed in a more restricted portion of the protoplanetary disk. 
Cumulative distribution functions showing the distribution of mass that builds the Earth in all simulations used in this study are shown in Figure~\ref{fig:3subplots_Earth_Mass_CDF_Grouped}.  

Each Earth-like planet accretes out of a population of protoplanets with a unique history due to the chaotic gravitational dynamics within the protoplanetary disk.
In other words, even within the same scenario, the number, timing, and characteristics of the accretion events (i.e., collisions) that occur during the growth of each simulated Earth-like planet are different.
A statistical approach is fraught, since one cannot assess whether the Solar System reflects an average outcome or a rare chance.
However, despite this stochasticity from one simulation to the next within a given planet formation scenario, past work has demonstrated that relationships are evident between the initial conditions and the final outcomes \citep[e.g.,][]{rubie2015accretion,gu2023comparisons}.
For this work, the initial radial compositional gradient through the protoplanetary disk is varied with the goal of creating Earth-like planets with mantles most similar to the Earth's, and then we compare both the quality of the match as well as the properties of the initial radial compositional gradient.

\subsection{Initial protoplanet compositions} \label{sec:protoplanet}

A bulk starting composition is assigned to all protoplanets in the protoplanetary disk based on a modified CI chondritic composition and their initial radial location in the disk.
\GIN{Elements in this version of the core-mantle differentiation model include the major elements: Fe, O, Ni, Si, Mg, Al, and Ca and the trace elements: Co, Nb, Ta, V, and Cr.}
Initial bulk planetary abundances (core and mantle combined) of refractory elements including Mg, Fe, Si, Ni, Co, Cr, Pd, and Mo are in CI-relative abundances (i.e., consistent with solar relative abundances), because these elements are not expected to fractionate with respect to one another within the protoplanetary disk \citep{Grossman1977,grossman2008redox}. 
\GIN{The Earth is enriched in ultra-refractory elements with respect to CI chondrites, indicated by the Al/Mg ratio (0.1072) of the primitive upper mantle, which is between that of CI chondrites (0.088) and CV chondrites (0.12) \citep{palme2014cosmochemical}.
\GIN{Following \citet{rubie2015accretion} and \citet{nathan2023constraining}, all initial protoplanets are therefore assigned ultra-refractory element abundances (e.g., Al, Ca, Nb, Ta) which are enhanced by 22\% relative to CI abundances.}
Vanadium, which is both volatile and refractory, is instead increased by 11\% because it is roughly half as enriched in CV chondrites compared to the other refractory elements \citep{rubie2011heterogeneous}.
This enrichment is necessary because there is no mix of chondritic material that can build the Earth; the refractory enrichment in the Earth mantle composition requires that a hidden reservoir must be invoked in order to produce the Earth \citep[e.g.,][]{drake2002determining}. 
An enstatite-like Earth is only possible when invoking a deep mantle reservoir that is highly fractionated with respect to the upper mantle \citep[e.g.,][]{javoy1995integral, javoy2010chemical}. 
Thus, we have chosen to use a refractory-enhanced CI composition because CI are the most abundant class of meteorites and some refractory enhancement is required no matter the starting reservoir composition. }
We do not include other minor or trace moderately volatile elements, because they are not primary building blocks of the terrestrial planets and even moderately volatile elements may be unpredictably affected by condensation processes within the protoplanetary disk; these elements are left for future work.

We separate the bulk composition into core and mantle reservoirs, and we calculate a modal mineralogy for every reservoir within each planetesimal and embryo according to mass balance and elemental partitioning. 
Oxygen content varies in the disk and affects the oxidation state of major rock-forming elements (e.g., Fe vs. FeO and Si vs. SiO$_2$) and is thus coupled with the mantle redox state of each planetary body. 
We report the major-rock forming elements as oxides within the mantle to track their presumed oxidation states and, therefore, the oxygen abundance of the mantle, but this modal mineralogy is not meant to reflect the mostly silicate mineralogy of the mantle, which will be multimineralic and vary with depth to first order.

From these bulk compositions, it is then straightforward to calculate the equilibration of silicate and metal liquids from the melted and interacting portions of planetary reservoirs: 
\begin{align}
\begin{split}
&\underbrace{[(\mathrm{FeO})_x (\mathrm{NiO})_y (\mathrm{SiO}_2)_z (\mathrm{Mg}_u \mathrm{Al}_m \mathrm{Ca}_n)\mathrm{O}]}_{\textit{unequilibrated silicate liquid}} \\
&\quad + \underbrace{[\mathrm{Fe}_a \mathrm{Ni}_b \mathrm{O}_c \mathrm{Si}_d]}_{\textit{unequilibrated metal liquid}} \\
&\longrightarrow \underbrace{[(\mathrm{FeO})_{x'} (\mathrm{NiO})_{y'} (\mathrm{SiO}_2)_{z'} (\mathrm{Mg}_{u} \mathrm{Al}_{m} \mathrm{Ca}_{n})\mathrm{O}]}_{\textit{equilibrated silicate liquid}} \\
&\quad + \underbrace{[\mathrm{Fe}_{a'} \mathrm{Ni}_{b'} \mathrm{O}_{c'} \mathrm{Si}_{d'}]}_{\textit{equilibrated metal liquid}}
\end{split}
\label{mass_balance_eq}
\end{align}
where the indices $a,\ b,\ c,\ d,\ x,\ y,\ z$ are the known starting quantities of these elements $u,\ m,\ n$ are constant and known from the starting bulk composition of the equilibrating materials because Mg, Al, and Ca are lithophile and will not partition to the metallic phase \citep{rubie2011heterogeneous}.
\GIN{Major elements (Fe, O, Ni, Si, Mg, Al, and Ca) described in Equation \ref{mass_balance_eq} are partitioned based on mass balance and laboratory measured partition coefficients (described later in this section).
Trace elements considered in the model (Co, Nb, Ta, V, and Cr) are partitioned based on laboratory measured partition coefficients alone, as their abundance is too small to affect mass balance calculations.
For these calculations, we follow the same procedures described in the supplementary material of  \citet{rubie2011heterogeneous}.}
All elements except Fe and Si are assigned to core and mantle based on likely silicate mantle valence states. 
Because iron and silicon are multivalent major rock forming elements, they require consideration of the redox state of the reaction, which is set by the abundance of oxygen.
This behavior motivates the oxidation gradient model we adopt here.

\GIN{
Diversity in the meteoritic record implies that early Solar System materials varied systematically by heliocentric distance.
The meteoritic record samples a wide range of oxidation states, commonly displayed in a Urey-Craig diagram \citep[e.g.,][]{Urey1953, Amano2025}, spanning three main chondrite types: enstatite chondrites, which are fully reduced; ordinary chondrites, which display a variety of relatively reduced redox states; and carbonaceous chondrites, which also span a range of relative oxidized redox states and additionally often show evidence of significant water-rock interaction.
\citet{gradie1982compositional} first showed the composition of objects in the asteroid belt varies systematically with heliocentric distance. 
Increased water and ice abundance at greater heliocentric distances in the protoplanetary disk is supported by meteoritic observations \citep{grossman2008redox} as well as theoretical frameworks \citep{rubie2015accretion, Krot2000}. 
Thus, the relative presence of water provides a natural mechanism for a redox gradient in which material becomes increasingly oxidized at greater distances from the Sun. }

\begin{figure}[pos=bt!]
    \centering
    \includegraphics[width=\columnwidth]{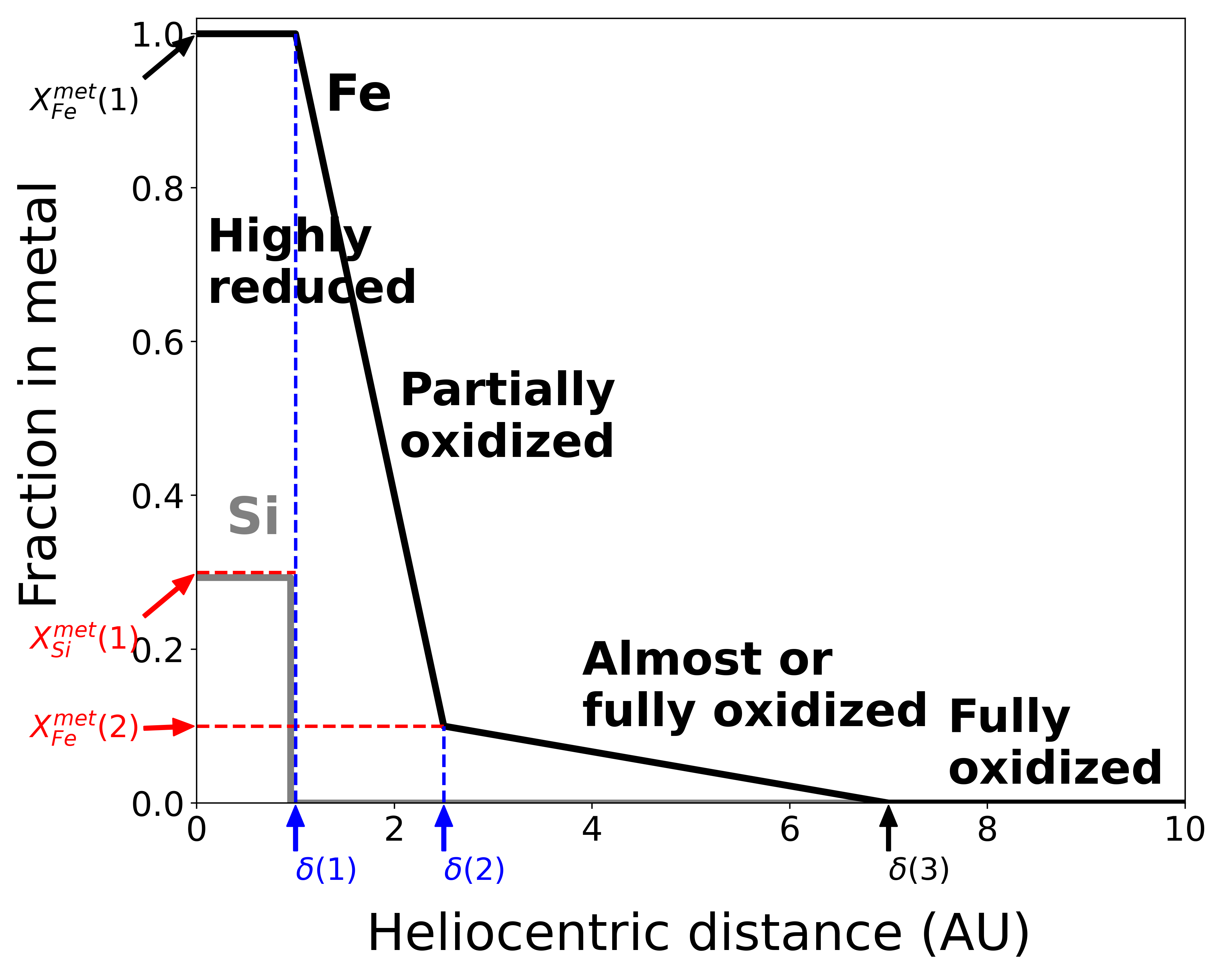}
    \caption{Various model parameters ($\delta(1)$, $\delta(2)$, $X^{met}_{Si}(1)$, and $X^{met}_{Fe}(2)$) control the oxidation gradient of Fe and Si in the initial protoplanetary disk for each Solar System formation simulation.
    This is the same oxidation gradient used in \citet{nathan2023constraining}.}
    \label{fig:example_oxidation_gradient}
\end{figure}

\GIN{
Motivated by this evidence, our model uses a radial oxidation gradient that is highly reduced close to the Sun and progressively more oxidized at greater heliocentric distances. 
Beyond 4.5 AU, we assume that bodies are not only fully oxidized but are also water-bearing. 
This interpretation is based primarily on the premise that bodies further from the Sun were fully oxidized, lacking any metallic material, and even retaining excess water. 
While the gradient we implement is most directly motivated by water abundance, it is ultimately agnostic as to whether an initial gradient was established by the addition of water \citep{Ciesla2006,Grewal2024}, self-oxidation during kinetic condensation \citep{Charnoz2026}, or some other process.
Here, the redox state gradient governs the fraction of iron and silicon in silicate and metallic reservoirs, as shown in Figure~\ref{fig:example_oxidation_gradient}. 
A gradient with variable iron and silicon oxidation states significantly influences the redox state of equilibrating materials in an accreting planetary body, and was shown by \citet{rubie2015accretion} to be compatible with forming Earth analogs whose mantle compositions match the bulk silicate Earth.
Because the structure of the redox gradient best suited to build an Earth analog is not precisely constrained by existing meteoritic data, we treat the features of this gradient as free parameters that may vary with formation scenario and individual simulations, an effect we explore here.
}

\subsection{Planetary accretion and differentiation}
Planetary mantles and cores are created through planetary differentiation, which is not a single-stage event, but rather a multi-stage process that ultimately establishes mantle and core reservoirs that are in chemical disequilibrium with one another. 
Core formation begins early in planetary formation and continues throughout accretion as the energy of accretion and impact events cause melting and re-equilibration between core-forming fluids and silicate melts. 
Because each subsequent differentiation event occurs at individual pressure, temperature, and redox conditions, the evolving mantle records a varied equilibration history instead of equilibrium at a single-stage.  

Initial differentiation of protoplanets occurs in all the simulated bodies due to their large initial sizes and early formation times. 
These planetesimals melt and differentiate as a result of the energy of accretion and heat of early short-lived radioisotopes such as $^{26}\text{Al}$ \citep{Hevey2006,Davis2022}.
Planetary mass and redox conditions of the initial body determine the mantle and core compositions during initial planetary differentiation after rapid accretion. 
Since most mass in the protoplanet population is expected to be in the largest objects due to runaway growth \citep{Wetherill1989,kokubo1996runaway,morbidelli2012building}, all the protoplanets in these simulations undergo differentiation due to short-lived radionuclides.
All protoplanets with Fe metal present form cores with elements partitioned between mantle and core according to a model of metal-silicate equilibration built from high-pressure laboratory experiments.
Thus, all bodies in the simulations are large enough to undergo complete internal melting and each body initially experiences an episode of core-mantle differentiation. 

Subsequent core formation occurs due to planetary impacts, as dictated by simulated \textit{N}-body accretion histories.
The high energies of planetary collisions can result in extensive melting and potentially global magma oceans \citep{Tonks1993,deVries2016,Nakajima2021}. 
An impact generates a melt plume in the target body, within which metallic liquid from the impactor core is entrained and emulsified with molten silicate from the target mantle \citep{Tonks1992,RUBIE201543}.
The extent of the target planet mantle that undergoes metal-silicate equilibration is determined by the experimental results of \citet{deguen2011experiments,deguen2014turbulent}.
Following equilibration, metal segregates to the core and the remaining equilibrated silicate is mixed into the portion of the mantle that did not undergo melting or re-equilibration.
This process results in re-equilibration and progressive compositional modification of a growing planet.

\begin{figure}[pos=bt!]
    \centering
    \includegraphics[width=0.9\columnwidth]{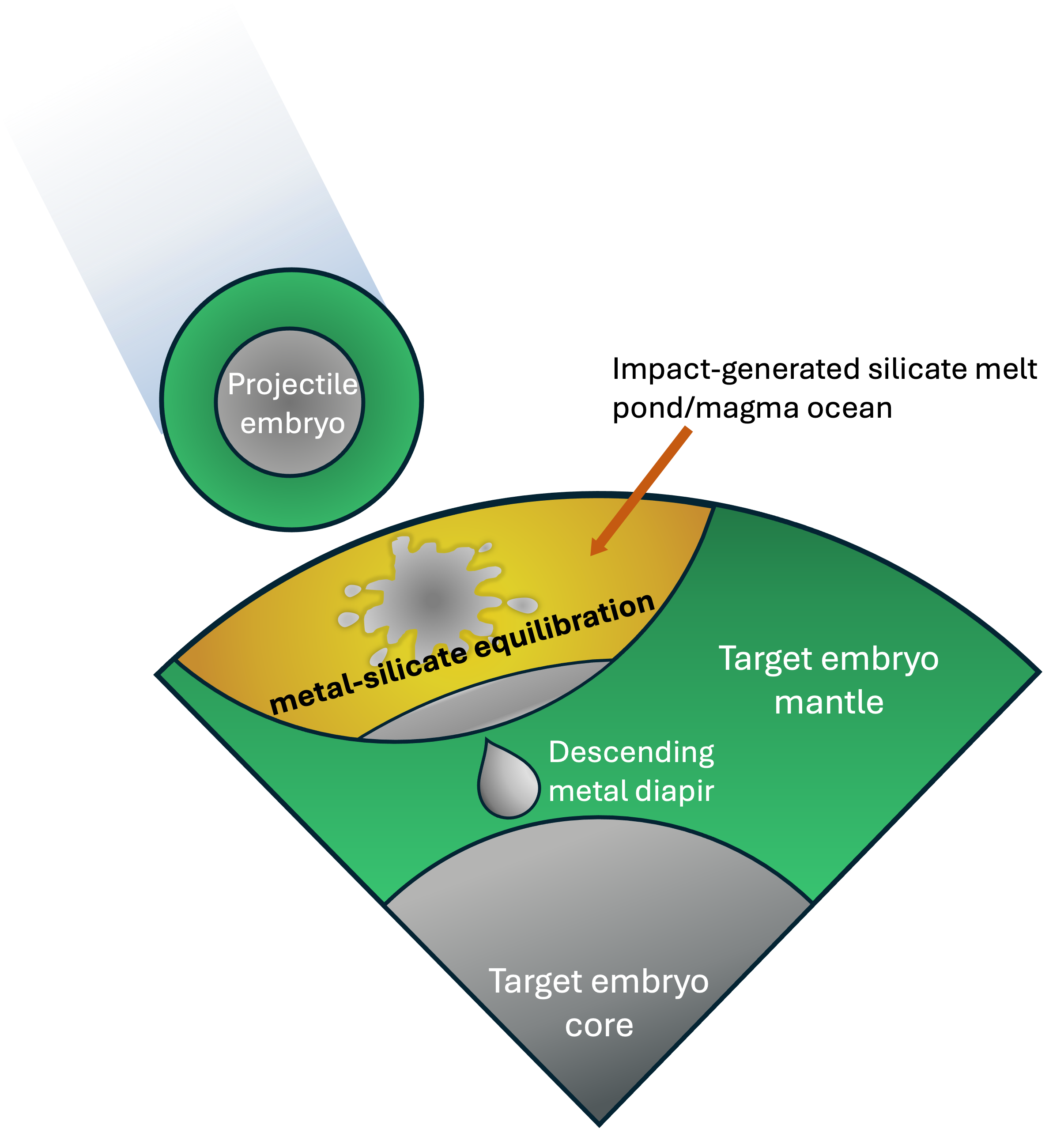}
    \caption{To calculate the geochemical effects of multi-stage planetary accretion, we use a core-mantle differentiation model first presented in \citet{rubie2015accretion}. In this model, planetary impacts produce silicate melt ponds in which a portion of the core and mantle of a projectile body equilibrate with the surrounding target body mantle. Mantle and core elemental compositions are updated according to mass balance and elemental partitioning behavior as measured in laboratory experiments. See similar illustrations and descriptions of other possible impact and equilibration regimes in \citet{Lichtenberg2023}.}
    \label{fig:illustration}
\end{figure}

Forming the Earth via multiple stages of impact-driven core formation is both consistent with astrophysical models of planet formation \citep{morbidelli2012building,Raymond2022} and is compatible with the known result that moderately siderophile abundances can be reproduced at mid-mantle pressures and temperatures; if formed over multiple episodes of re-equilibration, these mid-mantle conditions can represent an averaging of many equilibration events \citep{rubie2011heterogeneous, rubie2015accretion}. 
Thus, the model of multi-stage equilibration is compatible with astrophysical models of formation as well as compositional constraints in the mantle. 
A schematic illustration of this multi-stage accretion and differentiation model is shown in Figure~\ref{fig:illustration}.

During each differentiation event, major element partitioning is determined by metal-silicate partitioning behavior and mass balance of the moles of equilibrating metal and silicate.
Partitioning between mantle and core reservoirs is modeled for the moderately siderophile elements (Fe, Si, O, Ni, Co, V, Cr, Nb, and Ta) within the growing Earth. 
Partitioning behavior depends strongly on pressure ($P$), temperature ($T$), and oxygen fugacity ($f_{\text{O}2}$) conditions of equilibration.
The partitioning of an element $i$ is described by its partition coefficient: 
\begin{equation}
    D_i^\text{met-sil} = \frac{C_i^\text{met}}{C_i^\text{sil}}
\end{equation}
where $C_i^\text{met}$ and $C_i^\text{sil}$ are molar concentrations of element $i$ in metallic and silicate phase following equilibration. 
To incorporate the effects of $f_{\text{O}2}$, the partition coefficient of element $i$ is normalized to the partition coefficient of Fe, resulting in distribution coefficient $K_D^i$:
\begin{equation}
K_D^i = \frac{D^\text{met-sil}_i}{\left(D^\text{met-sil}_\text{Fe}\right)^{n/2}} = \frac{X^\text{metal}_i \left[X^\text{silicate}_\text{FeO}\right]^{n/2}}{X^\text{silicate}_{iO_{n/2}} \left[X^\text{metal}_\text{Fe}\right]^{n/2}}
\end{equation}
where $X$ represents molar fraction of a component $i$ in metal or silicate liquids following equilibration and $n$ is the valence state of element $i$.

\begin{table*}[ht]
    \centering
    \caption{\GIN{Partition coefficients for $\log K_D$ used in this study.
    Partitioning is parameterized as
    $\log K_D^i = a + b/T + cP/T$ unless noted otherwise.
    Ni and Co partitioning is from \citet{kegler2008new}; all other coefficients
    are from \citet{mann2009evidence}, as used in \citet{rubie2015accretion}.}}
    \label{tab:partitioning_parameters}

    \begin{tabular}{lcrrrrrr}
        \hline
        Element & Valence & $a$ & $\sigma_a$ & $b$ & $\sigma_b$ & $c$ & $\sigma_c$ \\
        \hline
        Si & $4+$ & 2.98  & 0.02 & -15934 & 0   & --\textsuperscript{\dag} & --\textsuperscript{\dag} \\
        Ni & $2+$ & 1.06  & 0.00 & 1553   & 0   & -98.0  & 7.66 \\
        Co & $2+$ & 0.13  & 0.00 & 2057   & 0   & -57.0  & 9.53 \\
        Nb & $5+$ & 2.66  & 0.11 & -14032 & 0   & -199.0 & 16.0 \\
        Ta & $5+$ & 0.84  & 0.09 & -13806 & 0   & -101.0 & 15.0 \\
        V  & $3+$ & -0.48 & 0.38 & -5063  & 894 & 0.00   & 0.00 \\
        Cr & $2+$ & -0.17 & 0.13 & -2730  & 312 & 0.00   & 0.00 \\
        \hline
    \end{tabular}

    \vspace{2pt}
    \begin{minipage}{0.95\textwidth}
        \footnotesize
        \textsuperscript{\dag}Regression term used for Si:
        $\log K_D = a + b/T + (c_1P + c_2P^2 + c_3P^3)/T$,
        with $c_1=-155\pm25$, $c_2=2.26\pm0.84$, and
        $c_3=-0.011\pm0.007$.
    \end{minipage}
\end{table*}

\GIN{Equilibration conditions are determined by the magnitude of planetary impacts and the evolving redox state of Fe and FeO available in the equilibrating material.
Thus, we use pressure, temperature and oxygen fugacity dependencies of distribution coefficient $K_D$ for element $i$, which is normalized to the partitioning behavior of Fe: 
\begin{equation}
\log K^{\text{i}}_{\text{D}} = a + b \cdot \frac{1}{T} + c \cdot \frac{P}{T}
\end{equation}
where values for parameters $a$, $b$, and $c$ used here are displayed in Table \ref{tab:partitioning_parameters}. 
As used in \citet{rubie2015accretion}, partition coefficients are from \citet{mann2009evidence} with the exception of Ni and Co, which are from \citet{kegler2008new}.
Other measured partition coefficients exist, but we have chosen to use the same as \citet{rubie2015accretion} for ease of comparison between studies, see further explanation of this choice in Section \ref{section:assumptions}.}
In this method, partitioning and redox state evolution of the core and mantle are self-consistent and mass balanced by absolute abundances, with no direct stipulations for the evolution of redox state, which evolves based on its initial conditions and subsequent core-mantle equilibration events. 

The core and mantle compositions of the Earth-like planet are determined by four inputs: published laboratory experiments that constrain elemental partitioning during metal-silicate equilibration \citep[e.g.,][]{cottrell2009metal,mann2009evidence,kegler2008new}, published laboratory and numerical fluid experiments that constrain how planetary impacts physically mix components \citep{deguen2011experiments,deguen2014turbulent}, published accretion histories from numerical simulations of planet formation \citep[e.g.,][]{nesvorny2021role}, and an initial compositional gradient in the disk. 
This last input, which controls the compositional gradient, is the only one for which the model is not explicitly constrained by prior experiments.
\GIN{To determine the compositional gradient, free parameters in the model are varied to reproduce a simulated Earth that matches the bulk and moderately siderophile element budget of the observed Earth. }
This is very similar to the approach in \citet{rubie2015accretion} which used the composition of the Earth as a constraint to determine conditions in the protoplanetary disk that best reproduce the Earth.

\subsection{Matching the mantle compositions of simulated Earth-like planets to Earth}
For each Solar System formation \textit{N}-body accretion history, we seek to determine the conditions of the initial disk that, in the model of multi-stage accretion and differentiation, produce an Earth analog that most closely resembles the actual Earth.
Thus, we solve an inverse problem, searching for conditions of the initial protoplanetary disk that result in simulated Earth analogs with accurate composition.
To assess the success of each simulation, we adopt a quantitative metric to directly compare the composition of a simulated Earth to geochemical estimates of the bulk silicate Earth.
Using a nonlinear least-squares fitting method, we explore free parameters to identify the initial conditions that most closely reproduce the bulk silicate Earth.
In this way, we are able to compare whether and how well various Solar System formation scenarios can produce an Earth analog.

The degree to which a simulated Earth reproduces the observed Earth is quantified with a goodness-of-fit metric that is based on the bulk silicate (mantle) of the Earth.
We compare to an estimate of the bulk silicate Earth presented in \citet{palme2014cosmochemical}, which combines geochemical constraints provided by analysis of Earth mantle rocks with cosmochemical constraints from the meteoritic record to determine a model composition of the bulk silicate Earth. 
This robust method provides a well-justified and self-consistent model of Earth composition to which we compare the simulated Earth analogs. 
\GIN{We use a subset of elements to fit the Earth-like planets' composition: Fe, Si, Ni, Co, Nb, Ta, V, Cr and the ratio of Nb/Ta.
These elements are chosen because they are well constrained, they partition between metal and silicate under core-forming conditions, and they trace the majority of a planet's accretion as opposed to highly siderophile elements which trace late stage accretion in particular \citep[e.g.,][]{dauphas2017isotopic, jacobson2014highly}.
The remaining lithophile major elements in the model (Mg, Al, and Ca) are not used as fitted elements because they do not partition into metal; their mantle abundances are therefore insensitive to the conditions of core–mantle equilibration and reflect only the assigned bulk compositions, see Section \ref{sec:protoplanet}.
Similarly, while we include oxygen in the model, it is not used as a fitted element because, as the anion pair to fitted elements (e.g., $\text{FeO}$, $\text{SiO}_2$, etc.), its abundance is correlated to successful matching of other elements.}
Goodness-of-fit is quantified by a reduced chi-squared value that describes how close a simulated Earth analog mantle matches the bulk silicate Earth:

\begin{equation}
\chi^2_\text{red} = \frac{1}{\nu}\sum \left(\frac{C^i_C - C^i_M}{\sigma^i_M}\right)^2
\end{equation}

where $C^i_C$ is the calculated mantle abundance of element $i$ in the simulated Earth analog, $C^i_M$ is the observed abundance of element $i$ in the bulk silicate Earth \citep[ref.][]{palme2014cosmochemical}, $\sigma^i_M$ is the uncertainty on the observed abundance.
$\nu$ is the number of degrees of freedom in the model, $\nu = N-n-1$, calculated by the number of observations $N$ (i.e. the elemental abundances in the BSE) minus the number of free parameters $n$, minus 1. 

Five free parameters describing the disk oxidation gradient and conditions of equilibration are varied in the accretion model. 
Four of the free parameters in the model control the iron and silicon oxidation state throughout the initial protoplanetary disk: $\delta(1)$, $\delta(2)$, $X^{met}_{\text{Si}}(1)$, and $X^{met}_{\text{Fe}}(2)$.
In this piecewise oxidation gradient model \citep{nathan2023constraining}, the initial amount of $\text{Fe}$ in metallic or $\text{FeO}$ silicate phase in the initial protoplanetary disk varies with a piecewise gradient and the amount of $\text{Si}$ in metallic or $\text{SiO}_2$ phase varies with a step function.
\GIN{In this way, the disk is chemically reduced close to the Sun and more oxidized at greater heliocentric distances.
Material initially close to the Sun will undergo initial differentiation in reducing conditions and contain all of its Fe as metal in its core (and some percentage of Si as metal in its core) and no $\text{FeO}$ in its mantle.
We note that this does not imply the material is an iron meteorite. 
Rather, it would resemble something akin to an Aubrite meteorite which is nearly FeO-free \citep{weisberg2006systematics}.
}
Free parameters $\delta(1)$ and $\delta(2)$ are heliocentric distances within the protoplanetary disk, measured in astronomical units (AU), at which the oxidation state of Si and Fe change from more reduced states (greater fraction of Fe or Si in metal) to more oxidized states (greater fraction of Fe or Si in silicate).
Free parameter $X^{met}_{\text{Si}}(1)$ determines the fraction of $\text{Si}$ in metal interior to $\delta(1)$. 
Beyond $\delta(1)$, there is no $\text{Si}$ in metallic phase ($X^{met}_{\text{Si}}=0.0$), it is all in silicate phase as $\text{SiO}_2$. 
Free parameter $X^{met}_{\text{Fe}}(2)$ controls the fraction of Fe in metallic phase at heliocentric distance $\delta(2)$.
Interior to heliocentric distance $\delta(1)$, all Fe is in metallic phase ($X^{met}_{\text{Fe}}=1.0$). 
Between $\delta(1)$ and $\delta(2)$, the fraction of Fe in metal decreases linearly until it reaches the value of free parameter $X^{met}_{\text{Fe}}(2)$ at heliocentric distance $\delta(2)$. 
From $\delta(2)$ to a fixed heliocentric distance $\delta(3) =\ 4.5 \text{ (AU)}$, the fraction of Fe in metal linearly decreases to zero ($X^{met}_{\text{Fe}}(\delta(>3))=0.0$ ). 
These four free parameters are shown visually in Figure~\ref{fig:example_oxidation_gradient}.

A fifth free parameter, $f_p$, controls the pressure at which metal-silicate equilibration occurs within the mantle of a planetary body. 
In the model, metal-silicate equilibration occurs in a plume of entrained metal and silicate as the projectile core descends to the target core; thus the equilibration does not occur at one single depth. 
Instead, we follow \citet{rubie2011heterogeneous,rubie2015accretion} and treat the pressure of equilibration as an average condition of equilibration within the impact generated melt. 
This averaged pressure of equilibration is expressed as a fraction of the core-mantle boundary pressure $P_\text{CMB}$ of the growing target planet:
\begin{equation}
    P_\text{equil} = f_p \times P_\text{CMB}
\end{equation}
where $f_p$ is a proportionality factor adjusting that pressure.
The oxygen fugacity at which the equilibration occurs is not imposed, but is self-consistently determined based on the molar abundances of Fe, FeO, Si, SiO$_2$ of the equilibrating materials. 
The temperature, $T_\text{equil}$, of metal-silicate equilibration corresponds to an intermediate value between the peridotite liquidus and solidus at the given equilibration pressure\GIN{, following the parameterization of \citet{rubie2015accretion}, which combines the peridotite melting curves of \citet{Herzberg1996} and \citet{Tronnes2002} with the FeO-corrected liquidus of \citet{Liebske2012}}: 
\begin{subequations}
\begin{align}
P &< 24\ \text{GPa}: \quad
\begin{aligned}[t]
T_\text{equil} ={}& 1874 + 55.43\,P \\
                 & - 1.74\,P^2 + 0.0193\,P^3
\end{aligned} \\
P &\geq 24\ \text{GPa}: \quad
\begin{aligned}[t]
T_\text{equil} ={}& 1249 + 58.28\,P \\
                 & - 0.395\,P^2 + 0.0011\,P^3
\end{aligned}
\end{align}
\end{subequations}

The optimal set of free parameters for each accretion simulation is found with a least-squares minimization. 
To minimize the goodness-of-fit chi-squared metric and find the free parameters that result in the closest match between simulated Earth-mantle and observed BSE composition, we compute a least-squares minimization routine with the downhill simplex method and a nonlinear minimization algorithm. 
To thoroughly search parameter space to achieve the best possible fit to the Earth for each individual simulation, we compute a least-squares minimization for a set of 200 randomly sampled initial simplexes for each \textit{N}-body accretion history input and then identify the best performing simplex of parameters. 
This iterative method and the use of hundreds of simplexes for each simulation improves the likelihood of finding a true global minimum best-fit set of free parameters for each simulation.  

\subsubsection{Identifying Earth and Theia}
The output of the model includes a simulated Earth with a composition that has evolved due to many impact and re-equilibration events. 
An Earth analog in a Solar System formation scenario is identified as a planet with mass $m$ between $0.75 M_\oplus$ and $1.55 M_\oplus$ that remains on a stable orbit at the end of a $\sim$200 Myr simulation, where $M_{\oplus}$ is one Earth mass. 
Because the composition of every object in each simulation is tracked, there is also a composition and history for Proto-Earth and Theia at the time of impact. 
In each simulation, Theia is defined as the final embryo impactor to hit the Earth analog. 
We determine which impact on the growing Earth is the last giant (embryo) impact and study the properties of Theia and the Proto-Earth.

Because Theia is a planetary embryo itself, which gravitationally interacts with all other objects in an \textit{N}-body simulation, it has an impact history of its own, before it strikes the Proto-Earth. 
This is in contrast to planetesimal objects in accretion simulations, which are smaller and only interact with larger bodies such as embryos and not with other planetesimals. 
If multiple embryos \GIN{collide} with Proto-Earth in a particular simulation, we identify the last of these impactors as Theia. 
We examine the time of impact, the mass ratio of Theia to Proto-Earth, and the FeO compositions of Proto-Earth and Theia at the time of impact.

\subsection{Assumptions and Uncertainties}
\label{section:assumptions}
Our approach requires several assumptions about the initial state of the protoplanetary disk, as well as how metal-silicate equilibration occurs during differentiation. 
We assume that the disk had a redox gradient that became more oxidized at greater heliocentric distances. 
This is grounded in meteoritic history and models of protoplanetary disk composition, as well as past work that has shown that the Earth is best reproduced when there is such a redox gradient \citep[e.g.,][]{rubie2015accretion}.
Whether or not the redox gradient model we use here is truly representative of the initial protoplanetary disk is unclear. 
In fact, it likely is not the true structure of the disk. 
However, it performs better at producing the Earth than constant compositional gradients or gradients that are oxidized close to the Sun and reduced at greater heliocentric distances (see \citet{rubie2011heterogeneous, rubie2015accretion}). 
Thus, it is a reasonable starting point for varying oxidation state within the disk. 
We examine this assumption in detail by conducting an experiment in which we adjust the history of accretion, see Section~\ref{results2}.

Further uncertainties arise from the mantle composition to which we compare Earth analogs, which mantle elemental compositions we are comparing to, as well as the partitioning behavior of elements. 
We use the bulk silicate Earth composition from \citet{palme2014cosmochemical}. 
While many model estimates of Earth mantle composition exist, they are generally in good agreement about the abundances of the bulk lithophile and moderately siderophile elements to which we fit the simulated Earth analogs.
Thus, the choice of \citet{palme2014cosmochemical} is reasonable and using a different model would not hold major consequences to the behavior of the model. 

\GIN{Similarly, different estimates of the partitioning coefficients for moderately siderophile elements exist \citep[e.g.,][]{fischer2015high}.
New measurements of Nb and Ta partitioning are presented in \citet{huang2020niobium} which, although we did not include these measurements in this study, merit inclusion in future core formation studies. 
\citet{huang2021high} provides new diamond anvil cell measurements of Mo and W partitioning that we have not incorporated, though we do not model or fit to Mo and W in this study. 
Including the effects of interaction parameters for Cr and V with O and Si \citep[e.g.,][]{kubik2021tracing, siebert2013terrestrial} would benefit a follow-up to this study, though we have not yet included interaction parameters in this model. 
\citet{Dale2026} investigated the effects of considering newer $\text{FeO}$ and $\text{SiO}_2$ partitioning models from \citet{sossi2025physicochemical} and found minimal differences with the \citet{rubie2015accretion} parameterization for equilibration pressures below 70 GPa.
Thus, we have chosen to use partition coefficients from \citet{mann2009evidence} and \citet{kegler2008new}, as is done in \citet{rubie2015accretion}, in part so that we can directly compare and validate this study with a prior model.
While newer experimental results have been able to make measurements at conditions closer to those expected during core formation, we do not expect the proposed relatively small partition coefficient changes for the elements considered here to have a substantial impact on our conclusions.}

\GIN{We also note that the degree of metal-silicate equilibration experienced by a body during its accretion has a strong control on its compositional evolution. 
Here, as in \citet{rubie2015accretion}, we continue to use the model of \citet{deguen2011experiments} to calculate the fraction of the target body mantle that equilibrates with the impactor core. 
While we acknowledge the strong effect of this choice in our model, we again follow the methods of \citet{rubie2015accretion} to ensure comparison between the current and prior study.
}

\section{Results}
\label{section_results}

%best-fit table
%These values are for the version with NO chi sq multiplier. I have another version with the chi sq multiplier for the lack of water. and I also am making a version with the delta(1) delta(2) AND delta(3) as free parameters for the publication version. 
\begin{table*}[bt!]
\caption{Chi-squared weighted average values ($\pm 1 \sigma$) of the best-fit free parameters ($f_p$, $\delta(1)$, $\delta(2)$, $X^{met}_{\text{Si}}(1)$, $X^{met}_{\text{Fe}}(2)$) for various Solar System formation scenario types. These free parameters control the oxidation state of initial bodies in the protoplanetary disk as well as the pressures of equilibration during core-mantle differentiation.}
\label{table_of_best_simplexes}
\begin{tabular*}{\linewidth}{@{\extracolsep{\fill}} lllllll}
\toprule
Formation Scenario                                                  & $f_p$ & $\delta(1)$ & $\delta(2)$ & $X^{met}_{\text{Si}}(1)$ & $X^{met}_{\text{Fe}}(2)$ & $\chi^2_{\text{red}}$ \\
\midrule
Grand Tack (GT) & $0.59 \pm 0.08$ & $1.14 \pm 0.19$ & $1.28 \pm 0.18$ & $0.19 \pm 0.05$ & $0.56 \pm 0.10$ & 2.78 \\
Circular Jupiter-Saturn (CJS) & $0.71 \pm 0.05$ & $1.55 \pm 0.33$ & $1.79 \pm 0.45$ & $0.14 \pm 0.02$ & $0.52 \pm 0.17$ & 3.22 \\
Eccentric Jupiter-Saturn (EJS) & $0.75 \pm 0.05$ & $1.21 \pm 0.20$ & $1.34 \pm 0.24$ & $0.14 \pm 0.01$ & $0.40 \pm 0.22$ & 4.50 \\
Truncated Disk (TD) & $0.68 \pm 0.03$ & $0.91 \pm 0.01$ & $0.95 \pm 0.03$ & $0.13 \pm 0.02$ & $0.15 \pm 0.13$ & 3.94 \\
Steep Profile (SP) & $0.52 \pm 0.02$ & $0.79 \pm 0.04$ & $0.88 \pm 0.10$ & $0.19 \pm 0.05$ & $0.49 \pm 0.12$ & 2.66 \\
Ring Model & $0.65 \pm 0.05$ & $0.93 \pm 0.05$ & $0.99 \pm 0.07$ & $0.16 \pm 0.02$ & $0.36 \pm 0.19$ & 4.60 \\
\bottomrule
\end{tabular*}
\label{table:bestfit}
\end{table*}

\subsection{\GIN{Many different formation scenarios can build Earth analogs with compositions similar to BSE}}
\GIN{All scenarios in the study are successful at producing Earth analogs with mantle compositions that match the bulk and the budget of notable moderately siderophile elements in the BSE.}
The model goodness-of-fit metric $\chi^2$ is uniformly low indicating that the simulated Earth analog mantle achieves a close match to the BSE cumulatively across all of the elements.
For the 164 simulations examined here, the mean $\chi^2$ goodness-of-fit value is $3.39 \pm 2.58$ (median $2.47$), indicating that most simulated Earth analogs reproduce the bulk silicate Earth abundances of the major rock forming elements and moderately siderophile elements abundances included in the model, within observational uncertainties.
The average values of the parameters and corresponding chi-squared metric of the best-fit simplex for each of the five fitting parameters \GIN{are} shown in Table~\ref{table_of_best_simplexes}. 
Thus, all scenarios in this study \GIN{achieve} average low chi-squared values ($\chi^2_\text{scenario} < 5$), and the chi-squared–weighted best-fit parameters are broadly consistent across solar-system formation scenarios.
While individual simulations vary in how closely they reproduce an Earth mantle composition, all are below $\chi^2 = 20$, with the average for each scenario below $\chi^2 = 5$. 
No scenario class systematically outperforms the others; differences arise from the stochastic nature of accretion histories in \textit{N}-body simulations.

Not just taken together, but moderately siderophile element abundances uniformly match well across individual elements within each Earth analog.
The final Earth analog mantle compositions of fitted elements (oxide abundances of Fe, Si, Ni, Co, Nb, Ta, V, Cr) and Nb/Ta ratio are shown in \GIN{Appendix Figures \ref{fig:plot_earth_information_Final_Earth_FeO}--\ref{fig:plot_earth_information_Final_Earth_NbTa}.}
The FeO and SiO$_2$ of the simulated Earth mantles are nearly all identical to that of the present-day bulk silicate Earth as defined by \citet{palme2014cosmochemical}, with the exception of a few Annulus model simulations (TD and Ring). 
Ni, Co, V, and Nb in the simulated Earth mantles closely match that of the bulk silicate Earth. 
Earth analogs are typically enriched in Cr compared to the Cr abundance in the bulk silicate Earth; this is particularly true for Grand Tack and Circular Jupiter-Saturn simulations. 
However, the Cr composition is typically overlapping in error with the composition of the Earth.
Ta is consistently depleted in the simulated Earth mantles, sometimes not within standard error of the measured bulk silicate Earth composition. 
However, the Nb/Ta ratio in simulated Earth bodies is nearly identical to the Nb/Ta ratio in the bulk silicate Earth.
\GIN{Additionally, we show the final O and Si composition of the cores of simulated Earth analogs in Appendix Figures \ref{fig:final_earth_core_oxygen} and \ref{fig:final_earth_core_silicon}; core O weight percent ranges from $\sim1\%$ to $\sim5\%$, while core Si weight percent ranges from $\sim7\%$ to $\sim9.5\%$.}

Moderately siderophile element abundances in the Earth analog mantles tend to increase over time.
Evolution of mantle composition for fitted elements (oxide abundances of Fe, Si, Ni, Co, Nb, Ta, V, Cr) and Nb/Ta ratio in simulations of each Solar System formation scenario can be seen in \GIN{Appendix Figures \ref{fig:plot_earth_information_Final_Earth_FeO}--\ref{fig:plot_earth_information_Final_Earth_NbTa}}. 
In nearly all simulations across all scenario types, mantle FeO starts at a very low wt.\%; the Earth initially is more reduced and is later more oxidized. 
Mantle SiO$_2$ wt.\% abundances typically decrease throughout accretion.

\subsection{Similar fitted model parameters across all terrestrial planet formation scenarios}
Across all simulations, we observe that moderately siderophile element abundances in the Earth's mantle are best obtained when core formation occurs at mid-mantle pressures and temperatures, matching prior work \citep[e.g.,][]{li1996geochemistry,righter2003partition}.
This is indicated by the average best-fit values for fractional pressure of equilibration ($f_p$) range from $\sim0.5$ to $\sim0.8$.
This means the average pressure of metal-silicate equilibration does not occur at the core-mantle boundary, but in the middle of the mantle (at 50\% to 80\% the pressure of the core-mantle boundary). 
Within simulation classes, the best fit values for fractional pressures of equilibration are similar: Grand Tack simulations (GT) have a best fit $f_p$ of $\sim0.59$, Classical simulations (CJS and EJS) have $f_p$ between $0.71$ and $0.75$, and in Annulus simulations (SP, TD, Ring), $f_p$ is between $0.52$ and $0.68$.

\begin{figure*}[pos=bt!]
    \centering
    \includegraphics[width=0.95\textwidth]{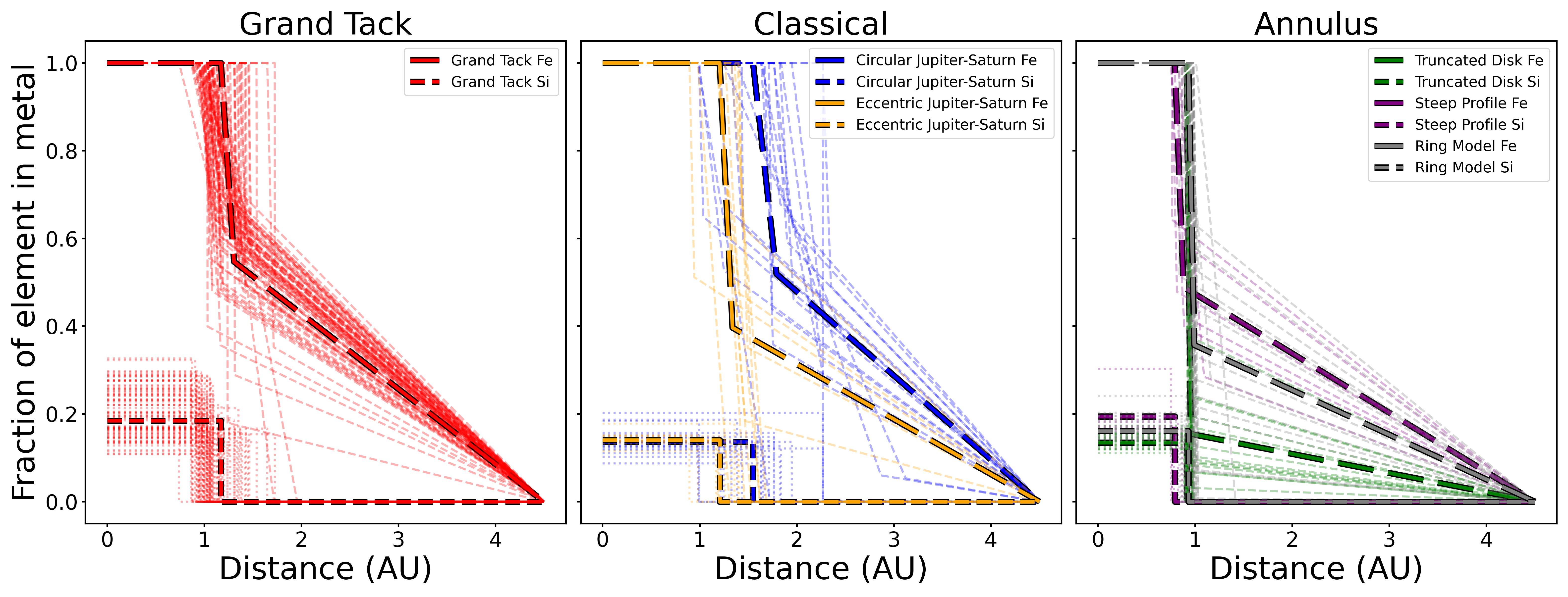}
    \caption{The initial disk gradients of metallic iron and silicon that best reproduce the Earth are similar between different types of simulations.
    Individual solid lines represent best fit redox conditions for individual simulations and dashed lines represent the average of the best fit conditions of a particular scenario type.
    Colors correspond to the type of scenario to which a particular simulation belongs: Grand Tack standard (GT), Circular Jupiter Saturn (CJS), Eccentric Jupiter Saturn (EJS), Truncated Disk (TD), Steep Profile (SP) or Ring Model.
    Scenarios are further grouped into three panels by Grand Tack, Classical, and Annulus-type.}
    \label{fig:3subplots_oxidation_gradients}
\end{figure*}

In general, the structure of the best-fit oxidation gradients is similar across all simulations, as shown in Figure~\ref{fig:3subplots_oxidation_gradients}: a steep transition from fully reduced to mostly (or fully) oxidized material in the region where most of the Earth's building blocks originate.
The best-fit piecewise redox-gradient inflection points, $\delta(1)$ and $\delta(2)$, are between 0.79–1.5 AU and 0.88–1.79 AU, respectively.
Most inferred oxidation gradients require an initial protoplanetary $\text{Fe-FeO}$ gradient to oxidize from 100\% Fe metal to $\sim$50\% Fe metal in a distance of 0.2 AU or less.
Similarly, the best-fit free parameters controlling the fractions of silicon and iron in metal ($X^{met}_{\text{Si}}(1)$ and $X^{met}_{\text{Fe}}(2)$, respectively) are consistent across scenario types. 
Beyond the inner disk, the Fe in the protoplanetary disk gradually is more oxidized until there is no Fe in metallic form (100\% Fe in silicates) beyond 4.5 AU.
$X^{met}{\text{Si}}(1)$ ranges between 0.13 and 0.19, while $X^{met}{\text{Fe}}(2)$ ranges between 0.155 and 0.59.
In other words, the $\text{Si-SiO}_2$ gradient drops from a reduced $\sim10-20$\% Si in metal to $0$\% Si in metal between 0.79 AU and 1.5 AU.
Across all terrestrial planet formation scenarios, Earth analogs grow from heterogeneously oxidized building materials rather than a homogeneous supply, reinforcing another conclusion from prior work \citep{rubie2011heterogeneous,rubie2015accretion,Dale2026}.

In particular, the Annulus terrestrial planet formation scenario requires notably steep oxidation gradients. 
While the gradients across all scenarios are qualitatively similar, differing mainly in steepness, Earth analogs built in the Annulus terrestrial planet formation scenario are best fit by particularly rapid transitions from reduced to oxidized material. 
This rapid transition allows the gradient to cut through the narrow region of material that builds the Earth analog, so that each Earth analog can still accrete from heterogeneous materials.

\subsection{Initial embryo size and location do not control final mantle composition}
\begin{figure*}[bt!]
    \centering

    \begin{subfigure}[t]{0.44\textwidth}
        \centering
        \includegraphics[width=\linewidth]{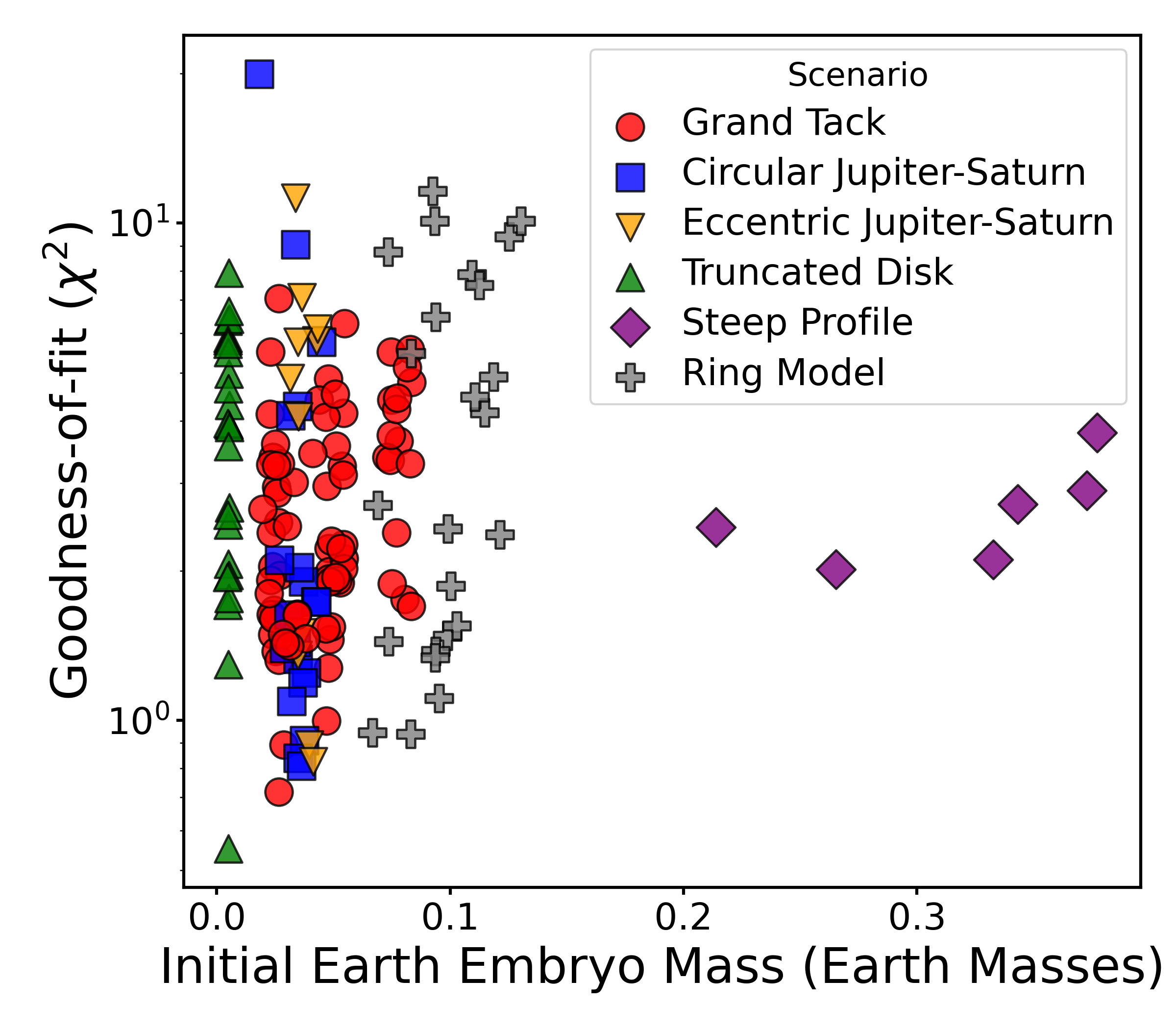}
        
        \phantomsubcaption
        \label{fig:chi_sq_vs_embryo_size}
        \vspace{0.2em}
        (a)
    \end{subfigure}
    \hfill
    \begin{subfigure}[t]{0.54\textwidth}
        \centering
        \includegraphics[width=\linewidth]{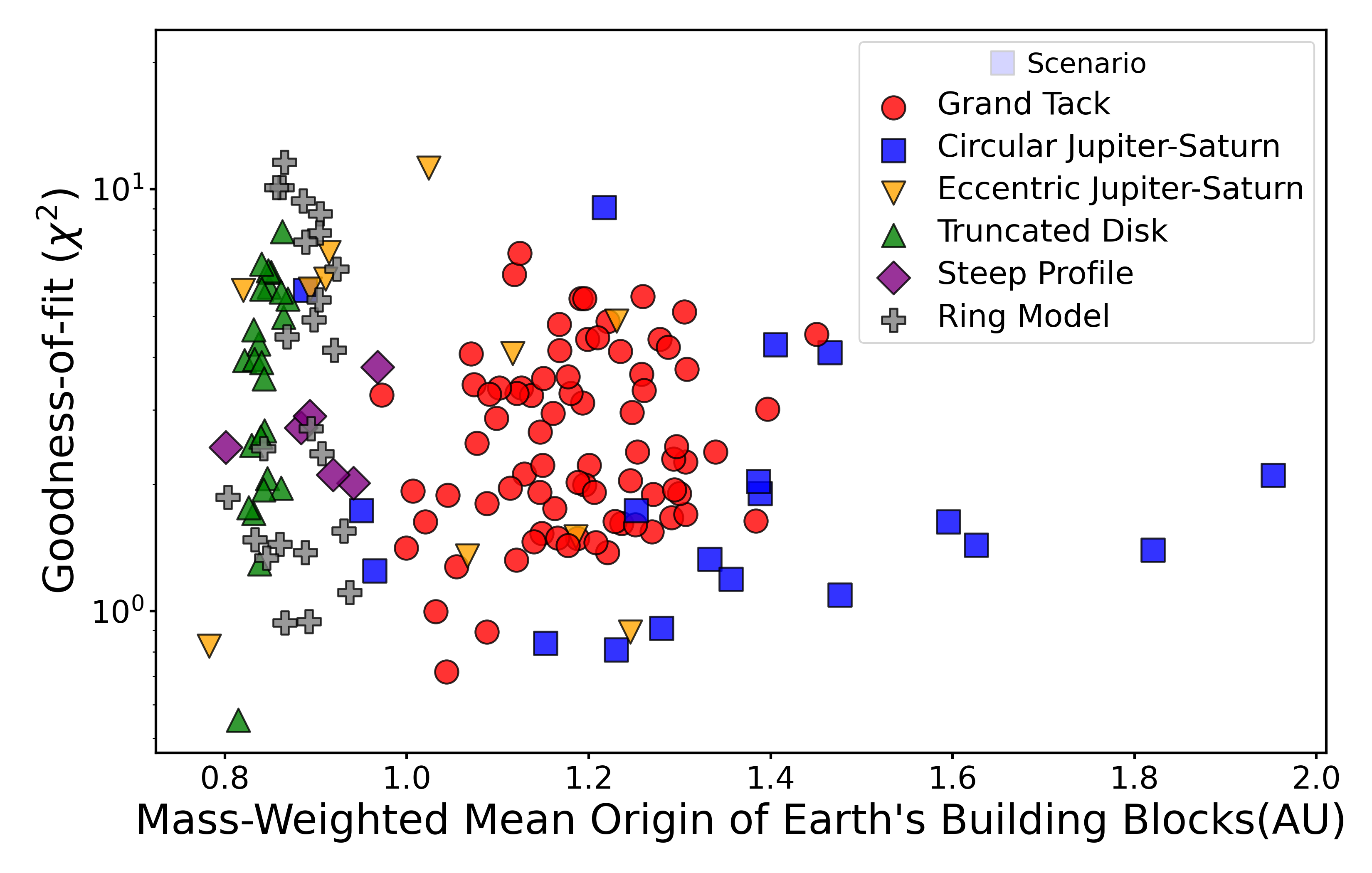}
        
        \phantomsubcaption
        \label{fig:average_accretion_location_vs_chi_sq}
        \vspace{0.2em}
        (b)
    \end{subfigure}

    \caption{\GIN{Goodness-of-fit metric $\chi^2$ is shown with respect to (a) the initial embryo mass of the Earth analog and (b) the average location of the origin of Earth's building blocks. A lower $\chi^2$ value represents a closer match of the simulated Earth analog mantle composition to the composition of the BSE. Each marker represents an Earth analog in a simulation, and the color represents from which scenario the simulation originates. There is no trend between the success of a simulated Earth analog to match the mantle composition of the Bulk Silicate Earth (BSE) and the size of the initial embryo mass of the Earth analog or the origin of its building blocks.} }
    \label{fig:chi_sq_combined}
\end{figure*}

There is no correlation between the initial mass of the Earth-forming embryo and the ability of the simulation to match the Earth's mantle composition, as shown in Figure~\ref{fig:chi_sq_vs_embryo_size}.
\GIN{Figure~\ref{fig:chi_sq_vs_embryo_size} shows a range of initial embryo masses, but most simulations have initial embryo masses of $\lesssim0.1M_\oplus$. 
Simulations with larger initial embryos achieve neither the best nor the worst fit to the Earth mantle. 
The small number of these larger initial embryo simulations prevents a more definitive assessment of their compatibility with reproducing the Earth mantle.}
This is notable because other studies that simulate planetary accretion have shown a strong dependence of planetary composition on initial planetary embryo size \citep{gu2023comparisons}.
Our methodology differs from \citet{gu2023comparisons}; that study explored a parameter space to determine effects of formation conditions on final planetary composition, whereas we instead work in reverse, searching the initial composition space for conditions that best reproduce the Earth for a given formation scenario. 
Here, while it is the case that the initial differentiation event is strongly affected by the characteristics of the initial planetary embryo, this does not serve as a barrier to eventually obtaining an Earth-like composition. 
This is, in part, because the composition of the Earth evolves significantly due to impact-driven core formation episodes. 
Additionally, our modeling approach allows a thorough search of parameter space for initial conditions that are conducive to building a compositionally accurate Earth analog across the elements of interest.

Similarly, there is also no trend between the average origin location of material that builds the Earth and the ability of that material to form an accurate Earth analog mantle, as shown in Figure~\ref{fig:average_accretion_location_vs_chi_sq}.
As each disk's redox gradient creates the conditions so that the Earth analog can accrete material with a variety of redox states, the Earth mantle composition can be reproduced.
Consequently, this means that the average source location of an Earth analog is not a major constraint on the ability to reproduce the Earth mantle analog composition.
In a different location, the Earth mantle will be reproduced if the oxidation gradient in the initial protoplanetary disk varies from reduced to oxidized at the location(s) of the formation of Proto-Earth and its constituent parts. 
What matters instead is that the oxidation gradient includes both reduced and oxidized materials across the region from which the Earth-building bodies originate.

\subsection{Earth analog mantles become increasingly oxidized during accretion}
\label{results2}
Given the best fit parameters obtained by the model, the early-accreted material in Earth history must be reduced (with higher fractions of Fe in metallic form) with later-accreted material becoming progressively more oxidized.
This reproduces a key result from \citet{rubie2015accretion}, which was based on Grand Tack scenario simulations, but now can be extended across many more proposed terrestrial planet formation scenarios.
A similar result was shown to be true for Ring model scenarios as well \citep{Dale2026} and \GIN{has} been argued on analytic grounds as well \citep{wood2006accretion}.

\begin{figure}[pos=bt!]
\centering
\includegraphics[width=0.95\columnwidth]{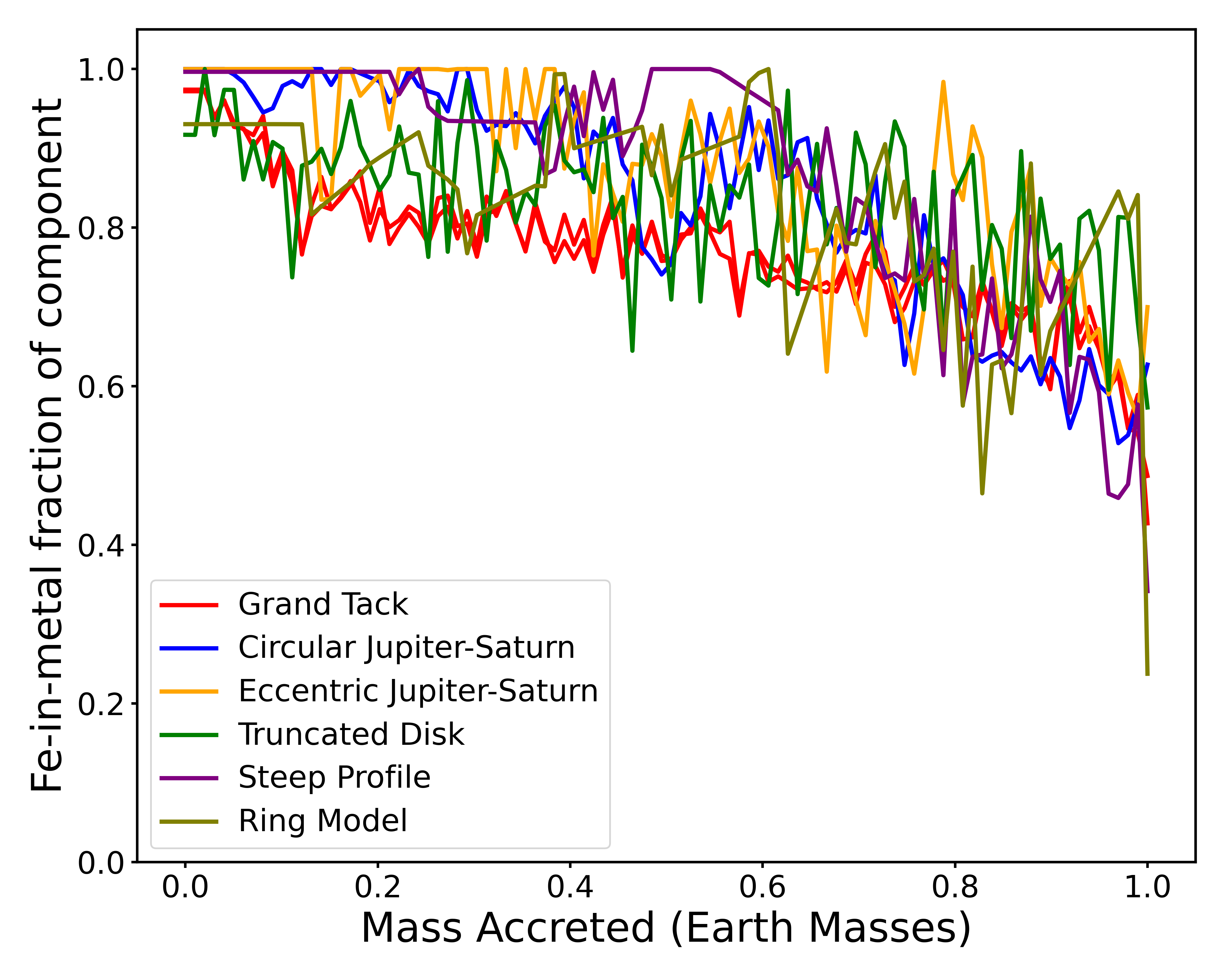}
\caption{
    The iron-in-metal fraction of the material added to the Earth analog over time decreases throughout its formation, meaning the Earth is built from successively more oxidized materials.
    Each dashed line is the average of many simulations from a particular scenario type (e.g., GT, CJS, EJS, SP, TD, ring), distinguished by color.
    The y-axis represents the fraction of Fe in metallic form of the mass delivered to Earth at a given moment in Earth's accretion.
    The x-axis shows the progressive accretion of the Earth, as a fraction of total mass accreted.}
    \label{fig:feo_building_blocks_changing_over_time}
\end{figure}

Each Earth-like planet accretes a mixture of bodies with diverse redox histories whose redox states vary greatly throughout the accretion history of a particular simulation. 
Many of the bodies that impact the growing Earth have their own compositional histories and may be partially oxidized to varying degrees. 
Thus, the Earth grows heterogeneously, and the average redox state of bodies delivered to the Earth \GIN{becomes} more oxidized over time; the fraction of Fe in metallic form of material accreted to Earth decreases over time.
This is a common trend across all scenarios in this study, as shown in Figure~\ref{fig:feo_building_blocks_changing_over_time}.
Since Fe in metal fraction has the strongest control on the redox state of the system, this trend controls the changing redox state of the Earth as it accretes.

\begin{figure}[pos=bt!]
    \centering
    \includegraphics[width=\columnwidth]{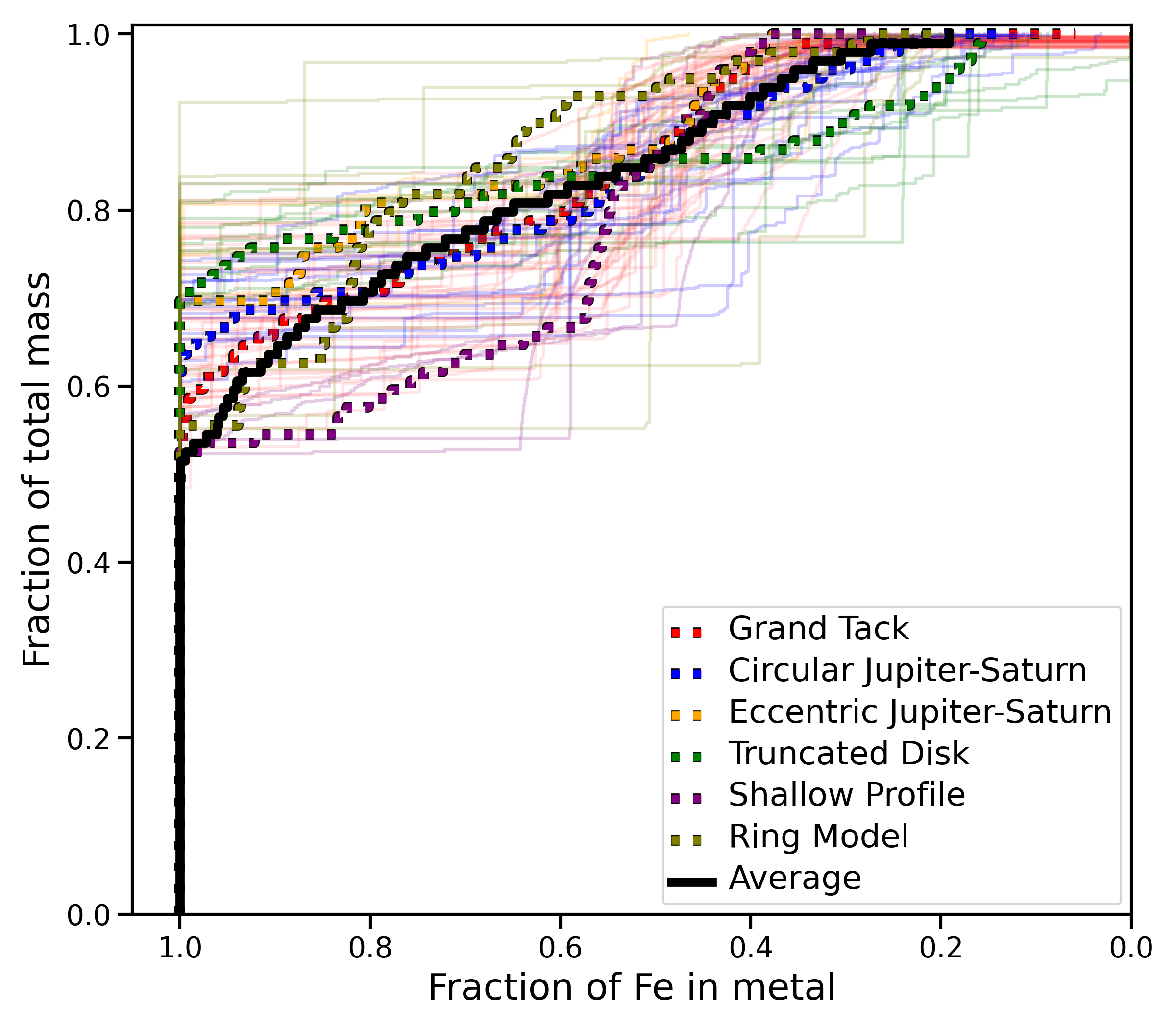}
    \caption{The cumulative distribution function of the fraction of iron initially in metal in all contributing planetesimals and embryos that build the Earth is similar across many individual simulations and formation scenarios.
    The y-axis in this plot shows the fraction of Fe in metallic form of a component that is delivered to the Earth.
    The x-axis shows the fractional contribution of a body with that given Fe redox state to the total mass of the simulated Earth analog. 
    Transparent lines correspond individual simulations, while dashed lines represent the cumulative average of iron oxidation state in each type of scenario.
    The solid black line represents the average of all scenarios. Only simulations that most closely match the Earth mantle ($\chi^2<3$) are included in this plot.}
    \label{fig:CDF-fe-in-metal}
\end{figure}

The initial disk redox gradients in Figure~\ref{fig:3subplots_oxidation_gradients} and the changing Fe in metal fraction of material delivered to Earth in Figure~\ref{fig:feo_building_blocks_changing_over_time} can together be thought of as the `recipe' to `bake' the Earth. % This originally said: "Figure~\ref{fig:feo_building_blocks_changing_over_time} is an average of the best distribution of iron oxidation states in order to build the Earth in each formation scenario type and so it can be thought of as the recipe..."
Using the best-fit simplexes (from Table~\ref{table:bestfit}) and accretion history of the Earth analog and its constituent parts in each simulation, we \GIN{calculate} a cumulative distribution of Fe redox states to build each Earth-like planet, as shown in Figure~\ref{fig:CDF-fe-in-metal}; this can be thought of as a list of raw ingredients needed to build the Earth.
The averaged curves for each scenario class are calculated by interpolating each function of iron abundance into one hundred discreet steps and finding the mean fraction of total mass at each fraction of Fe in metal. 
Roughly 50\% of the mass of the Earth must be delivered by planetesimals and embryos that possess 100\% of their Fe in metallic form, as shown in Figure  \ref{fig:CDF-fe-in-metal}.
Similarly, only around 2\% of each Earth-like planet is built from planetesimals and embryos whose Fe is in a fully oxidized state (100\% FeO).
\GIN{By this, we are not implying that the Earth accreted half of its mass from iron meteorites.
Rather, 50\% of the planetesimals and embryos accreted to the Earth only possess Fe in their core, and do not have FeO in their mantle. 
Examples of this type of material come from the Aubrite meteorites, which represent a differentiated parent body that equilibrated under reducing conditions (with very little free oxygen) and, as a result, have almost no FeO in the silicate phase, consisting primarily of nearly FeO-free enstatite \citep{weisberg2006systematics}. 
}

To test how sensitive the final bake (Earth) is to the recipe (order of accretion) relative to the list of ingredients (building blocks of varying redox state), we \GIN{explore} how changing the order of accretion \GIN{affects} the final composition of each Earth analog mantle.
We \GIN{create} three types of contrived formation scenarios, differing only by accretion order:
\begin{enumerate}
    \item Build an Earth analog from reduced to oxidized,
    \item Build an Earth analog from oxidized to reduced, and
    \item Build an Earth analog in a random order. 
\end{enumerate} 
\GIN{Each of these draws from the distribution of material found to produce the best-fitting Earth analogs, as shown in Figure~\ref{fig:CDF-fe-in-metal}.
This is the distribution of Fe fraction in metal in the building blocks of the Earth that best reproduces the Earth's mantle geochemistry for the elements we consider in this study. 
In the N-body simulations of Solar System formation from which we derive this idealized distribution of material, the material is delivered stochastically, though there is an overall trend of material becoming more oxidized through the course of accretion, as shown in Figure~\ref{fig:feo_building_blocks_changing_over_time}. 
In these contrived accretion scenario tests, we prescribe the order of accretion depending on the fraction of Fe in metal in the accreting bodies to be in specific ordering. 
When building an Earth analog in reduced to oxidized ordering, we begin with materials whose Fe metal fraction is 1.0 and progressively add the more oxidized materials to the growing Earth. 
When building an Earth analog in oxidized to reduced ordering, we start with materials whose Fe metal fraction is 0.0 and progressively add more reduced materials. 
The random ordering draws from the same distribution of compositions (shown visually in Figure~\ref{fig:CDF-fe-in-metal}), but the order of accretion is not in order of redox state.
In a reducing environment (i.e., when Fe metal fraction is 1.0), Si metal fraction corresponds to a value of 0.166, which is intermediate among the best-fit $X^{met}_\text{Si}(1)$ values in Table \ref{table_of_best_simplexes}; in an oxidizing environment (i.e., when Fe metal fraction is less than 1.0), Si metal fraction corresponds to a value of 0.0.}
We also \GIN{control} for the role of accretion event size by considering both equal-mass impactor accretion histories, in which the Earth grows in equal 1\% mass increments and ensembles of 10 variable-mass impactor accretion histories, in which impactor masses \GIN{are} randomly sampled from a distribution bounded by 0.89\% to 10\% of an Earth mass and then renormalized to ensure the final planet mass is 1 Earth mass.
These are not meant to be realistic growth histories of Earth, but that is not the goal of this test. 
Instead, the goal is to see if the order of accretion matters by testing which sequence of accreted materials most closely matches the composition of the bulk silicate Earth.

\begin{figure}[pos=bt!]
    \centering
    \includegraphics[width=0.95\columnwidth]{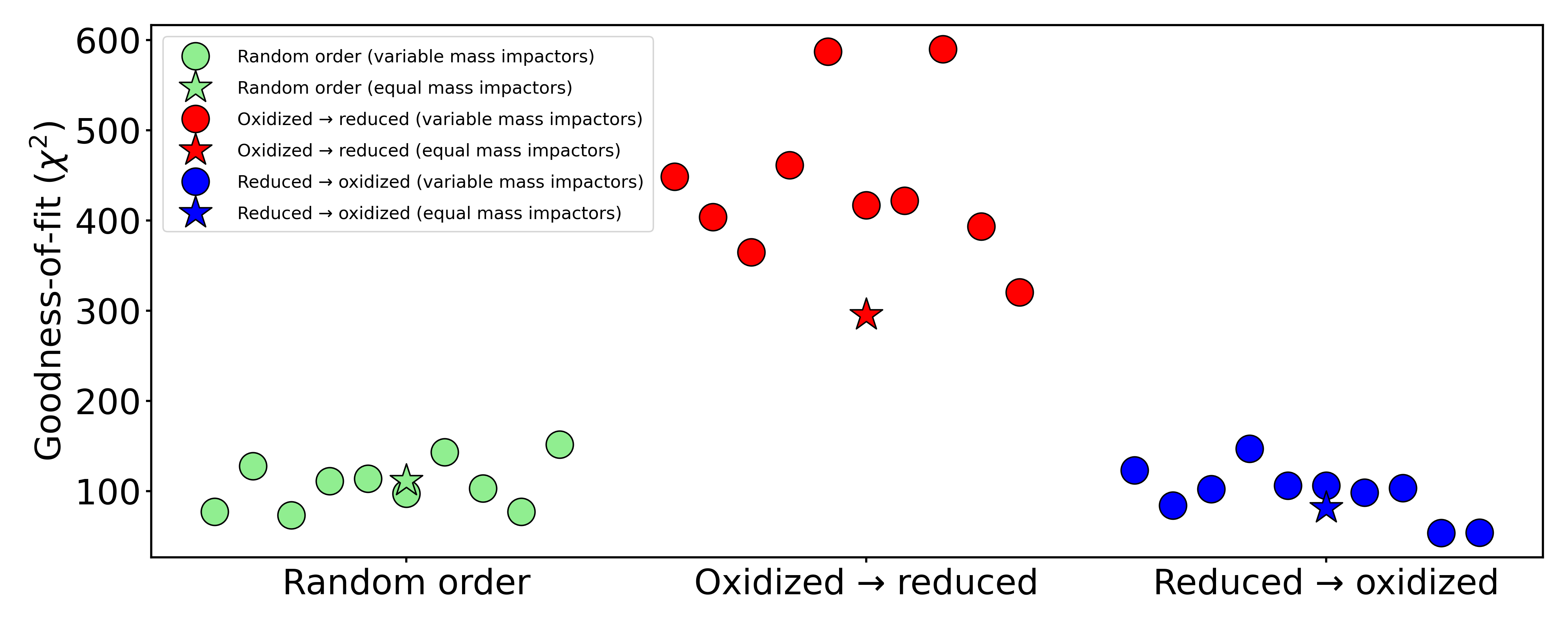}
    \caption{Three sets of contrived accretion scenarios and their corresponding chi-square values are shown.
    These simulations \GIN{are} created by using the average cumulative mass distribution function of Earth's building blocks and interpolating them into one hundred steps.
    The order of accretion is 1) randomized or 2) begins reduced and ends oxidized or 3) begins oxidized and ends reduced.
    Stars indicate simulations that build the Earth in 100 equal mass steps, while circle markers indicate simulations where the Earth accretes from impactors with randomly varied masses.
    Green, red, and blue represent the random-order, oxidized-to-reduced, and reduced-to-oxidized accretion scenarios respectively.}
    \label{fig:contrived-accretion-test}
\end{figure}

\GIN{We find the order of accretion is important: simulations with random accretion orders and those with reduced-to-oxidized accretion orders produce lower chi-squared values than simulations that begin with oxidized material and later add reduced material.  
This is shown in Figure~\ref{fig:contrived-accretion-test}, where star symbols indicate equal-mass impactor accretion histories and circular markers indicate variable-mass impactor histories.}
The scenario that begins with reduced building blocks and becomes more oxidized is shown by blue symbols and achieves the best fit to the Earth's mantle composition of these tests; the equal-mass impactor version of this scenario has $\chi^2=81.4$ and the average chi-squared of the variable-mass ensemble is $\chi^2=97.8$.
The scenario in which the ordering of oxidation states of accreted material during accretion are randomized is shown by green symbols and performs slightly worse, but comparably to the reduced-to-oxidized accretion histories; the equal-mass impactor version has $\chi^2=111.5$ and the average chi-squared of the variable-mass ensemble is $\chi^2=107.6$.
The scenario in which the building blocks of the Earth start oxidized and become increasingly reduced is shown by red symbols and produces an Earth analog mantle composition that is furthest from the true composition of the Earth mantle; the equal-mass impactor version has $\chi^2=295.2$ and the average chi-squared of the variable-mass ensemble is $\chi^2=440.9$.
\GIN{We note that while the reduced-to-oxidized accretion pathway achieves the best average goodness-of-fit for the Earth analogs produced, its success is comparable to that of the random accretion pathway. 
As such, there is not significant difference between accreting oxidized material later than reduced and a random ordering of accreted material. 
However, we note that the random accretion pathway still draws from the same compositional distribution and, because of the strongly reduced building blocks in that distribution, it is likely to accrete reduced material before oxidized material, even in a random ordering.}
\GIN{However, as indicated} by these high chi-squared values, none of these contrived scenarios \GIN{achieves} as good a fit to the composition of the Earth's mantle as \GIN{do} the earlier simulations drawn from \textit{N}-body accretion histories. 
This is expected because these are simplified, contrived accretion histories, and we \GIN{do} not vary any free model parameters to obtain a best fit. 

\subsection{Timing of the Moon-forming impact} \label{sec:timing}

Focusing now on the Moon-forming impact, we examine which simulations \GIN{perform} best according to the goodness-of-fit metric $\chi^2$ and what properties about Theia, the Proto-Earth, and late accretion can be inferred from these simulations.
In all simulations, the last giant (i.e., embryo-sized) impactor colliding with the Earth-like planet is considered to be the Moon-forming impactor for that simulated accretion history.
Depending on the particular simulation, Theia may be delivered as early as a few thousand years or as late as a couple of hundred million years into Earth's accretion.
Late accretion is defined as the mass accreted following that last giant impact.
All of late accretion must be from planetesimals with individual masses much less than 1\% the mass of Earth, since larger impactors would be considered giant impacts themselves. 
A late accretion mass of zero would indicate that the Moon-forming impact is the final impact in the accretion history of a particular Earth-like planet, and that no simulated planetesimals were accreted afterwards.

\begin{figure*}[bt!]
    \centering

    \begin{subfigure}[t]{0.49\textwidth}
        \centering
        \includegraphics[width=0.95\linewidth]{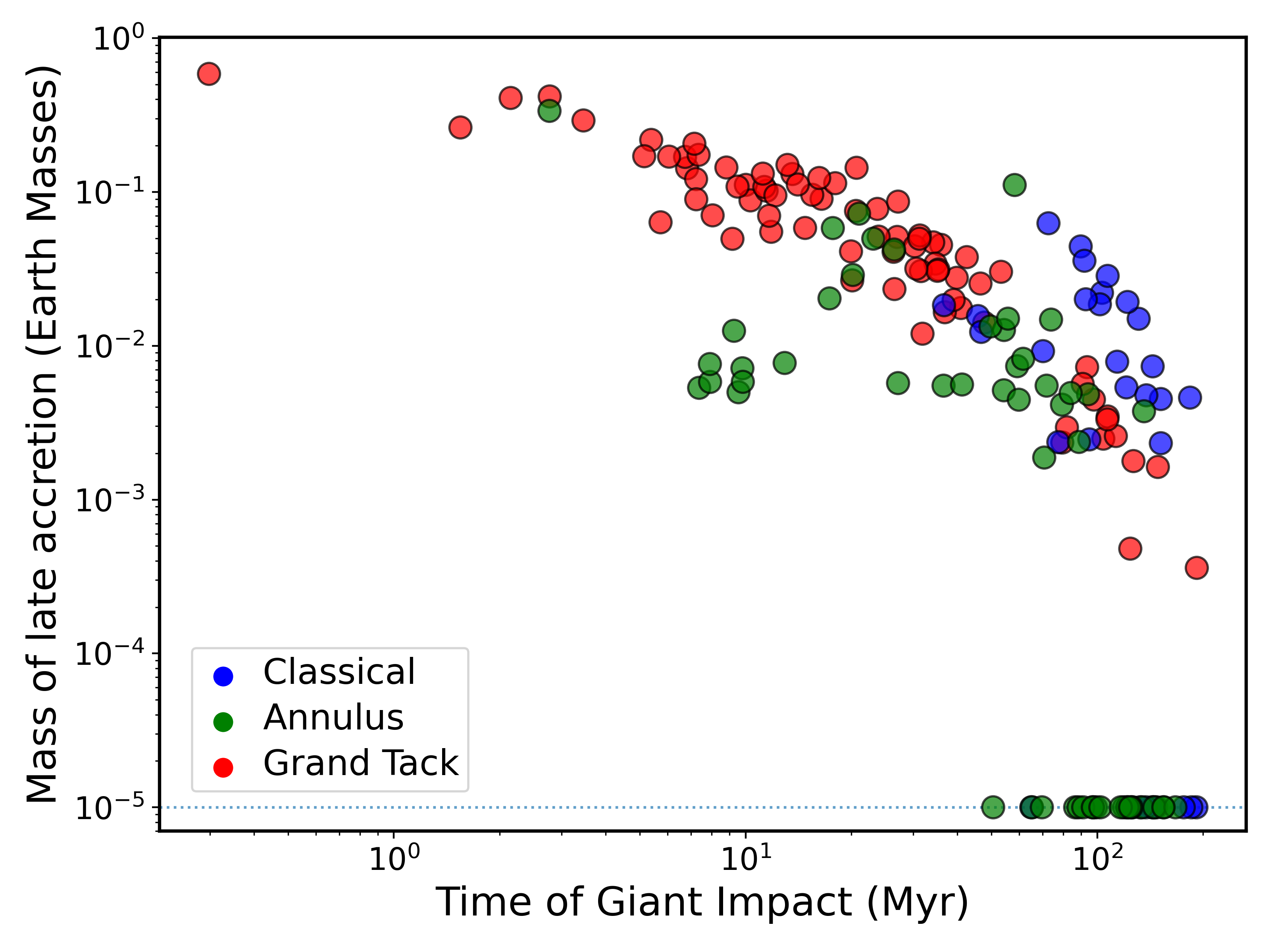}
        \phantomsubcaption
        \label{fig:late_accretion_vs_time_giant_impact}

        \vspace{0.2em}
        (a)
    \end{subfigure}
    \hfill
    \begin{subfigure}[t]{0.49\textwidth}
        \centering
        \includegraphics[width=0.95\linewidth]{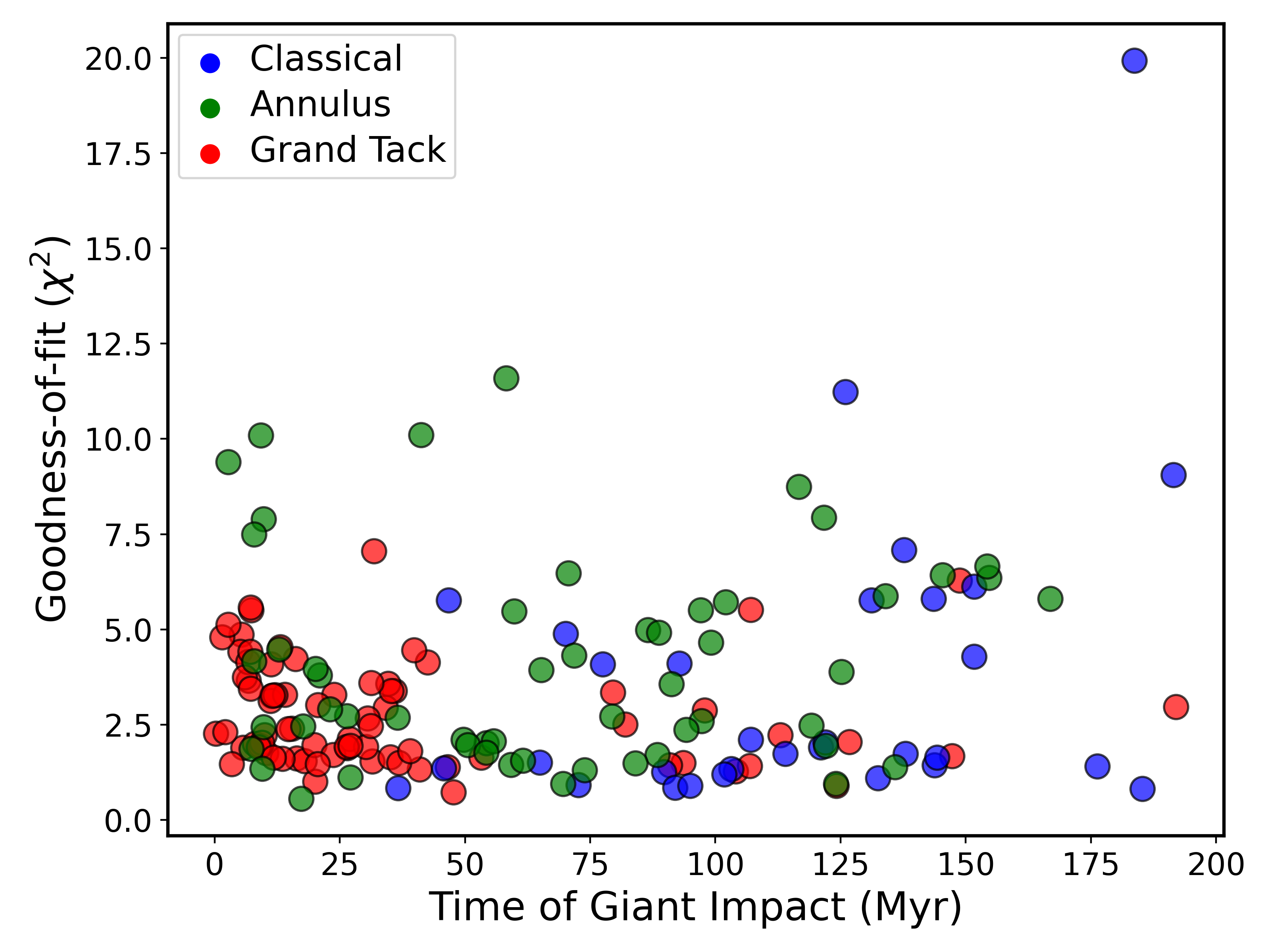}
        \phantomsubcaption
        \label{fig:chisq_vs_time_giant_impact}

        \vspace{0.2em}
        (b)
    \end{subfigure}

    \caption{
    \GIN{(a) Mass of late accretion to the Earth analog is shown with respect to the timing of the last giant impact. 
    % The late accreted mass by each Earth-like planet after the last giant (i.e., analog Moon-forming) impact is correlated with when the Moon-forming impact occurs. 
    Later giant impacts result in less material accreted to the Earth analog.
    Simulations in which Theia is the final impactor have no late-accreted mass and are shown on the dotted line at $10^{-5}\,M_\oplus$ for visualization in log space.
    (b) Goodness-of-fit metric ($\chi^2$) of Earth analogs is shown with respect to the timing of the last giant impact. 
    There is no trend between the timing of the giant impact and the goodness-of-fit metric.
    In each subfigure, the colors of markers denote from which scenario type the particular simulation originates.}
    }
    \label{fig:giant_impact_timing_combined}
\end{figure*}

The timing of delivery of the Theia impactor and the mass of material delivered to Earth after the Theia impact vary between simulations. 
In general, the later the Moon-forming giant impact on the Earth analog, the less material is added afterwards, a trend which is shown in Figure~\ref{fig:late_accretion_vs_time_giant_impact}.
Planetesimal impacts occur out of a large population of objects, and so their accretion rate follows a steady decline with time governed by large-number statistics, whereas giant embryo impacts occur out of a smaller population of objects, and so their accretion rate is not steady and is governed by small-number statistics \GIN{\citep[e.g.,][]{bottke2010stochastic}}.
This result was first reported for Classical and Grand Tack simulations \citep{jacobson2014highly} and has also been reported for Ring scenario simulations \citep{woo2024terrestrial}.
While this correlation appears in all scenarios, its strength and appearance vary between scenarios.

The mass of late accretion depends on broad features of formation scenario type as well as stochastic features of individual simulations. 
For instance, as shown in Figure~\ref{fig:late_accretion_vs_time_giant_impact}, Grand Tack scenario simulations exhibit a wide range in the timing of Moon-forming impacts and the mass of late accretion. 
In contrast, Annulus scenario simulations tend to deliver lower amounts of late accreted mass. 
This is a result of more uniform and closely packed material in the Annulus scenario of Solar System formation as compared to the Grand Tack scenario. 

The mantle highly siderophile element budget provides a direct constraint on the total mass of late accretion and, in turn, the timing of the Moon-forming impact \citep{jacobson2014highly}. 
Because highly siderophile elements (HSEs) partition extremely strongly into metallic phases, any significant core formation episode will efficiently remove HSEs from the mantle.
Accordingly, the nearly chondritic HSE abundance in the bulk silicate Earth may be interpreted as a 0.5wt\% addition of material added after the Moon-forming impact \GIN{\citep[e.g.,][]{walker2009highly,jacobson2014highly}}.
For Grand Tack and Classical scenario simulations, this requires a final giant impact close to 100 Myr into Solar System history, while some Annulus scenario simulations accrete much faster and could be consistent with a Moon-forming impact only 10 Myr after the start of Solar System formation.

\subsection{Moon-forming impact timing does not control final mantle composition} \label{sec:timingcomposition}

The ability to reproduce an Earth mantle with composition close to the bulk silicate Earth is unrelated to the timing of the delivery of Theia. 
This lack of correlation is shown in Figure~\ref{fig:chisq_vs_time_giant_impact}, which demonstrates that there is no trend between the timing of the Moon-forming impact and the goodness-of-fit of the Earth analog in a simulation. 
Furthermore, across all simulations, we \GIN{find} no relation between the goodness-of-fit metric ($\chi^2$) and how much late accretion is added after the Theia impactor is delivered to the Earth, as shown in Figure~\ref{fig:chisq_vs_mass_late_accretion}. 
Given that the timing of the last impact and the late accreted mass are correlated, and that no correlation exists between the timing of the giant impact and the goodness-of-fit, this result is expected.

\begin{figure}[pos=bt!]
    \centering
    \includegraphics[width=0.95\columnwidth]{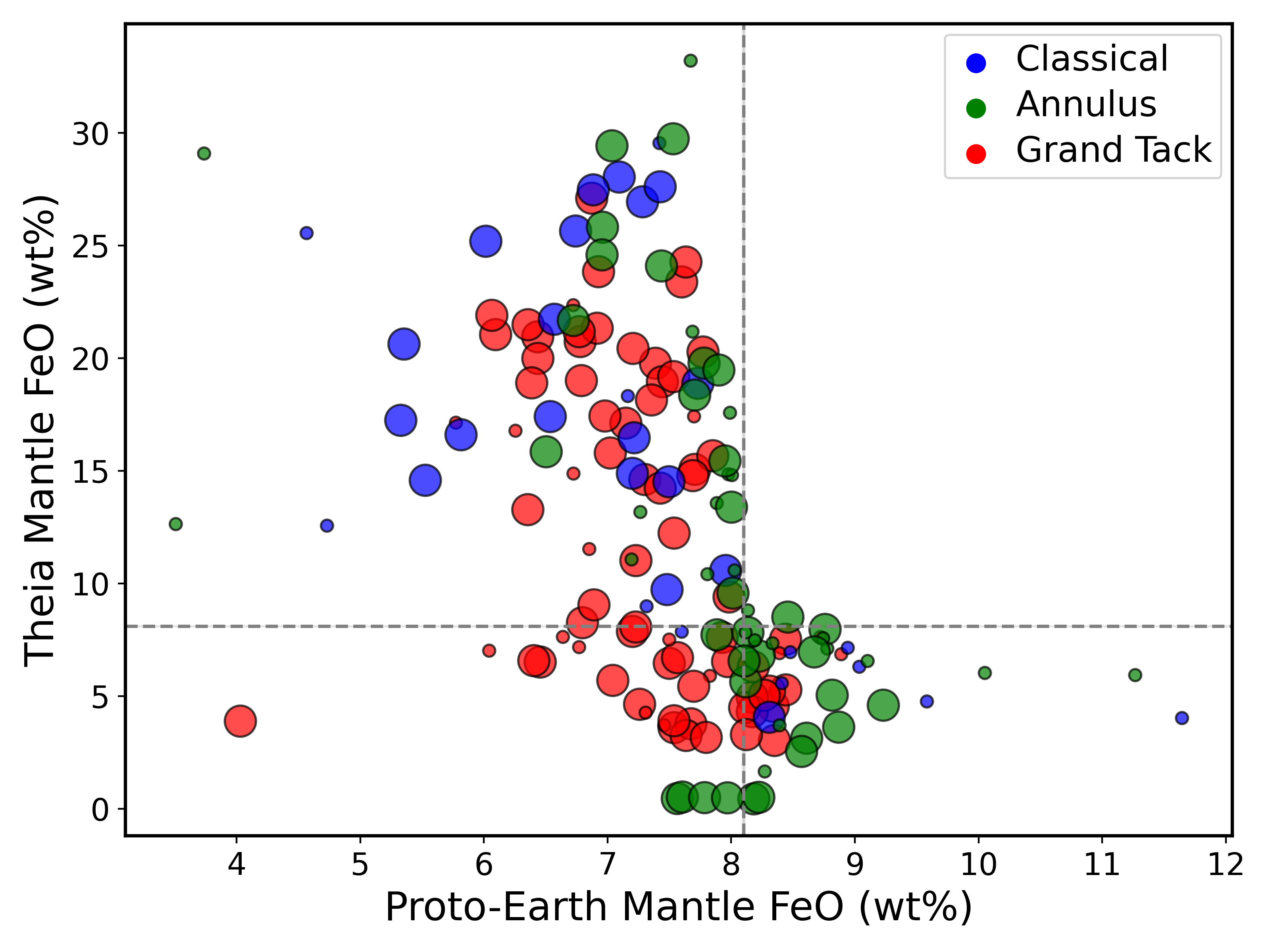}
    \caption{Either Theia or Proto-Earth (but not both) may have a mantle FeO content that exceeds that of the current Earth’s mantle. Theia and Proto-Earth mantle wt.\% FeO are shown for each simulated Earth analog (colors indicate the Earth analog’s formation scenario). The size of the marker represents the goodness-of-fit metric ($\chi^2$) of the final Earth analog in the simulation; a large marker indicates a better fit ($\chi^2<4$) and a small marker indicates a worse fit ($\chi^2>4$). The dashed lines show the final Earth FeO 8.1 wt.\%, forming four quadrants for combinations of Theia and Proto-Earth FeO compositions with mantle wt.\% FeO either higher or lower than that of the present-day Earth.}
    \label{fig:Theia_FeO_vs_protoEarth_FeO}
\end{figure}

\subsection{Many Proto-Earth and Theia compositions are viable, but depend on each other}

% There is a wide range of viable compositions for Earth and Theia
Proto-Earth and Theia analogs span a wide range of mantle compositions and mass ratios and yet, when combined, are able to reproduce Earth’s mantle composition. 
Figure~\ref{fig:Theia_FeO_vs_protoEarth_FeO} shows the range of mantle FeO compositions that Theia and Proto-Earth obtain in the simulations studied here. 
Although resulting Earth analogs reproduce the present-day Earth mantle FeO of $\sim8.1$ wt.\%, neither the Proto-Earth nor Theia need to match that mantle wt.\% FeO themselves. 
Proto-Earth analog mantle FeO compositions range from $\sim$4 to $\sim$12 wt.\% while Theia analog mantle FeO compositions range from $\sim$0 wt.\% (an initially fully or nearly fully reduced Theia) to as much as $\sim$30 wt.\%. 

\begin{figure*}[bt!]
    \centering

    % Top row: centered panel
    \begin{subfigure}[t]{0.49\textwidth}
        \centering
        \includegraphics[width=0.95\linewidth]{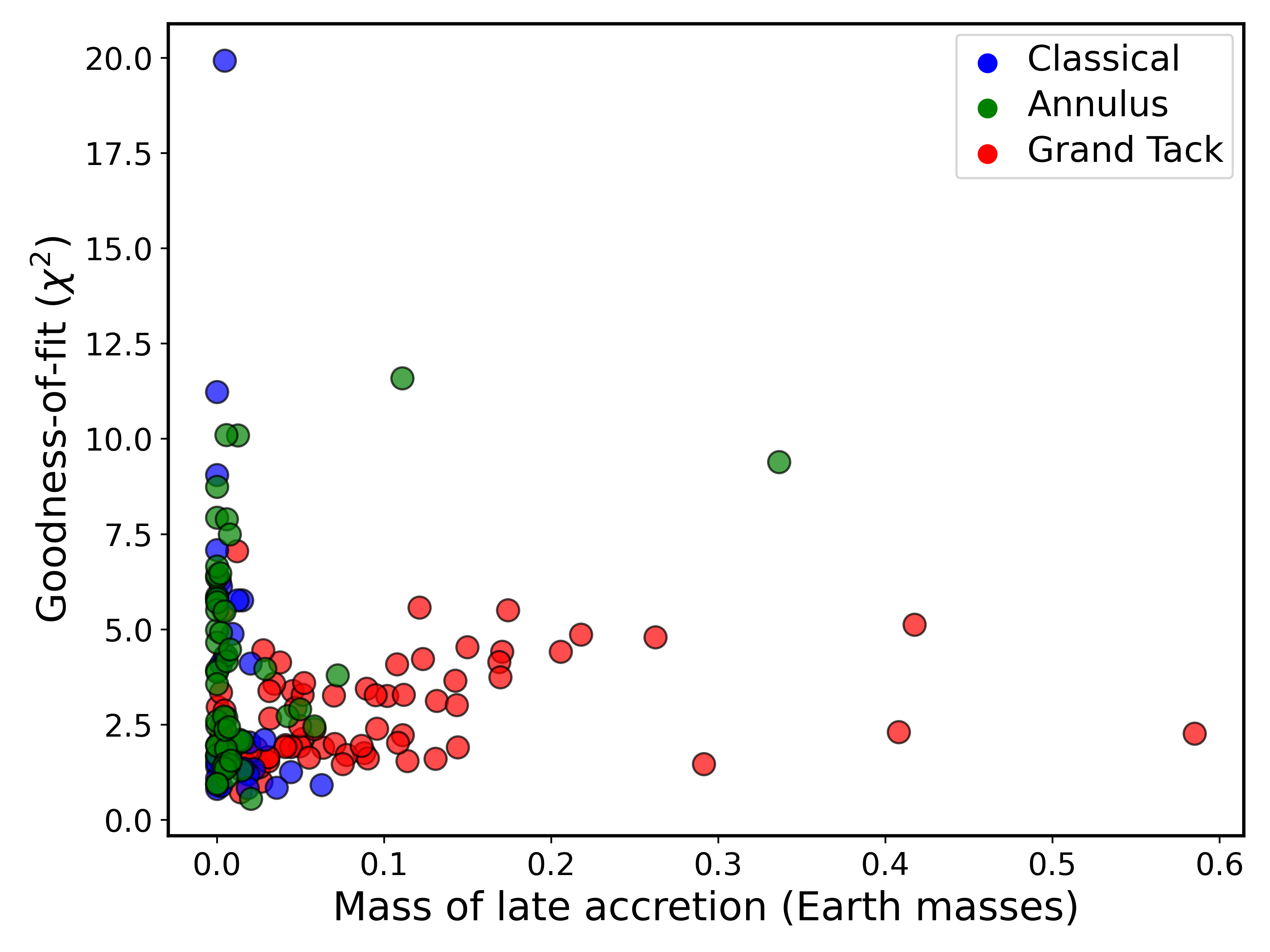}
        \phantomsubcaption
        \label{fig:chisq_vs_mass_late_accretion}

        \vspace{0.2em}
        (a)
    \end{subfigure}

    \vspace{0.5em}

    % Bottom row: two panels
    \begin{subfigure}[t]{0.49\textwidth}
        \centering
        \includegraphics[width=0.95\linewidth]{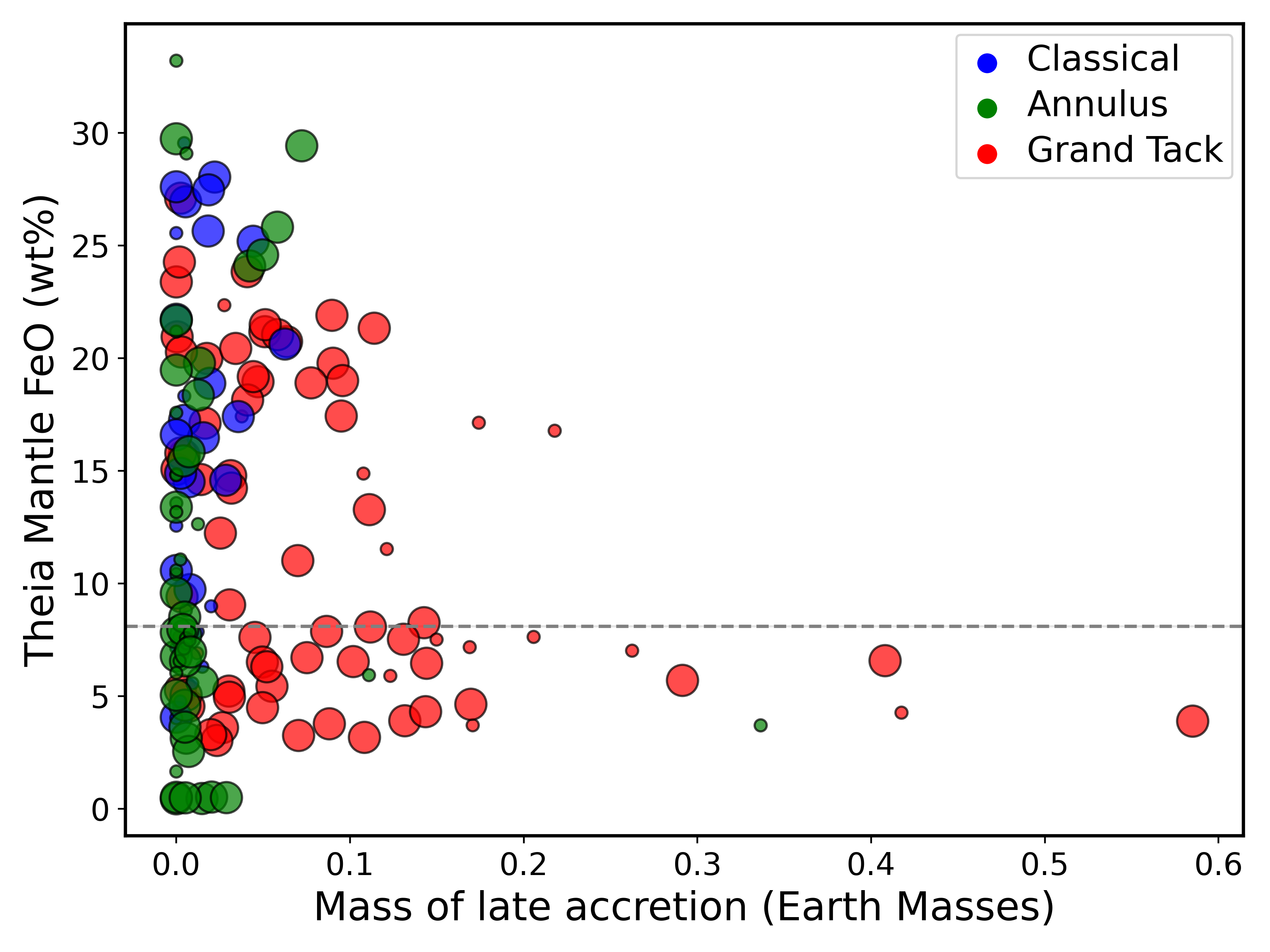}
        \phantomsubcaption
        \label{fig:Theia_FeO_vs_late_accretion}

        \vspace{0.2em}
        (b)
    \end{subfigure}
    \hfill
    \begin{subfigure}[t]{0.49\textwidth}
        \centering
        \includegraphics[width=0.95\linewidth]{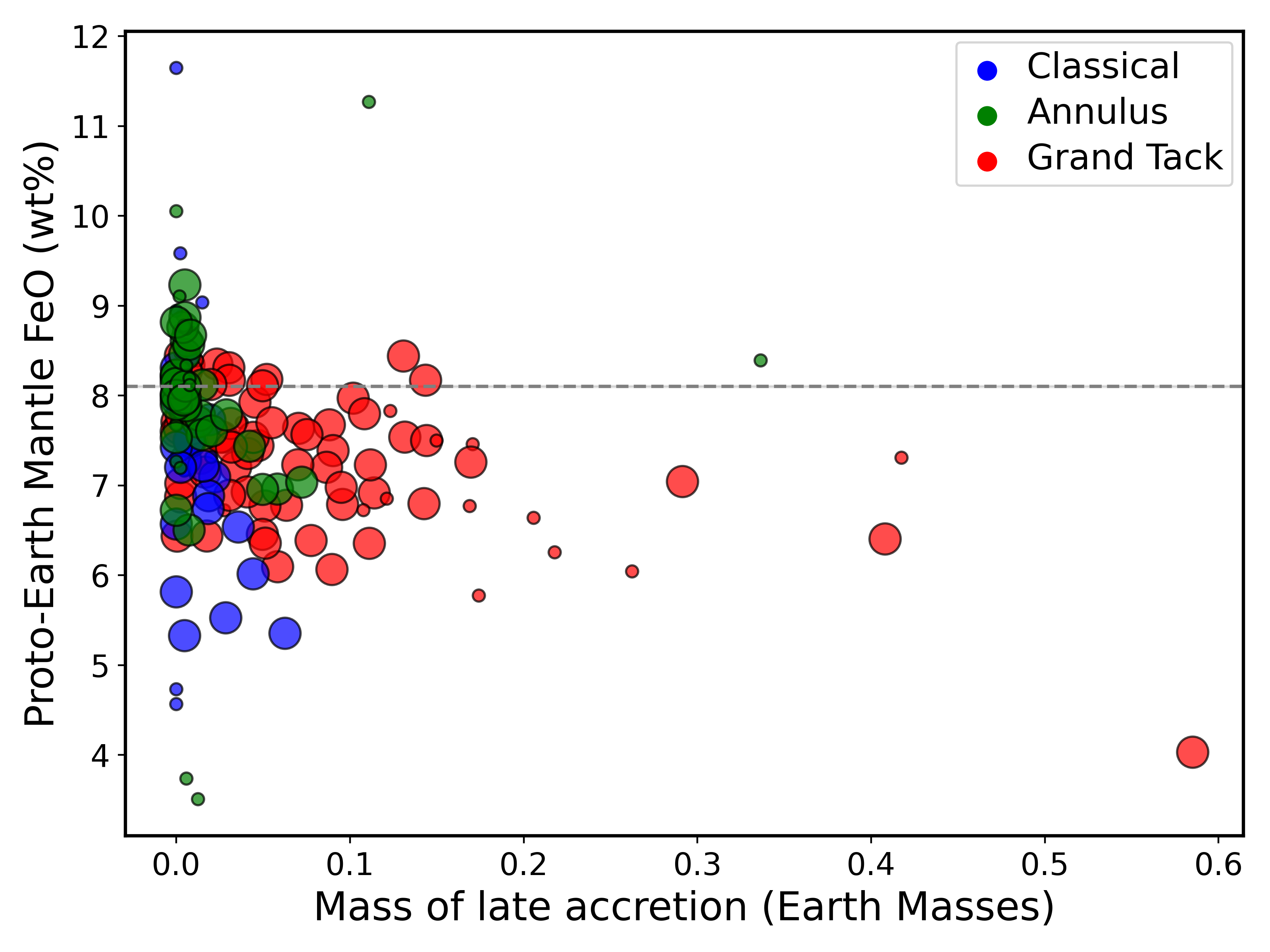}
        \phantomsubcaption
        \label{fig:protoEarth_FeO_vs_late_accretion}

        \vspace{0.2em}
        (c)
    \end{subfigure}

    \caption{
    \GIN{(a) Goodness-of-fit metric ($\chi^2$) of a simulated Earth analog mantle is shown with respect to the mass of late accretion delivered to the Earth analog. 
    Higher $\chi^2$ values indicate Earth analog mantles whose compositions are further from the observed bulk silicate Earth. 
    In each subfigure, markers represent individual Solar System formation simulations and colors correspond to the scenario type of that particular simulation.
    (b)Theia mantle FeO composition is shown with respect to the mass of late accretion delivered to the Earth analog. 
    In this subfigure and in (c), large markers indicate a better fit ($\chi^2<4$) and small markers indicate a worse fit ($\chi^2>4$) of the Earth analog mantle composition to the present-day Earth mantle composition. 
    The dashed line shows the FeO composition of the present-day Earth mantle (8.1 wt.\%).
    (c) Proto-Earth mantle FeO composition is shown with respect to the mass of late accretion delivered to the Earth analog. 
    The mantle wt.\% FeO of a Proto-Earth analog is typically near or below the current bulk silicate Earth's wt.\% FeO, even when there is a significant mass of late accretion. 
    }
    }
    \label{fig:late_accretion_FeO_combined}
\end{figure*}

\begin{figure*}[bt!]
    \centering

    % ============================================================
    % Top row
    % ============================================================

    % Top row: centered panel
    \begin{subfigure}[t]{0.49\textwidth}
        \centering
        \includegraphics[width=0.95\linewidth]{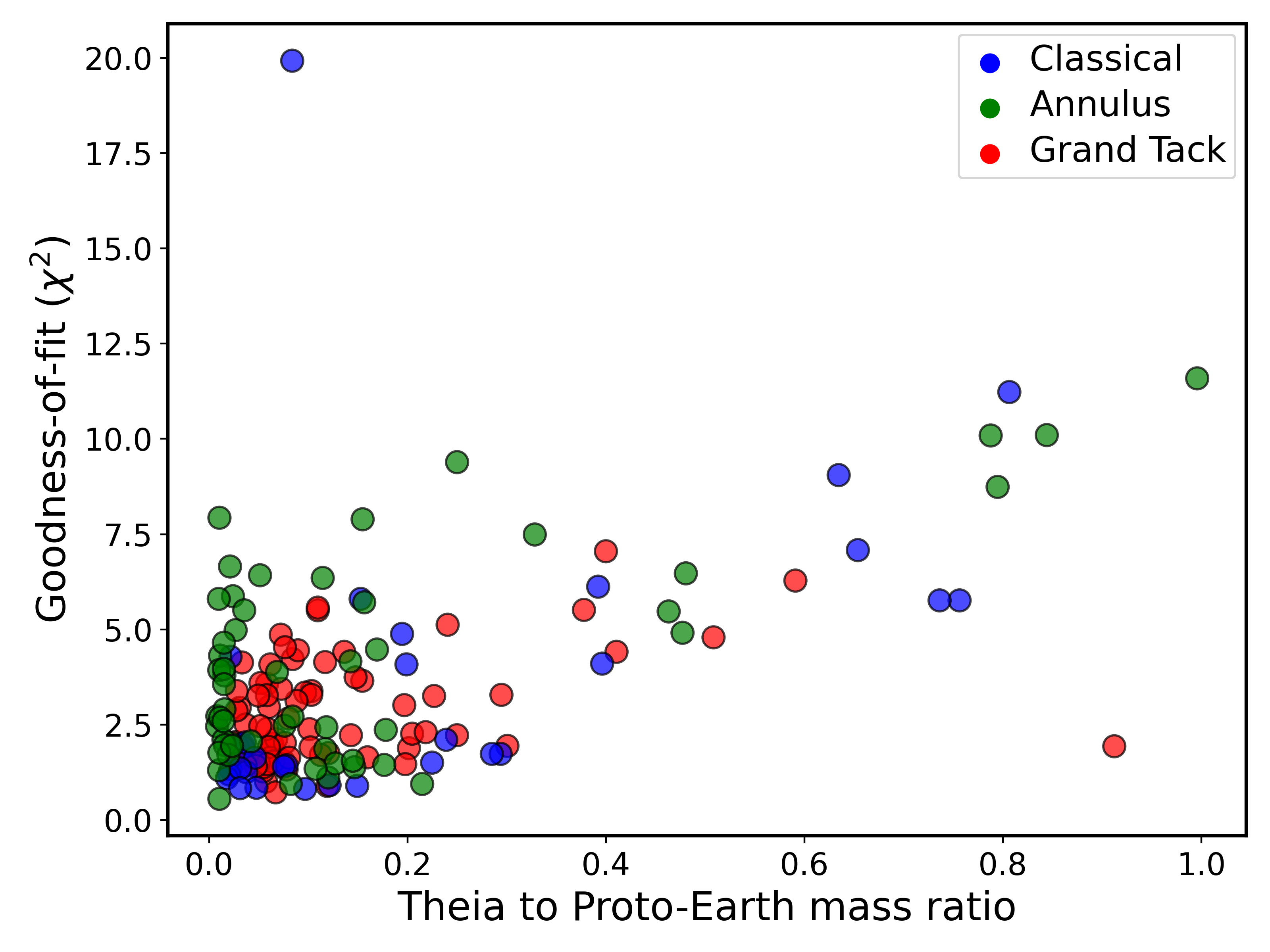}
        \phantomsubcaption
        \label{fig:chisq_vs_mass_ratio}

        \vspace{0.2em}
        (a)
    \end{subfigure}

    \vspace{0.5em}

    % Bottom row: two panels
    \begin{subfigure}[t]{0.49\textwidth}
        \centering
        \includegraphics[width=0.95\linewidth]{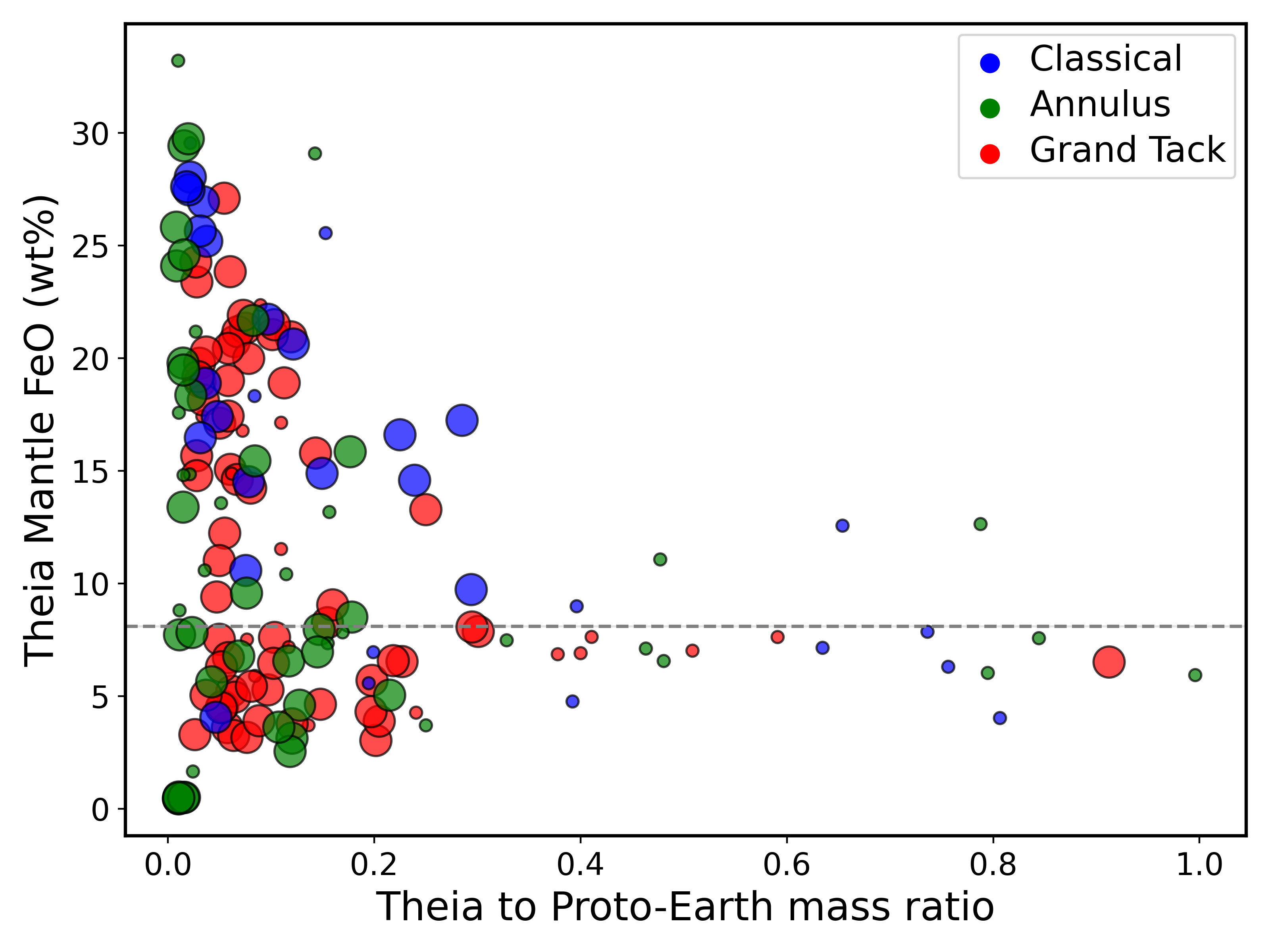}
        \phantomsubcaption
        \label{fig:Theia_FeO_vs_mass_ratio}

        \vspace{0.2em}
        (b)
    \end{subfigure}
    \hfill
    \begin{subfigure}[t]{0.49\textwidth}
        \centering
        \includegraphics[width=0.95\linewidth]{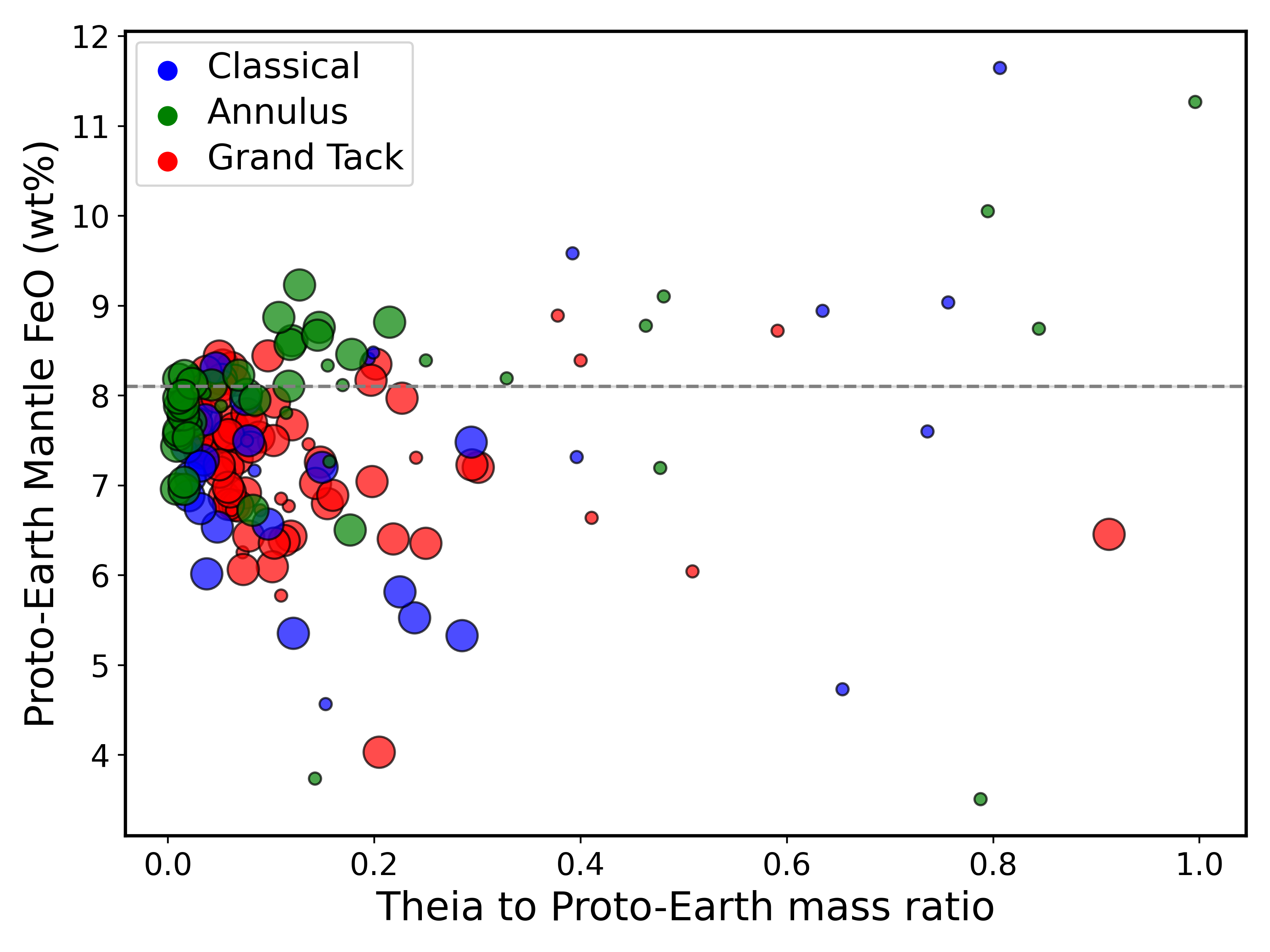}
        \phantomsubcaption
        \label{fig:protoEarth_FeO_vs_mass_ratio}

        \vspace{0.2em}
        (c)
    \end{subfigure}

    \caption{
    \GIN{
    (a) Goodness-of-fit metric ($\chi^2$) of a simulated Earth analog mantle is shown with respect to the Theia to Proto-Earth mass ratio. 
    Larger Theia to Proto-Earth mass ratios result in higher $\chi^2$ values, indicating Earth analog mantles whose compositions are further from the observed bulk silicate Earth. 
    In each subfigure, markers represent individual Solar System formation simulations and colors correspond to the scenario type of that particular simulation.
    (b) Theia mantle FeO composition is shown with respect to the Theia-to-Proto-Earth mass ratio.  
    In this subfigure and in (c), large markers indicate a better fit ($\chi^2<4$) and small markers indicate a worse fit ($\chi^2>4$) of the Earth analog mantle composition to the present-day Earth mantle composition. 
    The dashed line shows the FeO composition of the present-day Earth mantle (8.1 wt.\%).
    (c) Proto-Earth mantle FeO composition is shown with respect to the Theia-to-Proto-Earth mass ratio.  
    }
    }
    \label{fig:Theia_protoEarth_mass_ratio_combined}
\end{figure*}

In Figures~\ref{fig:Theia_FeO_vs_protoEarth_FeO}, \ref{fig:Theia_FeO_vs_late_accretion}, \ref{fig:protoEarth_FeO_vs_late_accretion}, \ref{fig:Theia_FeO_vs_mass_ratio}, and \ref{fig:protoEarth_FeO_vs_mass_ratio}, markers are reduced in size if the $\chi^2$ goodness-of-fit metric is greater than four for that simulation. 
Because the mean $\chi^2$ goodness-of-fit value achieved for the simulations in this study is $3.39 \pm 2.58$ and the median $2.47$, we \GIN{choose} a cutoff of $\chi^2 > 4.0$ because it is a threshold above the mean, and so helps to visually display which simulations perform better than the mean and produce the \GIN{Earth analogs whose bulk and moderately siderophile element composition is closest to the observed BSE}.

The mantles of Proto-Earth analogs tend to be more reduced than that of present-day Earth, primarily because material accreted earlier in Earth's growth is, on average, more reduced than material accreted later, as discussed in Section \ref{results2}.
As a result, the mantle wt.\% FeO of a simulated Proto-Earth analog is typically near or below the present-day Earth mantle wt.\% FeO, as shown in Figure~\ref{fig:protoEarth_FeO_vs_late_accretion}.
Because Theia and late accretion both tend to be more oxidized than Proto-Earth, the Proto-Earth itself must be comparatively reduced in order to reproduce the mantle composition of the present-day Earth.
Theia, on the other hand, can exhibit a much wider range of mantle wt.\% FeO, even when very little mass is delivered after its impact, as shown in Figure~\ref{fig:Theia_FeO_vs_late_accretion}. 
Because the mass of late accretion correlates with timing of Theia's delivery and earlier delivered material is more reduced, Theia analogs tend to be more reduced when delivered early and followed by a large amount of late accretion.

Larger Theia to Proto-Earth mass ratios result in less accurate mantle compositions for the final Earth analog. 
If the mass of the Theia analog approaches the mass of the Proto-Earth, it is harder to achieve a composition for the simulated Earth mantle that matches the BSE, a trend that is shown in Figure~\ref{fig:chisq_vs_mass_ratio}. 
\GIN{The most straightforward solution might seem to be the case where the mass of Theia is vanishingly small, therefore causing the least deviation of the Earth body from its composition. 
This, however, is not shown by the trend in Figure~\ref{fig:chisq_vs_mass_ratio}; while there is a strong correlation between Theia to Proto-Earth mass ratio and goodness of fit ($\chi^2$), this correlation appears only above mass ratios of $\sim0.3$. 
Below this value, there is little correlation between the size of Theia compared to the Proto-Earth and the ability to reproduce the bulk and trace element composition of the Earth mantle.}

Theia can have a wide range of FeO compositions so long as Theia is small relative to Proto-Earth, as shown in Figure~\ref{fig:Theia_FeO_vs_mass_ratio}.
In this Figure, simulations with higher (worse) goodness-of-fit metric ($\chi^2\ge4$) are shown by smaller markers.
If Theia is larger than approximately 0.3 the mass of the Proto-Earth, then the final Earth analog mantle composition is not as good a match to that of the present-day Earth mantle.
When the mass ratio of Theia to Proto-Earth increases, the wt.\% FeO in Theia's mantle is lower and closer to that of the present-day Earth mantle. 
This is because the larger Theia is, the more direct influence it exerts on the final Earth composition. 
Thus, the larger Theia is, the closer to the present-day Earth composition it needs to be in order to build an Earth analog that actually resembles the BSE.
The simulation that plots near 0.9 Theia to Proto-Earth mass ratio while still achieving a chi-squared in the better 50\% is an outlier Earth analog that accretes nearly 90\% of its mass after the delivery of Theia.

The Proto-Earth mantle displays a relatively more restricted range of mantle FeO wt.\% ($\sim$4 to 9 wt.\%) while still successfully matching the final Earth BSE if the Theia contribution is less than $\sim$0.3 the mass of Proto-Earth. 
Contributions of Theia greater than $\sim30$\% of the mass of Proto-Earth rarely result in good fits to the Earth mantle, regardless of the Proto-Earth FeO wt.\%. 
This can be seen in Figure \ref{fig:protoEarth_FeO_vs_mass_ratio}, where simulations
with higher (worse) goodness-of-fit metric ($\chi^2 \ge4$) are
shown by smaller markers.

\clearpage
\twocolumn
\section{Discussion} \label{discussion}

\subsection{Compositional redox gradient in the protoplanetary disk}

Using a core–mantle differentiation model, we studied several different classes of Solar System formation scenarios and demonstrated that it is possible to self-consistently build a simulated Earth analog whose mantle composition matches that of the bulk silicate Earth in all the formation scenarios studied.
This is true if two conditions are fulfilled:
\begin{samepage}
\begin{enumerate}
\item There must be a redox gradient in the protoplanetary disk that allows the Earth to be built from planetesimals and embryos with a wide range of oxidation states.
\item The Earth must accrete material that is (on average) initially reduced and later material that is more oxidized.
\end{enumerate}
\end{samepage}
If we impose a compositional redox gradient that cuts across the building materials of the Earth, a good match to the Earth mantle is achievable, regardless of the particular growth history of the planets in the simulation. 

We found that reproducing the Earth requires a rapidly oxidizing gradient near 1 AU. 
That gradient must become oxidizing over a distance of 0.1 to 0.2 AU in nearly all simulations to reproduce the Earth (see the values for $\delta(1)$ and $\delta(2)$ in Table~\ref{table_of_best_simplexes}).
This means that to build the Earth mantle analog with lowest $\chi^2$, a sharp change from reduced in the innermost part of the protoplanetary disk to oxidized further out is typically required.
This is true even in scenarios where initial mass is distributed widely throughout the protoplanetary disk (such as Classical scenarios like EJS and CJS), where a sharp change from reducing to oxidizing provides the best conditions to reproduce the Earth mantle most closely.

The material accreted to the growing Earth becomes gradually more oxidized over the course of its formation.
This reproduces results of \citet{rubie2015accretion}, but in a more robust manner because we include a wide range of Solar System formation scenarios, beyond just the Grand Tack \citep{walsh2011effect}.  
The Earth cannot have accreted with the reduced material added at the very end of its accretion.
The material that formed the Earth must have been material accreted in order from reduced to oxidized.
\citet{siebert2013terrestrial} note that the partitioning behavior of several siderophile elements (e.g. V, Cr) is evidence that the Earth accreted in relatively oxidizing conditions. 
However, such a scenario is not compatible with the model presented in \citet{rubie2015accretion}, which demonstrated that Earth must first accrete reduced material and then more oxidized material because Si strongly partitions into metal even under oxidizing conditions.
That same effect applies here:
When a large amount of Si partitions to the core, it leaves excess oxygen in the mantle.
That oxygen is incorporated into FeO, pushing the mantle to an overabundance of FeO.
This follows from the metal silicate equilibration reaction which is valid if equilibration occurs at the Fe-FeO buffer \citep{javoy1995integral, rubie2011heterogeneous}: 
\begin{equation}
\ce{2Fe (met) + SiO2 (sil) -> 2FeO (sil) + Si (met)}
\end{equation} 

If accretion begins too oxidized, later core formation cannot efficiently reduce the mantle FeO composition.
Moreover, if mantle oxygen fugacity becomes too high, partitioning of moderately siderophile elements can change, which makes it difficult to match their observed mantle abundances.
Because Si partitioning to the core drives up the mantle FeO composition at a 2:1 rate, this process exerts a strong control over the accretion of the Earth. 
The challenge is keeping the FeO weight percent low enough that it does not exceed $\sim$8 wt\%; if it does, subsequent partial melting and continued Si partitioning act as a pump that further increases FeO, making later addition of reduced material ineffective at lowering mantle FeO.
This dynamic governs much of the behavior in the simulations and is the key reason that the Earth accretes initially reduced material followed by increasingly oxidized material in successful simulations.

To test the robustness of this result, we conducted additional simulations that varied only the order in which reduced and oxidized building blocks were delivered, independent of any \textit{N}-body accretion history. 
Using Earth-building blocks of idealized oxidation states, we reordered the accretion sequence and found that scenarios in which Earth accretes oxidized material early and then accretes increasingly reduced material produce significantly poorer matches to the bulk silicate Earth, see Figure~\ref{fig:contrived-accretion-test}).
When the only varied parameter is the order in which the Earth accretes oxidized and reduced material, we find that the ordering affects the final composition of the Earth. 
This conclusion supports the broad observation that starting with initially more oxidized material consistently results in mantle compositions that fail to match the abundances of bulk elements and the trace abundances of moderately siderophile elements in the Earth's mantle. 

\GIN{While we do not explicitly consider the non-carbonaceous-carbonaceous (NC-CC) meteorite dichotomy \citep{budde2016molybdenum, warren2011stable, kleine2020non} in this study because we did not investigate isotopic compositions in our simulations, we nonetheless consider our results to be compatible with the NC-CC dichotomy. 
Here, we have focused on the Earth's accretion and show that the majority of its accreted material is from reduced planetesimals and embryos, which would correspond most directly to non-carbonaceous material. 
The contribution of mostly or fully oxidized bodies (analogous to CC material) to the Earth is less than $\sim10\%$ (see Figure \ref{fig:CDF-fe-in-metal}).
We show that more reduced material is the main contributor to Earth, consistent with recent estimates that the CC contribution was minimal \citep{sossi2026homogeneous}, and at most 2-10\% the mass of Earth \citep{nimmo2024mechanisms}.
This is in agreement with additional studies arguing for low CC contribution to the Earth \citep[e.g.,][]{dauphas2017isotopic,burkhardt2021terrestrial, dauphas2024bayesian}.}

\subsection{Moon-formation scenarios compatible with producing the bulk silicate Earth}

Prior studies of the formation of the Earth-Moon system have historically followed two parallel approaches: dynamical modeling of the Moon-forming impact \citep[e.g.,][]{canup2001origin} and determining geochemical constraints derived from terrestrial and lunar samples.
Properties such as the angular momentum of the Earth-Moon system and the mass ratio of the Moon to the Earth constrain smoothed particle hydrodynamics simulations of planetary impacts \citep[e.g.,][]{canup2012forming,kegerreis2022immediate}.
Geochemical and isotopic studies comparing lunar and terrestrial samples have shown remarkable similarity between the Earth and the Moon across a wide range of isotopic systems \citep[e.g.,][]{wiechert2001oxygen}, which complicates the origin history of the Earth and Moon.
This study integrates traditionally separate approaches by combining geochemical modeling with simulations of terrestrial planet formation to constrain the masses and compositions of Proto-Earth and Theia.
Using many simulations of Solar System formation, we constrained the viable compositions and mass ratios for Theia and Proto-Earth that successfully reproduce planetary analogs with the bulk composition of the present-day Earth mantle. 

\GIN{Across all Solar System formation scenarios and simulations we examined, material accreted later to the Earth is consistently more oxidized than earlier accreted material, which has consequences for the compositions of Theia and Proto-Earth.
Accretion that begins reduced and later is more oxidizing is also compatible with a natural result of the orbital dynamics at play during planet formation.
Materials that originate farther from the Sun interact less frequently with embryos at 1 AU, so if outer material is more oxidized, its later delivery is consistent with accretion dynamics \citep[e.g.,][]{Raymond2017};
more oxidized material will naturally be delivered later during the accretion of the Earth.}

We find that smaller Theia analogs produce Earth analogs with mantles that are closer matches to the composition of the bulk silicate Earth, and accordingly, we consider Moon-formation scenarios that involve a large, oxidized Theia \citep[e.g.,][]{Desch2019} to be less compatible with producing the bulk silicate Earth.
Several scenarios have been proposed where Theia is 0.4$M_\oplus$ (see \citet{Desch2019, johansen2021pebble,Herwartz2014}). 
\citet{Desch2019} proposed that a large Theia (40\% the mass of Earth) could incorporate hydrogen while it is molten and in order to account for the Moon's anomalous $\delta$D composition. 
This scenario involves a very large Theia with 15 wt.\% FeO in its mantle. 
Such a scenario is rare in the simulations we examined, and when it occurs, it does not result in a subsequent Earth analog that is among the most successful matches to the BSE.
Successful Earth analogs occur with a wide range of Theia mantle FeO abundances ($\sim0-30$ wt.\%), as long as Theia is small. 
Proto-Earth displays a range of FeO abundances ($\sim$4 wt.\% to 10 wt.\%), but the mantles of Theia and Proto-Earth cannot both exceed the present-day Earth's mantle FeO abundance.
The best Earth analogs occur when Theia is small and the mantles of both Theia and Proto-Earth possess equal to or less than the FeO  wt.\% of the present-day Earth's mantle.
\GIN{This restricted parameter space for Theia and Proto-Earth is shown in Figures \ref{fig:Theia_FeO_vs_protoEarth_FeO}, \ref{fig:late_accretion_FeO_combined}, and \ref{fig:Theia_protoEarth_mass_ratio_combined}. }
\GIN{
Figure~\ref{fig:Theia_FeO_vs_protoEarth_FeO} indicates the solution space for Theia and Proto-Earth mantle FeO compositions divided into four quadrants by the present-day Earth-mantle FeO composition; solutions are present in quadrants two, three, and four but nearly absent in quadrant one. 
Solutions in the first quadrant indicate a Proto-Earth and Theia with more oxidizing mantles than that of the present-day Earth.
Solutions in the second and fourth quadrant indicate one or the other of Theia and Proto-Earth has a more oxidizing mantle than the present-day Earth.
Solutions in the third quadrant of Figure~\ref{fig:Theia_FeO_vs_protoEarth_FeO} represent simulations in which the mantles of both Theia and Proto-Earth were more reduced than the present-day Earth mantle. 
In this case, this indicates a reducing environment for the formation of Theia and Proto-Earth followed by a more oxidizing late accretion delivered to the Earth after the last embryo impactor (Theia) in the form of smaller relatively oxidized planetesimals.
}

A smaller, reduced Theia is generally compatible with the scenario described in \citet{Wade2016}, in which a $\sim 0.1-0.2 M_\oplus$ impactor with $\sim0.3$ mantle wt.\% FeO strikes a moderately oxidized Proto-Earth ($\sim10.7$ mantle wt.\% FeO).
This scenario is on the upper bound of the model's observed Proto-Earth mantle FeO abundance.
As discussed in \citet{Wade2016}, a highly reduced Theia would also result in the delivery of sulfur to the lunar mantle, providing conditions for a lunar Hadean matte and, in turn, segregating lunar chalcophile elements to the Moon's core.
Nucleosynthetic Fe isotopic measurements of lunar material also support an inner Solar System origin for Theia \citep{hopp2025moon}, which is consistent with the finding that Theia cannot be both large and highly oxidized.

The isotopic similarity of the Earth and Moon suggests that some mixing or equilibration must have occurred during or after the Moon-forming impact.
Our mass and compositional constraints show that a small, high-velocity Theia is more compatible with producing the Earth than an equal-mass impact model.
Prior literature attempts to reconcile the isotopic similarity of the Earth and Moon with the delivery of significant mass to the Earth via Theia include the following general approaches: 
\begin{enumerate}
    \item Some form of post-impact thermodynamic equilibration in the proto-lunar disk \citep[e.g.,][]{pahlevan2007equilibration} or in a high-energy synestia \citep[e.g,][]{lock2018origin}.
    \item Physical mixing between Proto-Earth and Theia to recreate nearly identical Earth and Moon, which plausibly involves a Proto-Earth and Theia of nearly similar masses \citep[e.g.,][]{canup2012forming}.
    \item Initial isotopic similarity between Proto-Earth and Theia, arising from similar formation locations in the disk \citep[e.g.,][]{Mastrobuono-Battisti2015} or homogeneous pebble flux \citep[e.g.,][]{Johansen2021}. 
\end{enumerate}
 
While the simulations in this study do not definitively rule out any of these mechanisms, they instead show that equal-mass mixing models of Proto-Earth and Theia are less compatible with producing the Earth and favor equilibration processes that involve a smaller Theia.
Because late accretion cannot reduce Earth’s mantle FeO after the Moon-forming impact, the Proto-Earth must have had a mantle FeO abundance lower than that of the present-day Earth at the time of the collision. 
Combined with the compositional flexibility afforded by a smaller Theia, this indicates that initial similarity scenarios involving a small, high-velocity Theia are the most compatible with the model.
Equal-mass impact scenarios remain possible but require particularly fine-tuned initial conditions for Proto-Earth and Theia and therefore appear less likely within the framework of the simulations.

\begin{figure}[pos=bt!]
    \centering
    \includegraphics[width=0.95\columnwidth]{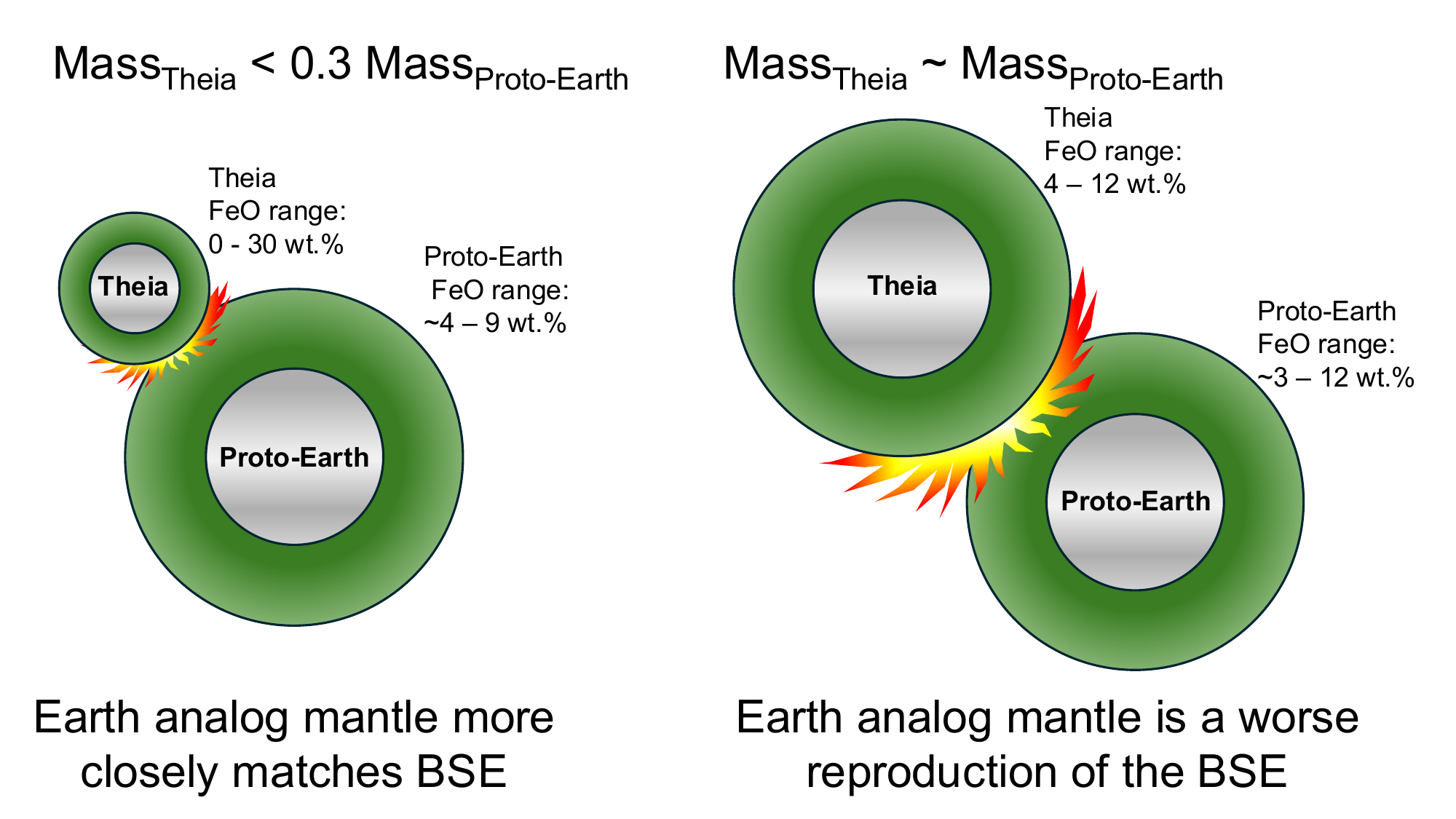}
    \caption{This schematic shows the relationship between Theia-to-Proto-Earth mass ratio, ranges of FeO mantle abundances in Theia and the Proto-Earth, and the outcome of the Earth analog in these two scenarios.}
    \label{fig:ch3illustration1}
\end{figure}

When Theia is small, there are a wide range of mantle FeO wt.\% abundances for both Theia and the Proto-Earth that produce an Earth analog with mantle composition close to that of the BSE. 
In contrast, if Theia is close in size to the Proto-Earth, the range of acceptable mantle FeO wt.\% abundance for Theia becomes much narrower, and the resulting Earth analogs do not as accurately reproduce the BSE.
A visual summary of these constraints is shown in Figure~\ref{fig:ch3illustration1}. 

\subsection{Caveats}
While we have shown that many different types of formation scenarios produce an Earth analog whose mantle composition matches the bulk silicate Earth, there is a major caveat to this result. 
\GIN{Our model }allows substantial flexibility in adjusting the redox gradient of the solids in the protoplanetary disk. 
This helps achieve good fits, but does not solely explain why nearly all successful simulations require a sharp redox transition over only 0.1–0.2 AU in the protoplanetary disk.
That feature, and the observation that the Earth accretes initially reduced and later oxidized material, is due to the geochemical constraints of the Earth mantle and is not an outcome of overfitting to a particular formation scenario. 

\GIN{This suggests that, instead of asking if a redox gradient can reproduce the Earth's mantle, it may be more informative to determine if the redox gradient required by a particular simulation is physically plausible.}
For example, narrow-annulus models demand a particularly rapid transition from reduced to oxidized material over a small radial distance, which may not be feasibly obtained in a protoplanetary disk.
Alternatively, some dynamical Solar System formation scenarios may not naturally deliver oxidized material later than reduced material.
Isotopic constraints and the compositions of the other terrestrial planets can serve as further constraints; perhaps it is not possible to match the composition of the other inner terrestrial planets in certain formation scenarios.

\GIN{
Additionally, we note that our treatment of equilibration temperature does not consider superheating following planetary impacts.
As noted in Section \ref{sec:methods}, we consider the temperature of equilibration to be an intermediate value between the liquidus and solidus of peridotite at a given pressure. However, kinetic energy from large planetary impacts can concentrate and result in temperatures heated significantly above the liquidus \citep[e.g.,][]{lock2017structure}.  
This superheating could affect chemical partitioning, though we do not anticipate significant differences between Solar System formation scenarios due to this process.
}

\section{Conclusions}
We tested the ability of many different dynamical Solar System formation scenarios to produce the mantle composition of the Earth, and what implications this accretion history had on the composition and size of Proto-Earth and Theia.
The six types of Solar System formation scenarios we tested include: Grand Tack, Circular Jupiter-Saturn, Eccentric Jupiter-Saturn, Truncated Disk, Steep Profile, and Ring Model. 
To do this, we post processed \textit{N}-body accretion histories with a core-mantle differentiation simulation (\texttt{NB-Differentiate}) first presented in \citet{rubie2011heterogeneous, rubie2015accretion}. 
These coupled \textit{N}-body accretion and core-mantle geochemical models provided us with simulated Earth analog compositional evolutions. 
Free parameters in this model control the initial redox state of materials in the protoplanetary disk and the pressures of metal-silicate equilibration during core formation until an Earth analog in a simulation matches the bulk and trace element composition of the Bulk Silicate Earth (BSE) as accurately as possible.

We found that: 
\begin{enumerate}
    \item The mantle of the Earth can be accurately reproduced via many different types of dynamical pathways, if a redox gradient exists in the protoplanetary disk that cuts across the building block materials of Earth and thus the Earth is built from a range of reduced and oxidized material. 
    \item The delivery of increasingly oxidized materials throughout accretion is a consistent outcome of simulations that form \GIN{Earth analogs with bulk and moderately siderophile element abundances matching that of the BSE}.
    \item The order of accretion matters; late delivery of oxidized material results in Earth analogs that more closely resemble the observed bulk silicate Earth.  
    \item Late accretion (mass added after the Moon-forming impactor, Theia) tends to be more oxidizing than earlier stages of accretion, which indirectly restricts the viable combinations of mass and composition of Theia.
    \item If Theia and Proto-Earth were similar in size, they both must have had a mantle FeO abundance similar to that of the present-day Earth.
    \item If Theia was small compared to Proto-Earth, it could have had a wider range of mantle FeO wt.\% abundances (ranging from $\sim$0 wt.\% to $\sim$30 wt.\%).
    \item \GIN{The Earth analog is more accurately reproduced in Solar System simulations when Theia is smaller with respect to Proto-Earth and worse when Theia is larger; this trend is most pronounced at Theia to Proto-Earth mass ratios greater than 0.3.}
\end{enumerate}

Our finding that the Earth formed with increasingly more oxidized material throughout accretion is in agreement with past studies, particularly with the results of \citet{rubie2015accretion}.
\GIN{Alternate formation pathways for terrestrial formation where oxidation does not increase throughout accretion exist \citep[e.g.,][]{badro2015core,Johansen2023}.
Notably, such pathways assume substantially more extensive metal-silicate equilibration after each accretion event than estimated from laboratory fluid experiments \citep[e.g.,][]{deguen2011experiments,deguen2014turbulent} and numerical simulations \citep[e.g.,][]{deVries2016,Nakajima2021}, so reducing an initially oxidized mantle is likely not possible because even very large core forming events do not equilibrate with the entire mantle \citep{rubie2011heterogeneous,rubie2015accretion}.}
With these simulations, we demonstrate that it is possible to form Earth analogs with BSE composition in a wide range of different accretion scenarios, not just the Grand Tack scenario \citep[e.g.,][]{rubie2015accretion} or a narrow ring formation model \citep[e.g.,][]{dale2025compositional}, as in prior studies. 
Further studies of these simulations and the consequences of these redox gradients on the other non-Earth analog planets are reasonable next steps to investigate. 

These findings have a range of implications for the viable formation scenarios for the Earth-Moon system.
Given the relationship between late accreted mass and the timing of the Moon-forming impact, the results are generally agnostic to when the Moon-forming impact occurred.
We demonstrate that a smaller Theia has a wider range of reasonable FeO compositions than a large Theia, which requires an FeO composition closer to that of the present-day Earth. 
To reconcile the Earth-Moon isotopic similarity, some degree of equilibration must have occurred. 
If chemical equilibration and resultant isotopic equilibration occurred due to an energetic impact, then the energy must have come from a high-velocity Theia rather than from an extremely massive Theia. 
Because a large Theia less frequently yields the correct Earth mantle composition, and if it does it is a worse recreation of the mantle than when Theia is small, an equal-mass impact scenario is less feasible.
Our results do not rule out a scenario in which Proto-Earth and Theia had identical chemical and isotopic compositions; in such a case, it is possible for a large Theia to produce the Earth-Moon system. 
However, such a scenario requires a unique set of coincidences in compositions and accretion histories for Proto-Earth and Theia. 
Further work will be required to reconcile the isotopic similarity between the Earth and Moon with this type of modeling approach.

\section*{Acknowledgments}

GIN acknowledges support from the Future Investigators in NASA Earth and Space Science and Technology (FINESST) program Grant No. 80NSSC23K1376.
HDN acknowledges support from the MSU Professorial Assistantship (PA) Program and the Department of Earth and Environmental Sciences Undergraduate Research Fellowship. 
This work was also supported in part through computational resources and services provided by the Institute for Cyber-Enabled Research at Michigan State University.

\section*{\GIN{Data Availability}}
\GIN{
All data, code, and inputs relevant to this study are publicly available through Zenodo at   
\href{https://doi.org/10.5281/zenodo.22717203}{10.5281/zenodo.22717203}.
The Zenodo archive mirrors the GitHub repository
\url{https://github.com/gijnathan/Earth_Theia_accretion_ICARUS2026}.
}

\printcredits

\bibliographystyle{cas-model2-names}
\bibliography{references}

@article{akram2016zirconium,
	title        = {{Zirconium isotope constraints on the composition of Theia and current Moon-forming theories}},
	author       = {{Akram}, W. and {Sch{\"o}nb{\"a}chler}, M.},
	year         = 2016,
	month        = sep,
	journal      = {Earth and Planetary Science Letters},
	volume       = 449,
	pages        = {302--310},
	doi          = {10.1016/j.epsl.2016.05.022},
	adsurl       = {https://ui.adsabs.harvard.edu/abs/2016E&PSL.449..302A}
}

@article{Amano2025,
	title        = {{Updating the Urey-Craig diagram: The iron redox states of the building blocks of the outer solar system}},
	author       = {{Amano}, Kana and {Viennet}, Jean-Christophe and {Beck}, Pierre and {Guyot}, Fran{\c{c}}ois and {Yaroslavtsev}, Sergey and {Bessas}, Dimitrios and {Chumakov}, Aleksandr and {Delbes}, Ludovic and {Poch}, Olivier and {Roskosz}, Mathieu},
	year         = 2025,
	month        = nov,
	journal      = {Earth and Planetary Science Letters},
	volume       = 669,
	pages        = 119587,
	doi          = {10.1016/j.epsl.2025.119587},
	eid          = 119587,
	adsurl       = {https://ui.adsabs.harvard.edu/abs/2025E&PSL.66919587A}
}

@article{Andrews2018,
	title        = {{The Disk Substructures at High Angular Resolution Project (DSHARP). I. Motivation, Sample, Calibration, and Overview}},
	author       = {{Andrews}, Sean M. and {Huang}, Jane and {P{\'e}rez}, Laura M. and {Isella}, Andrea and {Dullemond}, Cornelis P. and {Kurtovic}, Nicol{\'a}s T. and {Guzm{\'a}n}, Viviana V. and {Carpenter}, John M. and {Wilner}, David J. and {Zhang}, Shangjia and {Zhu}, Zhaohuan and {Birnstiel}, Tilman and {Bai}, Xue-Ning and {Benisty}, Myriam and {Hughes}, A. Meredith and {{\"O}berg}, Karin I. and {Ricci}, Luca},
	year         = 2018,
	month        = dec,
	journal      = {The Astrophysical Journal},
	volume       = 869,
	number       = 2,
	pages        = {L41},
	doi          = {10.3847/2041-8213/aaf741},
	eid          = {L41},
	archiveprefix = {arXiv},
	eprint       = {1812.04040},
	primaryclass = {astro-ph.SR},
	adsurl       = {https://ui.adsabs.harvard.edu/abs/2018ApJ...869L..41A}
}

@article{armstrong2019deep,
	title        = {Deep magma ocean formation set the oxidation state of Earth's mantle},
	author       = {Armstrong, Katherine and Frost, Daniel J and McCammon, Catherine A and Rubie, David C and Boffa Ballaran, Tiziana},
	year         = 2019,
	journal      = {Science},
	publisher    = {American Association for the Advancement of Science},
	volume       = 365,
	number       = 6456,
	pages        = {903--906}
}

@article{armytage2011silicon,
	title        = {Silicon isotopes in meteorites and planetary core formation},
	author       = {Armytage, RMG and Georg, RB and Savage, PS and Williams, HM and Halliday, AN},
	year         = 2011,
	journal      = {Geochimica et Cosmochimica Acta},
	publisher    = {Elsevier},
	volume       = 75,
	number       = 13,
	pages        = {3662--3676}
}

@article{armytage2012silicon,
	title        = {{Silicon isotopes in lunar rocks: Implications for the Moon{\textquoteright}s formation and the early history of the Earth}},
	author       = {{Armytage}, R.~M.~G. and {Georg}, R.~B. and {Williams}, H.~M. and {Halliday}, A.~N.},
	year         = 2012,
	month        = jan,
	journal      = {Geochimica et Cosmochimica Acta},
	volume       = 77,
	pages        = {504--514},
	doi          = {10.1016/j.gca.2011.10.032},
	adsurl       = {https://ui.adsabs.harvard.edu/abs/2012GeCoA..77..504A}
}

@article{badro2015core,
	title        = {{Core formation and core composition from coupled geochemical and geophysical constraints}},
	author       = {{Badro}, James and {Brodholt}, John P. and {Piet}, H{\'e}l{\`e}ne and {Siebert}, Julien and {Ryerson}, Frederick J.},
	year         = 2015,
	month        = oct,
	journal      = {Proceedings of the National Academy of Science},
	volume       = 112,
	number       = 40,
	pages        = {12310--12314},
	doi          = {10.1073/pnas.1505672112},
	adsurl       = {https://ui.adsabs.harvard.edu/abs/2015PNAS..11212310B}
}

@article{Benz1986,
	title        = {{The origin of the moon and the single-impact hypothesis I}},
	author       = {{Benz}, W. and {Slattery}, W.~L. and {Cameron}, A.~G.~W.},
	year         = 1986,
	month        = jun,
	journal      = {Icarus},
	volume       = 66,
	number       = 3,
	pages        = {515--535},
	doi          = {10.1016/0019-1035(86)90088-6},
	adsurl       = {https://ui.adsabs.harvard.edu/abs/1986Icar...66..515B}
}

@article{Bizzarro2025,
	title        = {The cosmochemistry of planetary systems.},
	author       = {Bizzarro, Martin and Johansen, Anders and Dorn, Caroline},
	year         = 2025,
	journal      = {Nat Rev Chem},
	location     = {Centre for Star and Planet Formation, Globe Institute, University of Copenhagen, Copenhagen, Denmark. Institut de Physique du Globe de Paris, Universit{\'e} de Paris, Paris, France. bizzarro\@sund.ku.dk. Centre for Star and Planet Formation, Globe Institute, University of Copenhagen, Copenhagen, Denmark. Lund Observatory, Department of Astronomy and Theoretical Physics, Lund University, Lund, Sweden. ETH Zurich, Institute for Particle Physics and Astrophysics, Zurich, Switzerland.}
}

@article{blanchard2022metal,
	title        = {The metal--silicate partitioning of carbon during Earth's accretion and its distribution in the early solar system},
	author       = {Blanchard, I and Rubie, DC and Jennings, Eleanor S and Franchi, IA and Zhao, X and Petitgirard, S and Miyajima, N and Jacobson, SA and Morbidelli, A},
	year         = 2022,
	journal      = {Earth and Planetary Science Letters},
	publisher    = {Elsevier},
	volume       = 580,
	pages        = 117374
}

@article{bottke2010stochastic,
	title        = {Stochastic late accretion to Earth, the Moon, and Mars},
	author       = {Bottke, William F and Walker, Richard J and Day, James MD and Nesvorny, David and Elkins-Tanton, Linda},
	year         = 2010,
	journal      = {science},
	publisher    = {American Association for the Advancement of Science},
	volume       = 330,
	number       = 6010,
	pages        = {1527--1530}
}

@article{brasser2016analysis,
	title        = {{Analysis of Terrestrial Planet Formation by the Grand Tack Model: System Architecture and Tack Location}},
	author       = {{Brasser}, R. and {Matsumura}, S. and {Ida}, S. and {Mojzsis}, S.~J. and {Werner}, S.~C.},
	year         = 2016,
	month        = apr,
	journal      = {The Astrophysical Journal},
	volume       = 821,
	number       = 2,
	pages        = 75,
	doi          = {10.3847/0004-637X/821/2/75},
	eid          = 75,
	archiveprefix = {arXiv},
	eprint       = {1603.01009},
	primaryclass = {astro-ph.EP},
	adsurl       = {https://ui.adsabs.harvard.edu/abs/2016ApJ...821...75B}
}

@article{budde2016molybdenum,
	title        = {Molybdenum isotopic evidence for the origin of chondrules and a distinct genetic heritage of carbonaceous and non-carbonaceous meteorites},
	author       = {Budde, Gerrit and Burkhardt, Christoph and Brennecka, Gregory A and Fischer-G{\"o}dde, Mario and Kruijer, Thomas S and Kleine, Thorsten},
	year         = 2016,
	journal      = {Earth and Planetary Science Letters},
	publisher    = {Elsevier},
	volume       = 454,
	pages        = {293--303}
}

@article{burkhardt2021terrestrial,
	title        = {Terrestrial planet formation from lost inner solar system material},
	author       = {Burkhardt, Christoph and Spitzer, Fridolin and Morbidelli, Alessandro and Budde, Gerrit and Render, Jan H and Kruijer, Thomas S and Kleine, Thorsten},
	year         = 2021,
	journal      = {Science advances},
	publisher    = {American Association for the Advancement of Science},
	volume       = 7,
	number       = 52,
	pages        = {eabj7601}
}

@inproceedings{cameron1976origin,
	title        = {The origin of the Moon},
	author       = {Cameron, Alastair G W and Ward, William R},
	year         = 1976,
	booktitle    = {Proceeedings of the Lunar and Planetary Science Conference},
	volume       = 7,
	pages        = {120--122}
}

@article{canup2001origin,
	title        = {Origin of the Moon in a giant impact near the end of the Earth's formation},
	author       = {Canup, Robin M and Asphaug, Erik},
	year         = 2001,
	journal      = {Nature},
	publisher    = {Nature Publishing Group UK London},
	volume       = 412,
	number       = 6848,
	pages        = {708--712}
}

@article{Canup2004,
	title        = {{Simulations of a late lunar-forming impact}},
	author       = {{Canup}, Robin M.},
	year         = 2004,
	month        = apr,
	journal      = {Icarus},
	volume       = 168,
	number       = 2,
	pages        = {433--456},
	doi          = {10.1016/j.icarus.2003.09.028},
	adsurl       = {https://ui.adsabs.harvard.edu/abs/2004Icar..168..433C}
}

@article{canup2008lunar,
	title        = {Lunar-forming collisions with pre-impact rotation},
	author       = {Canup, Robin M},
	year         = 2008,
	journal      = {Icarus},
	publisher    = {Elsevier},
	volume       = 196,
	number       = 2,
	pages        = {518--538}
}

@article{canup2012forming,
	title        = {Forming a Moon with an Earth-like composition via a giant impact},
	author       = {Canup, Robin M},
	year         = 2012,
	journal      = {Science},
	publisher    = {American Association for the Advancement of Science},
	volume       = 338,
	number       = 6110,
	pages        = {1052--1055}
}

@article{chambers2001making,
	title        = {Making more terrestrial planets},
	author       = {Chambers, JE},
	year         = 2001,
	journal      = {Icarus},
	publisher    = {Academic Press},
	volume       = 152,
	number       = 2,
	pages        = {205--224}
}

@article{Charnoz2026,
	title        = {{Non-equilibrium condensation of the first Solar System solids}},
	author       = {{Charnoz}, S{\'e}bastien and {Al{\'e}on}, J{\'e}r{\^o}me and {Chaussidon}, Marc and {Sossi}, Paolo A. and {Marrocchi}, Yves and {Franco}, Patrick},
	year         = 2026,
	month        = apr,
	journal      = {Nature},
	volume       = 652,
	number       = 8111,
	pages        = {925--930},
	doi          = {10.1038/s41586-026-10257-5},
	adsurl       = {https://ui.adsabs.harvard.edu/abs/2026Natur.652..925C}
}

@article{Ciesla2006,
	title        = {{The evolution of the water distribution in a viscous protoplanetary disk}},
	author       = {{Ciesla}, Fred J. and {Cuzzi}, Jeffrey N.},
	year         = 2006,
	month        = mar,
	journal      = {Icarus},
	volume       = 181,
	number       = 1,
	pages        = {178--204},
	doi          = {10.1016/j.icarus.2005.11.009},
	archiveprefix = {arXiv},
	eprint       = {astro-ph/0511372},
	primaryclass = {astro-ph},
	adsurl       = {https://ui.adsabs.harvard.edu/abs/2006Icar..181..178C}
}

@article{clement2018mars,
	title        = {Mars' growth stunted by an early giant planet instability},
	author       = {Clement, Matthew S and Kaib, Nathan A and Raymond, Sean N and Walsh, Kevin J},
	year         = 2018,
	journal      = {Icarus},
	publisher    = {Elsevier},
	volume       = 311,
	pages        = {340--356}
}

@article{clement2019early,
	title        = {The early instability scenario: Terrestrial planet formation during the giant planet instability, and the effect of collisional fragmentation},
	author       = {Clement, Matthew S and Kaib, Nathan A and Raymond, Sean N and Chambers, John E and Walsh, Kevin J},
	year         = 2019,
	journal      = {Icarus},
	publisher    = {Elsevier},
	volume       = 321,
	pages        = {778--790}
}

@article{cottrell2009metal,
	title        = {Metal--silicate partitioning of tungsten at high pressure and temperature: implications for equilibrium core formation in Earth},
	author       = {Cottrell, Elizabeth and Walter, Michael J and Walker, David},
	year         = 2009,
	journal      = {Earth and Planetary Science Letters},
	publisher    = {Elsevier},
	volume       = 281,
	number       = {3-4},
	pages        = {275--287}
}

@article{cuk2012making,
	title        = {Making the Moon from a fast-spinning Earth: A giant impact followed by resonant despinning},
	author       = {{\'C}uk, Matija and Stewart, Sarah T},
	year         = 2012,
	journal      = {science},
	publisher    = {American Association for the Advancement of Science},
	volume       = 338,
	number       = 6110,
	pages        = {1047--1052}
}

@article{cuk2016tidal,
	title        = {Tidal evolution of the Moon from a high-obliquity, high-angular-momentum Earth},
	author       = {{\'C}uk, Matija and Hamilton, Douglas P and Lock, Simon J and Stewart, Sarah T},
	year         = 2016,
	journal      = {Nature},
	publisher    = {Nature Publishing Group UK London},
	volume       = 539,
	number       = 7629,
	pages        = {402--406}
}

@article{cuk2021tidal,
	title        = {{Tidal Evolution of the Earth-Moon System with a High Initial Obliquity}},
	author       = {{{\'C}uk}, Matija and {Lock}, Simon J. and {Stewart}, Sarah T. and {Hamilton}, Douglas P.},
	year         = 2021,
	month        = aug,
	journal      = {The Planetary Science Journal},
	volume       = 2,
	number       = 4,
	pages        = 147,
	doi          = {10.3847/PSJ/ac12d1},
	eid          = 147,
	archiveprefix = {arXiv},
	eprint       = {2107.03353},
	primaryclass = {astro-ph.EP},
	adsurl       = {https://ui.adsabs.harvard.edu/abs/2021PSJ.....2..147C}
}

@article{Dale2023,
	title        = {{An improved model of metal/silicate differentiation during Earth's accretion}},
	author       = {{Dale}, K.~I. and {Rubie}, D.~C. and {Nakajima}, M. and {Jacobson}, S. and {Nathan}, G. and {Golabek}, G.~J. and {Cambioni}, S. and {Morbidelli}, A.},
	year         = 2023,
	month        = dec,
	journal      = {Icarus},
	volume       = 406,
	pages        = 115739,
	doi          = {10.1016/j.icarus.2023.115739},
	eid          = 115739,
	archiveprefix = {arXiv},
	eprint       = {2308.04476},
	primaryclass = {astro-ph.EP},
	adsurl       = {https://ui.adsabs.harvard.edu/abs/2023Icar..40615739D}
}

@article{dale2025compositional,
	title        = {Compositional outcomes of Earth formation from a narrow ring},
	author       = {Dale, Katherine I and Morbidelli, Alessandro and Rubie, David C and Nesvorn{\`y}, David},
	year         = 2025,
	journal      = {Earth and Planetary Science Letters},
	publisher    = {Elsevier},
	volume       = 658,
	pages        = 119334
}

@article{Dale2026,
	title        = {{Oxidation Constraints on Terrestrial Planet Formation from a Ring}},
	author       = {{Dale}, Katherine I. and {Morbidelli}, Alessandro and {Nathan}, Gabriel and {Woo}, Jason and {Nesvorn{\'y}}, David and {Rubie}, David C.},
	year         = 2026,
	month        = may,
	journal      = {arXiv e-prints},
	pages        = {arXiv:2605.31081},
	doi          = {10.48550/arXiv.2605.31081},
	eid          = {arXiv:2605.31081},
	archiveprefix = {arXiv},
	eprint       = {2605.31081},
	primaryclass = {astro-ph.EP},
	adsurl       = {https://ui.adsabs.harvard.edu/abs/2026arXiv260531081D}
}

@article{dauphas2011hf,
	title        = {{Hf}--{W}--{Th} evidence for rapid growth of Mars and its status as a planetary embryo},
	author       = {Dauphas, N and Pourmand, Ali},
	year         = 2011,
	journal      = {Nature},
	publisher    = {Nature Publishing Group UK},
	volume       = 473,
	number       = 7348,
	pages        = {489--492}
}

@article{dauphas2016mass,
	title        = {{Mass Fractionation Laws, Mass-Independent Effects, and Isotopic Anomalies}},
	author       = {{Dauphas}, Nicolas and {Schauble}, Edwin A.},
	year         = 2016,
	month        = jun,
	journal      = {Annual Review of Earth and Planetary Sciences},
	volume       = 44,
	pages        = {709--783},
	doi          = {10.1146/annurev-earth-060115-012157},
	adsurl       = {https://ui.adsabs.harvard.edu/abs/2016AREPS..44..709D}
}

@article{dauphas2017isotopic,
	title        = {The isotopic nature of the Earth's accreting material through time},
	author       = {Dauphas, Nicolas},
	year         = 2017,
	journal      = {Nature},
	publisher    = {Nature Publishing Group},
	volume       = 541,
	number       = 7638,
	pages        = {521--524}
}

@article{dauphas2024bayesian,
	title        = {Bayesian inference on the isotopic building blocks of Mars and Earth},
	author       = {Dauphas, Nicolas and Hopp, Timo and Nesvorn{\`y}, David},
	year         = 2024,
	journal      = {Icarus},
	publisher    = {Elsevier},
	volume       = 408,
	pages        = 115805
}

@article{Davis2022,
	title        = {{Short-Lived Nuclides in the Early Solar System: Abundances, Origins, and Applications}},
	author       = {{Davis}, Andrew M.},
	year         = 2022,
	month        = sep,
	journal      = {Annual Review of Nuclear and Particle Science},
	volume       = 72,
	doi          = {10.1146/annurev-nucl-010722-074615},
	adsurl       = {https://ui.adsabs.harvard.edu/abs/2022ARNPS..72..339D}
}

@article{day2016highly,
	title        = {{Highly Siderophile Elements in Earth, Mars, the Moon, and Asteroids}},
	author       = {{Day}, James M.~D. and {Brandon}, Alan D. and {Walker}, Richard J.},
	year         = 2016,
	month        = jan,
	journal      = {Reviews in Mineralogy and Geochemistry},
	volume       = 81,
	number       = 1,
	pages        = {161--238},
	doi          = {10.2138/rmg.2016.81.04},
	adsurl       = {https://ui.adsabs.harvard.edu/abs/2016RvMG...81..161D}
}

@article{deguen2011experiments,
	title        = {Experiments on turbulent metal-silicate mixing in a magma ocean},
	author       = {Deguen, Renaud and Olson, Peter and Cardin, Philippe},
	year         = 2011,
	journal      = {Earth and Planetary Science Letters},
	publisher    = {Elsevier},
	volume       = 310,
	number       = {3-4},
	pages        = {303--313}
}

@article{deguen2014turbulent,
	title        = {Turbulent metal–silicate mixing, fragmentation, and equilibration in magma oceans},
	author       = {Renaud Deguen and Maylis Landeau and Peter Olson},
	year         = 2014,
	journal      = {Earth and Planetary Science Letters},
	volume       = 391,
	pages        = {274--287},
	doi          = {https://doi.org/10.1016/j.epsl.2014.02.007},
	issn         = {0012-821X},
	url          = {https://www.sciencedirect.com/science/article/pii/S0012821X14000806}
}

@article{deienno2016grand,
	title        = {{Is the Grand Tack model compatible with the orbital distribution of main belt asteroids?}},
	author       = {{Deienno}, Rogerio and {Gomes}, Rodney S. and {Walsh}, Kevin J. and {Morbidelli}, Alessandro and {Nesvorn{\'y}}, David},
	year         = 2016,
	month        = jul,
	journal      = {Icarus},
	volume       = 272,
	pages        = {114--124},
	doi          = {10.1016/j.icarus.2016.02.043},
	archiveprefix = {arXiv},
	eprint       = {1701.02775},
	primaryclass = {astro-ph.EP},
	adsurl       = {https://ui.adsabs.harvard.edu/abs/2016Icar..272..114D}
}

@article{demeo2013taxonomic,
	title        = {{The taxonomic distribution of asteroids from multi-filter all-sky photometric surveys}},
	author       = {{DeMeo}, F.~E. and {Carry}, B.},
	year         = 2013,
	month        = sep,
	journal      = {Icarus},
	volume       = 226,
	number       = 1,
	pages        = {723--741},
	doi          = {10.1016/j.icarus.2013.06.027},
	archiveprefix = {arXiv},
	eprint       = {1307.2424},
	primaryclass = {astro-ph.EP},
	adsurl       = {https://ui.adsabs.harvard.edu/abs/2013Icar..226..723D}
}

@article{Desch2019,
	title        = {{A unified model for hydrogen in the Earth and Moon: No one expects the Theia contribution}},
	author       = {{Desch}, Steven J. and {Robinson}, Katharine L.},
	year         = 2019,
	month        = dec,
	journal      = {Chemie der Erde / Geochemistry},
	volume       = 79,
	number       = 4,
	pages        = 125546,
	doi          = {10.1016/j.chemer.2019.125546},
	adsurl       = {https://ui.adsabs.harvard.edu/abs/2019ChEG...79l5546D}
}

@article{deVries2016,
	title        = {{Impact-induced melting during accretion of the Earth}},
	author       = {{de Vries}, Jellie and {Nimmo}, Francis and {Melosh}, H. Jay and {Jacobson}, Seth A. and {Morbidelli}, Alessandro and {Rubie}, David C.},
	year         = 2016,
	month        = dec,
	journal      = {Progress in Earth and Planetary Science},
	volume       = 3,
	pages        = 7,
	doi          = {10.1186/s40645-016-0083-8},
	eid          = 7,
	archiveprefix = {arXiv},
	eprint       = {1603.08682},
	primaryclass = {astro-ph.EP},
	adsurl       = {https://ui.adsabs.harvard.edu/abs/2016PEPS....3....7D}
}

@article{drake2002determining,
	title        = {Determining the composition of the Earth},
	author       = {Drake, Michael J and Righter, Kevin},
	year         = 2002,
	journal      = {Nature},
	publisher    = {Nature Publishing Group},
	volume       = 416,
	number       = 6876,
	pages        = {39--44}
}

@article{fischer2014dynamics,
	title        = {{Dynamics of the terrestrial planets from a large number of N-body simulations}},
	author       = {{Fischer}, Rebecca A. and {Ciesla}, Fred J.},
	year         = 2014,
	month        = apr,
	journal      = {Earth and Planetary Science Letters},
	volume       = 392,
	pages        = {28--38},
	doi          = {10.1016/j.epsl.2014.02.011},
	adsurl       = {https://ui.adsabs.harvard.edu/abs/2014E&PSL.392...28F}
}

@article{fischer2015high,
	title        = {High pressure metal--silicate partitioning of Ni, Co, V, Cr, Si, and O},
	author       = {Fischer, Rebecca A and Nakajima, Yoichi and Campbell, Andrew J and Frost, Daniel J and Harries, Dennis and Langenhorst, Falko and Miyajima, Nobuyoshi and Pollok, Kilian and Rubie, David C},
	year         = 2015,
	journal      = {Geochimica et Cosmochimica Acta},
	publisher    = {Elsevier},
	volume       = 167,
	pages        = {177--194}
}

@article{fitoussi2012silicon,
	title        = {Silicon isotope evidence against an enstatite chondrite Earth},
	author       = {Fitoussi, Caroline and Bourdon, Bernard},
	year         = 2012,
	journal      = {Science},
	publisher    = {American Association for the Advancement of Science},
	volume       = 335,
	number       = 6075,
	pages        = {1477--1480}
}

@article{gaffney2023magmatic,
	title        = {Magmatic evolution I: Initial differentiation of the Moon},
	author       = {Gaffney, Amy M and Gross, Juliane and Borg, Lars E and Hanna, Kerri L Donaldson and Draper, David S and Dygert, Nick and Elkins-Tanton, Lindy T and Prissel, Kelsey B and Prissel, Tabb C and Steenstra, Edgar S and others},
	year         = 2023,
	journal      = {Reviews in Mineralogy and Geochemistry},
	publisher    = {Mineralogical Society of America},
	volume       = 89,
	number       = 1,
	pages        = {103--145}
}

@article{galimov2005redox,
	title        = {Redox evolution of the Earth caused by a multi-stage formation of its core},
	author       = {Galimov, EM},
	year         = 2005,
	journal      = {Earth and Planetary Science Letters},
	publisher    = {Elsevier},
	volume       = 233,
	number       = {3-4},
	pages        = {263--276}
}

@article{georg2007silicon,
	title        = {Silicon in the Earth's core},
	author       = {Georg, R Bastian and Halliday, Alex N and Schauble, Edwin A and Reynolds, Ben C},
	year         = 2007,
	journal      = {Nature},
	publisher    = {Nature Publishing Group},
	volume       = 447,
	number       = 7148,
	pages        = {1102--1106}
}

@article{gradie1982compositional,
	title        = {Compositional structure of the asteroid belt},
	author       = {Gradie, J and Tedesco, E},
	year         = 1982,
	journal      = {Science},
	publisher    = {American Association for the Advancement of Science},
	volume       = 216,
	number       = 4553,
	pages        = {1405--1407}
}

@inproceedings{gradie1989distribution,
	title        = {{Distribution of taxonomic classes and the compositional structure of the asteroid belt.}},
	author       = {{Gradie}, Jonathan C. and {Chapman}, Clark R. and {Tedesco}, Edward F.},
	year         = 1989,
	month        = jan,
	booktitle    = {Asteroids II},
	pages        = {316--335},
	editor       = {{Binzel}, Richard P. and {Gehrels}, Tom and {Matthews}, Mildred Shapley},
	adsurl       = {https://ui.adsabs.harvard.edu/abs/1989aste.conf..316G}
}

@article{Grewal2024,
	title        = {{Accretion of the earliest inner Solar System planetesimals beyond the water snowline}},
	author       = {{Grewal}, Damanveer S. and {Nie}, Nicole X. and {Zhang}, Bidong and {Izidoro}, Andre and {Asimow}, Paul D.},
	year         = 2024,
	month        = mar,
	journal      = {Nature Astronomy},
	volume       = 8,
	pages        = {290--297},
	doi          = {10.1038/s41550-023-02172-w},
	archiveprefix = {arXiv},
	eprint       = {2408.17032},
	primaryclass = {astro-ph.EP},
	adsurl       = {https://ui.adsabs.harvard.edu/abs/2024NatAs...8..290G}
}

@incollection{Grossman1977,
	title        = {{Chemical fractionation in the solar nebula.}},
	author       = {{Grossman}, L.},
	year         = 1977,
	booktitle    = {The Soviet-American Conference on Cosmochemistry of the Moon and Planets},
	publisher    = {NASA Special Publications},
	volume       = 370,
	pages        = {787--796},
	editor       = {{Pomeroy}, John H. and {Hubbard}, Norman J.},
	adsurl       = {https://ui.adsabs.harvard.edu/abs/1977NASSP.370..787G}
}

@article{grossman2008redox,
	title        = {Redox conditions in the solar nebula: Observational, experimental, and theoretical constraints},
	author       = {Grossman, Lawrence and Beckett, John R and Fedkin, Alexei V and Simon, Steven B and Ciesla, Fred J},
	year         = 2008,
	journal      = {Reviews in Mineralogy and Geochemistry},
	publisher    = {Mineralogical Society of America},
	volume       = 68,
	number       = 1,
	pages        = {93--140}
}

@article{gu2023comparisons,
	title        = {Comparisons of the core and mantle compositions of earth analogs from different terrestrial planet formation scenarios},
	author       = {Gu, Jesse T and Fischer, Rebecca A and Brennan, Matthew C and Clement, Matthew S and Jacobson, Seth A and Kaib, Nathan A and O'Brien, David P and Raymond, Sean N},
	year         = 2023,
	journal      = {Icarus},
	publisher    = {Elsevier},
	volume       = 394,
	pages        = 115425
}

@article{halliday2004mixing,
	title        = {{Mixing, volatile loss and compositional change during impact-driven accretion of the Earth}},
	author       = {{Halliday}, Alex N.},
	year         = 2004,
	month        = feb,
	journal      = {Nature},
	volume       = 427,
	number       = 6974,
	pages        = {505--509},
	doi          = {10.1038/nature02275},
	adsurl       = {https://ui.adsabs.harvard.edu/abs/2004Natur.427..505H}
}

@article{hallis2010oxygen,
	title        = {{The oxygen isotope composition, petrology and geochemistry of mare basalts: Evidence for large-scale compositional variation in the lunar mantle}},
	author       = {{Hallis}, L.~J. and {Anand}, M. and {Greenwood}, R.~C. and {Miller}, M.~F. and {Franchi}, I.~A. and {Russell}, S.~S.},
	year         = 2010,
	month        = dec,
	journal      = {Geochimica et Cosmochimica Acta},
	volume       = 74,
	number       = 23,
	pages        = {6885--6899},
	doi          = {10.1016/j.gca.2010.09.023},
	adsurl       = {https://ui.adsabs.harvard.edu/abs/2010GeCoA..74.6885H}
}

@article{hansen2009formation,
	title        = {Formation of the terrestrial planets from a narrow annulus},
	author       = {Hansen, Brad MS},
	year         = 2009,
	journal      = {The Astrophysical Journal},
	publisher    = {IOP Publishing},
	volume       = 703,
	number       = 1,
	pages        = 1131
}

@article{hartmann1975lunar,
	title        = {Lunar ``cataclysm'': A misconception?},
	author       = {Hartmann, William K},
	year         = 1975,
	journal      = {Icarus},
	publisher    = {Elsevier},
	volume       = 24,
	number       = 2,
	pages        = {181--187}
}

@article{hartmann1975satellite,
	title        = {Satellite-sized planetesimals and lunar origin},
	author       = {Hartmann, William K and Davis, Donald R},
	year         = 1975,
	journal      = {Icarus},
	publisher    = {Elsevier},
	volume       = 24,
	number       = 4,
	pages        = {504--515}
}

@article{Hauck2013,
	title        = {{The curious case of Mercury's internal structure}},
	author       = {{Hauck}, Steven A. and {Margot}, Jean-Luc and {Solomon}, Sean C. and {Phillips}, Roger J. and {Johnson}, Catherine L. and {Lemoine}, Frank G. and {Mazarico}, Erwan and {McCoy}, Timothy J. and {Padovan}, Sebastiano and {Peale}, Stanton J. and {Perry}, Mark E. and {Smith}, David E. and {Zuber}, Maria T.},
	year         = 2013,
	month        = jun,
	journal      = {Journal of Geophysical Research (Planets)},
	volume       = 118,
	number       = 6,
	pages        = {1204--1220},
	doi          = {10.1002/jgre.20091},
	adsurl       = {https://ui.adsabs.harvard.edu/abs/2013JGRE..118.1204H}
}

@article{hayashi1981structure,
	title        = {{Structure of the Solar Nebula, Growth and Decay of Magnetic Fields and Effects of Magnetic and Turbulent Viscosities on the Nebula}},
	author       = {{Hayashi}, C.},
	year         = 1981,
	month        = jan,
	journal      = {Progress of Theoretical Physics Supplement},
	volume       = 70,
	pages        = {35--53},
	doi          = {10.1143/PTPS.70.35},
	adsurl       = {https://ui.adsabs.harvard.edu/abs/1981PThPS..70...35H}
}

@article{Herwartz2014,
	title        = {{Identification of the giant impactor Theia in lunar rocks}},
	author       = {{Herwartz}, Daniel and {Pack}, Andreas and {Friedrichs}, Bjarne and {Bischoff}, Addi},
	year         = 2014,
	month        = jun,
	journal      = {Science},
	volume       = 344,
	number       = 6188,
	pages        = {1146--1150},
	doi          = {10.1126/science.1251117},
	adsurl       = {https://ui.adsabs.harvard.edu/abs/2014Sci...344.1146H}
}

@article{Herzberg1996,
	title        = {{Melting experiments on anhydrous peridotite KLB-1: Compositions of magmas in the upper mantle and transition zone}},
	author       = {{Herzberg}, Claude and {Zhang}, Jianzhong},
	year         = 1996,
	month        = apr,
	journal      = {Journal of Geophysical Research},
	volume       = 101,
	number       = {B4},
	pages        = {8271--8295},
	doi          = {10.1029/96JB00170},
	adsurl       = {https://ui.adsabs.harvard.edu/abs/1996JGR...101.8271H}
}

@article{Hevey2006,
	title        = {{A model for planetesimal meltdown by $^{26}$Al and its implications for meteorite parent bodies}},
	author       = {{Hevey}, Phonsie J. and {Sanders}, Ian S.},
	year         = 2006,
	month        = jan,
	journal      = {Meteoritics and Planetary Science},
	volume       = 41,
	number       = 1,
	pages        = {95--106},
	doi          = {10.1111/j.1945-5100.2006.tb00195.x},
	adsurl       = {https://ui.adsabs.harvard.edu/abs/2006M&PS...41...95H}
}

@incollection{Hood2000,
	title        = {{Recent Refinements in Geophysical Constraints on Lunar Origin and Evolution}},
	author       = {{Hood}, L.~L. and {Zuber}, M.~T.},
	year         = 2000,
	booktitle    = {Origin of the Earth and Moon},
	publisher    = {University of Arizona Press},
	pages        = {397--409},
	editor       = {{Canup}, R.~M. and {Righter}, K. and {et al.}},
	adsurl       = {https://ui.adsabs.harvard.edu/abs/2000orem.book..397H}
}

@article{hopp2025moon,
	title        = {The Moon-forming impactor Theia originated from the inner Solar System},
	author       = {Hopp, Timo and Dauphas, Nicolas and Boyet, Maud and Jacobson, Seth A and Kleine, Thorsten},
	year         = 2025,
	journal      = {Science},
	publisher    = {American Association for the Advancement of Science},
	volume       = 390,
	number       = 6775,
	pages        = {819--823}
}

@article{huang2020niobium,
	title        = {The niobium and tantalum concentration in the mantle constrains the composition of Earth’s primordial magma ocean},
	author       = {Huang, Dongyang and Badro, James and Siebert, Julien},
	year         = 2020,
	journal      = {Proceedings of the National Academy of Sciences},
	publisher    = {National Academy of Sciences},
	volume       = 117,
	number       = 45,
	pages        = {27893--27898}
}

@article{huang2021high,
	title        = {High pressure partitioning behavior of Mo and W and late sulfur delivery during Earth’s core formation},
	author       = {Huang, Dongyang and Siebert, Julien and Badro, James},
	year         = 2021,
	journal      = {Geochimica et Cosmochimica Acta},
	publisher    = {Elsevier},
	volume       = 310,
	pages        = {19--31}
}

@article{izidoro2014terrestrial,
	title        = {Terrestrial planet formation in a protoplanetary disk with a local mass depletion: A successful scenario for the formation of Mars},
	author       = {Izidoro, Andr{\'e} and Haghighipour, N and Winter, OC and Tsuchida, M},
	year         = 2014,
	journal      = {The Astrophysical Journal},
	publisher    = {IOP Publishing},
	volume       = 782,
	number       = 1,
	pages        = 31
}

@article{izidoro2015terrestrial,
	title        = {Terrestrial planet formation constrained by Mars and the structure of the asteroid belt},
	author       = {Izidoro, Andr{\'e} and Raymond, Sean N and Morbidelli, Alessandro and Winter, Othon C},
	year         = 2015,
	journal      = {Monthly Notices of the Royal Astronomical Society},
	publisher    = {Oxford University Press},
	volume       = 453,
	number       = 4,
	pages        = {3619--3634}
}

@article{izidoro2021effect,
	title        = {The effect of a strong pressure bump in the Sun's natal disk: terrestrial planet formation via planetesimal accretion rather than pebble accretion},
	author       = {Izidoro, Andr{\'e} and Bitsch, Bertram and Dasgupta, Rajdeep},
	year         = 2021,
	journal      = {The Astrophysical Journal},
	publisher    = {IOP Publishing},
	volume       = 915,
	number       = 1,
	pages        = 62
}

@article{izidoro2022planetesimal,
	title        = {{Planetesimal rings as the cause of the Solar System's planetary architecture}},
	author       = {{Izidoro}, Andre and {Dasgupta}, Rajdeep and {Raymond}, Sean N. and {Deienno}, Rogerio and {Bitsch}, Bertram and {Isella}, Andrea},
	year         = 2022,
	month        = mar,
	journal      = {Nature Astronomy},
	volume       = 6,
	pages        = {357--366},
	doi          = {10.1038/s41550-021-01557-z},
	archiveprefix = {arXiv},
	eprint       = {2112.15558},
	primaryclass = {astro-ph.EP},
	adsurl       = {https://ui.adsabs.harvard.edu/abs/2022NatAs...6..357I}
}

@article{Jacobsen2005hf,
	title        = {{The Hf-W Isotopic System and the Origin of the Earth and Moon}},
	author       = {{Jacobsen}, Stein B.},
	year         = 2005,
	month        = jan,
	journal      = {Annual Review of Earth and Planetary Sciences},
	volume       = 33,
	pages        = {531--570},
	doi          = {10.1146/annurev.earth.33.092203.122614},
	adsurl       = {https://ui.adsabs.harvard.edu/abs/2005AREPS..33..531J}
}

@article{jacobson2014highly,
	title        = {Highly siderophile elements in Earth's mantle as a clock for the Moon-forming impact},
	author       = {Jacobson, Seth A and Morbidelli, Alessandro and Raymond, Sean N and O'Brien, David P and Walsh, Kevin J and Rubie, David C},
	year         = 2014,
	journal      = {Nature},
	publisher    = {Nature Publishing Group},
	volume       = 508,
	number       = 7494,
	pages        = {84--87}
}

@article{jacobson2014lunar,
	title        = {Lunar and terrestrial planet formation in the Grand Tack scenario},
	author       = {Jacobson, Seth A and Morbidelli, Alessandro},
	year         = 2014,
	journal      = {Philosophical Transactions of the Royal Society A: Mathematical, Physical and Engineering Sciences},
	publisher    = {The Royal Society Publishing},
	volume       = 372,
	number       = 2024,
	pages        = 20130174
}

@article{jacobson2015earth,
	title        = {Earth and terrestrial planet formation},
	author       = {Jacobson, Seth A and Walsh, Kevin J},
	year         = 2015,
	month        = sep,
	journal      = {The early Earth: Accretion and differentiation},
	publisher    = {Wiley Online Library},
	volume       = 212,
	pages        = {49--70},
	doi          = {10.1002/9781118860359.ch3},
	archiveprefix = {arXiv},
	eprint       = {1502.03852},
	primaryclass = {astro-ph.EP},
	adsurl       = {https://ui.adsabs.harvard.edu/abs/2015GMS...212...49J}
}

@article{javoy1995integral,
	title        = {The integral enstatite chondrite model of the Earth},
	author       = {Javoy, M},
	year         = 1995,
	journal      = {Geophysical Research Letters},
	publisher    = {Wiley Online Library},
	volume       = 22,
	number       = 16,
	pages        = {2219--2222}
}

@article{javoy2010chemical,
	title        = {{The chemical composition of the Earth: Enstatite chondrite models}},
	author       = {{Javoy}, M. and {Kaminski}, E. and {Guyot}, F. and {Andrault}, D. and {Sanloup}, C. and {Moreira}, M. and {Labrosse}, S. and {Jambon}, A. and {Agrinier}, P. and {Davaille}, A. and {Jaupart}, C.},
	year         = 2010,
	month        = may,
	journal      = {Earth and Planetary Science Letters},
	volume       = 293,
	number       = {3-4},
	pages        = {259--268},
	doi          = {10.1016/j.epsl.2010.02.033},
	adsurl       = {https://ui.adsabs.harvard.edu/abs/2010E&PSL.293..259J}
}

@article{jennings2021metal,
	title        = {Metal--silicate partitioning of {W} and {Mo} and the role of carbon in controlling their abundances in the bulk silicate earth},
	author       = {Jennings, Eleanor S and Jacobson, SA and Rubie, David C and Nakajima, Y and Vogel, AK and Rose-Weston, LA and Frost, Daniel J},
	year         = 2021,
	journal      = {Geochimica et Cosmochimica Acta},
	publisher    = {Elsevier},
	volume       = 293,
	pages        = {40--69}
}

@article{Johansen2021,
	title        = {{A pebble accretion model for the formation of the terrestrial planets in the Solar System}},
	author       = {{Johansen}, Anders and {Ronnet}, Thomas and {Bizzarro}, Martin and {Schiller}, Martin and {Lambrechts}, Michiel and {Nordlund}, {\r{A}}ke and {Lammer}, Helmut},
	year         = 2021,
	month        = feb,
	journal      = {Science Advances},
	volume       = 7,
	number       = 8,
	pages        = {eabc0444},
	doi          = {10.1126/sciadv.abc0444},
	archiveprefix = {arXiv},
	eprint       = {2102.08611},
	primaryclass = {astro-ph.EP},
	adsurl       = {https://ui.adsabs.harvard.edu/abs/2021SciA....7..444J}
}

@article{johansen2021pebble,
	title        = {A pebble accretion model for the formation of the terrestrial planets in the Solar System},
	author       = {Johansen, Anders and Ronnet, Thomas and Bizzarro, Martin and Schiller, Martin and Lambrechts, Michiel and Nordlund, {\AA}ke and Lammer, Helmut},
	year         = 2021,
	journal      = {Science Advances},
	publisher    = {American Association for the Advancement of Science},
	volume       = 7,
	number       = 8,
	pages        = {eabc0444}
}

@article{Johansen2023,
	title        = {{Anatomy of rocky planets formed by rapid pebble accretion. II. Differentiation by accretion energy and thermal blanketing}},
	author       = {{Johansen}, Anders and {Ronnet}, Thomas and {Schiller}, Martin and {Deng}, Zhengbin and {Bizzarro}, Martin},
    journal={Astronomy \& Astrophysics},
	year         = 2023,
	month        = mar,
	volume       = 671,
	pages        = {A75},
	doi          = {10.1051/0004-6361/202142142},
	eid          = {A75},
	archiveprefix = {arXiv},
	eprint       = {2207.09804},
	primaryclass = {astro-ph.EP},
	adsurl       = {https://ui.adsabs.harvard.edu/abs/2023A\&A...671A..75J}
}

@article{jones1986geochemical,
	title        = {{Geochemical constraints on core formation in the Earth}},
	author       = {{Jones}, J.~H. and {Drake}, M.~J.},
	year         = 1986,
	month        = jul,
	journal      = {Nature},
	volume       = 322,
	number       = 6076,
	pages        = {221--228},
	doi          = {10.1038/322221a0},
	adsurl       = {https://ui.adsabs.harvard.edu/abs/1986Natur.322..221J}
}

@article{jones2012petrographic,
	title        = {Petrographic constraints on the diversity of chondrule reservoirs in the protoplanetary disk},
	author       = {Jones, Rhian H},
	year         = 2012,
	journal      = {Meteoritics \& Planetary Science},
	publisher    = {Wiley Online Library},
	volume       = 47,
	number       = 7,
	pages        = {1176--1190}
}

@article{Kaib2015,
	title        = {{Brief follow-up on recent studies of Theia's accretion}},
	author       = {{Kaib}, Nathan A. and {Cowan}, Nicolas B.},
	year         = 2015,
	month        = sep,
	journal      = {Icarus},
	volume       = 258,
	pages        = {14--17},
	doi          = {10.1016/j.icarus.2015.06.019},
	archiveprefix = {arXiv},
	eprint       = {1506.06132},
	primaryclass = {astro-ph.EP},
	adsurl       = {https://ui.adsabs.harvard.edu/abs/2015Icar..258...14K}
}

@article{kaib2016fragility,
	title        = {{The fragility of the terrestrial planets during a giant-planet instability}},
	author       = {{Kaib}, Nathan A. and {Chambers}, John E.},
	year         = 2016,
	month        = feb,
	journal      = {Monthly Notices of the Royal Astronomical Society},
	volume       = 455,
	number       = 4,
	pages        = {3561--3569},
	doi          = {10.1093/mnras/stv2554},
	archiveprefix = {arXiv},
	eprint       = {1510.08448},
	primaryclass = {astro-ph.EP},
	adsurl       = {https://ui.adsabs.harvard.edu/abs/2016MNRAS.455.3561K}
}

@article{kegerreis2022immediate,
	title        = {Immediate Origin of the Moon as a Post-impact Satellite},
	author       = {Kegerreis, Jacob A and Ruiz-Bonilla, Sergio and Eke, Vincent R and Massey, Richard J and Sandnes, Thomas D and Teodoro, Lu{\'\i}s FA},
	year         = 2022,
	journal      = {The Astrophysical Journal Letters},
	publisher    = {American Astronomical Society},
	volume       = 937,
	number       = 2,
	pages        = {L40}
}

@article{kegler2008new,
	title        = {New Ni and Co metal-silicate partitioning data and their relevance for an early terrestrial magma ocean},
	author       = {Kegler, Ph and Holzheid, A and Frost, DJ and Rubie, DC and Dohmen, R and Palme, H},
	year         = 2008,
	journal      = {Earth and Planetary Science Letters},
	publisher    = {Elsevier},
	volume       = 268,
	number       = {1-2},
	pages        = {28--40}
}

@article{kleine2009hf,
	title        = {Hf--W chronology of the accretion and early evolution of asteroids and terrestrial planets},
	author       = {Kleine, Thorsten and Touboul, Mathieu and Bourdon, Bernard and Nimmo, Francis and Mezger, Klaus and Palme, Herbert and Jacobsen, Stein B and Yin, Qing-Zhu and Halliday, Alexander N},
	year         = 2009,
	journal      = {Geochimica et Cosmochimica Acta},
	publisher    = {Elsevier},
	volume       = 73,
	number       = 17,
	pages        = {5150--5188}
}

@article{kleine2020non,
	title        = {The non-carbonaceous--carbonaceous meteorite dichotomy},
	author       = {Kleine, Thorsten and Budde, G and Burkhardt, C and Kruijer, TS and Worsham, EA and Morbidelli, A and Nimmo, F},
	year         = 2020,
	journal      = {Space Science Reviews},
	publisher    = {Springer},
	volume       = 216,
	pages        = {1--27}
}

@article{kley2019planet,
	title        = {{Planet Formation and Disk-Planet Interactions}},
	author       = {{Kley}, Wilhelm},
	year         = 2019,
	month        = jan,
	journal      = {Saas-Fee Advanced Course},
	volume       = 45,
	pages        = 151,
	doi          = {10.1007/978-3-662-58687-7_2},
	adsurl       = {https://ui.adsabs.harvard.edu/abs/2019SAAS...45..151K}
}

@article{kokubo1996runaway,
	title        = {{On Runaway Growth of Planetesimals}},
	author       = {{Kokubo}, Eiichiro and {Ida}, Shigeru},
	year         = 1996,
	month        = sep,
	journal      = {Icarus},
	volume       = 123,
	number       = 1,
	pages        = {180--191},
	doi          = {10.1006/icar.1996.0148},
	adsurl       = {https://ui.adsabs.harvard.edu/abs/1996Icar..123..180K}
}

@article{kokubo1998oligarchic,
	title        = {Oligarchic growth of protoplanets},
	author       = {Kokubo, Eiichiro and Ida, Shigeru},
	year         = 1998,
	journal      = {Icarus},
	publisher    = {Elsevier},
	volume       = 131,
	number       = 1,
	pages        = {171--178}
}

@inproceedings{Krot2000,
	title        = {{Meteoritical and Astrophysical Constraints on the Oxidation State of the Solar Nebula}},
	author       = {{Krot}, A.~N. and {Fegley}, Jr., B. and {Lodders}, K. and {Palme}, H.},
	year         = 2000,
	month        = may,
	booktitle    = {Protostars and Planets IV},
	pages        = 1019,
	editor       = {{Mannings}, V. and {Boss}, A.~P. and {Russell}, S.~S.},
	adsurl       = {https://ui.adsabs.harvard.edu/abs/2000prpl.conf.1019K}
}

@article{kruijer2015lunar,
	title        = {{Lunar tungsten isotopic evidence for the late veneer}},
	author       = {{Kruijer}, Thomas S. and {Kleine}, Thorsten and {Fischer-G{\"o}dde}, Mario and {Sprung}, Peter},
	year         = 2015,
	month        = apr,
	journal      = {Nature},
	volume       = 520,
	number       = 7548,
	pages        = {534--537},
	doi          = {10.1038/nature14360},
	adsurl       = {https://ui.adsabs.harvard.edu/abs/2015Natur.520..534K}
}

@article{kubik2021tracing,
	title        = {Tracing Earth's Volatile Delivery With Tin},
	author       = {Kubik, E and Siebert, J and Mahan, B and Creech, J and Blanchard, I and Agranier, A and Shcheka, S and Moynier, F},
	year         = 2021,
	journal      = {Journal of Geophysical Research: Solid Earth},
	publisher    = {Wiley Online Library},
	volume       = 126,
	number       = 10,
	pages        = {e2021JB022026}
}

@article{Lambrechts2019,
	title        = {{Formation of planetary systems by pebble accretion and migration. How the radial pebble flux determines a terrestrial-planet or super-Earth growth mode}},
	author       = {{Lambrechts}, Michiel and {Morbidelli}, Alessandro and {Jacobson}, Seth A. and {Johansen}, Anders and {Bitsch}, Bertram and {Izidoro}, Andre and {Raymond}, Sean N.},
	year         = 2019,
	month        = jul,
	journal      = {Astronomy and Astrophysics},
	volume       = 627,
	pages        = {A83},
	doi          = {10.1051/0004-6361/201834229},
	eid          = {A83},
	archiveprefix = {arXiv},
	eprint       = {1902.08694},
	primaryclass = {astro-ph.EP},
	adsurl       = {https://ui.adsabs.harvard.edu/abs/2019A\&A...627A..83L}
}

@article{li1996geochemistry,
	title        = {Geochemistry of mantle--core differentiation at high pressure},
	author       = {Li, Jie and Agee, Carl B},
	year         = 1996,
	journal      = {Nature},
	publisher    = {Nature Publishing Group UK London},
	volume       = 381,
	number       = 6584,
	pages        = {686--689}
}

@article{lichtenberg2021bifurcation,
	title        = {{Bifurcation of planetary building blocks during Solar System formation}},
	author       = {{Lichtenberg}, Tim and {Dr{\k{a}}{\.z}kowska}, Joanna and {Sch{\"o}nb{\"a}chler}, Maria and {Golabek}, Gregor J. and {Hands}, Thomas O.},
	year         = 2021,
	month        = jan,
	journal      = {Science},
	volume       = 371,
	number       = 6527,
	pages        = {365--370},
	doi          = {10.1126/science.abb3091},
	archiveprefix = {arXiv},
	eprint       = {2101.08571},
	primaryclass = {astro-ph.EP},
	adsurl       = {https://ui.adsabs.harvard.edu/abs/2021Sci...371..365L}
}

@inproceedings{Lichtenberg2023,
	title        = {{Geophysical Evolution During Rocky Planet Formation}},
	author       = {{Lichtenberg}, Tim and {Schaefer}, Laura K. and {Nakajima}, Miki and {Fischer}, Rebecca A.},
	year         = 2023,
	month        = jul,
	booktitle    = {Protostars and Planets VII},
	series       = {Astronomical Society of the Pacific Conference Series},
	volume       = 534,
	pages        = 907,
	doi          = {10.48550/arXiv.2203.10023},
	editor       = {{Inutsuka}, S. and {Aikawa}, Y. and {Muto}, T. and {Tomida}, K. and {Tamura}, M.},
	archiveprefix = {arXiv},
	eprint       = {2203.10023},
	primaryclass = {astro-ph.EP},
	adsurl       = {https://ui.adsabs.harvard.edu/abs/2023ASPC..534..907L}
}

@article{Liebske2012,
	title        = {{Melting phase relations in the MgO-MgSiO$_{3}$ system between 16 and 26 GPa: Implications for melting in Earth's deep interior}},
	author       = {{Liebske}, Christian and {Frost}, Daniel J.},
	year         = 2012,
	month        = sep,
	journal      = {Earth and Planetary Science Letters},
	volume       = 345,
	pages        = {159--170},
	doi          = {10.1016/j.epsl.2012.06.038},
	adsurl       = {https://ui.adsabs.harvard.edu/abs/2012E&PSL.345..159L}
}

@article{Liu2022,
	title        = {{Early Solar System instability triggered by dispersal of the gaseous disk}},
	author       = {{Liu}, Beibei and {Raymond}, Sean N. and {Jacobson}, Seth A.},
	year         = 2022,
	month        = apr,
	journal      = {Nature},
	volume       = 604,
	number       = 7907,
	pages        = {643--646},
	doi          = {10.1038/s41586-022-04535-1},
	archiveprefix = {arXiv},
	eprint       = {2205.02026},
	primaryclass = {astro-ph.EP},
	adsurl       = {https://ui.adsabs.harvard.edu/abs/2022Natur.604..643L}
}

@article{lock2017structure,
	title        = {The structure of terrestrial bodies: Impact heating, corotation limits, and synestias},
	author       = {Lock, Simon J and Stewart, Sarah T},
	year         = 2017,
	journal      = {Journal of Geophysical Research: Planets},
	publisher    = {Wiley Online Library},
	volume       = 122,
	number       = 5,
	pages        = {950--982}
}

@article{lock2018origin,
	title        = {The origin of the Moon within a terrestrial synestia},
	author       = {Lock, Simon J and Stewart, Sarah T and Petaev, Michail I and Leinhardt, Zo{\"e} and Mace, Mia T and Jacobsen, Stein B and Cuk, Matija},
	year         = 2018,
	journal      = {Journal of Geophysical Research: Planets},
	publisher    = {Wiley Online Library},
	volume       = 123,
	number       = 4,
	pages        = {910--951}
}

@article{lugmair1998early,
	title        = {{Early solar system timescales according to $^{53}$Mn- $^{53}$Cr systematics}},
	author       = {{Lugmair}, G.~W. and {Shukolyukov}, A.},
	year         = 1998,
	month        = aug,
	journal      = {Geochimica et Cosmochimica Acta},
	volume       = 62,
	number       = 16,
	pages        = {2863--2886},
	doi          = {10.1016/S0016-7037(98)00189-6},
	adsurl       = {https://ui.adsabs.harvard.edu/abs/1998GeCoA..62.2863L}
}

@article{lykawka2023terrestrial,
	title        = {{Terrestrial planet and asteroid belt formation by Jupiter-Saturn chaotic excitation}},
	author       = {{Lykawka}, Patryk Sofia and {Ito}, Takashi},
	year         = 2023,
	month        = mar,
	journal      = {Scientific Reports},
	volume       = 13,
	pages        = 4708,
	doi          = {10.1038/s41598-023-30382-9},
	eid          = 4708,
	archiveprefix = {arXiv},
	eprint       = {2208.13647},
	primaryclass = {astro-ph.EP},
	adsurl       = {https://ui.adsabs.harvard.edu/abs/2023NatSR..13.4708L}
}

@article{mann2009evidence,
	title        = {{Evidence for high-pressure core-mantle differentiation from the metal-silicate partitioning of lithophile and weakly-siderophile elements}},
	author       = {{Mann}, Ute and {Frost}, Daniel J. and {Rubie}, David C.},
	year         = 2009,
	month        = dec,
	journal      = {Geochimica et Cosmochimica Acta},
	volume       = 73,
	number       = 24,
	pages        = {7360--7386},
	doi          = {10.1016/j.gca.2009.08.006},
	adsurl       = {https://ui.adsabs.harvard.edu/abs/2009GeCoA..73.7360M}
}

@article{masset2001reversing,
	title        = {{Reversing type II migration: resonance trapping of a lighter giant protoplanet}},
	author       = {{Masset}, F. and {Snellgrove}, M.},
	year         = 2001,
	month        = feb,
	journal      = {Monthly Notices of the Royal Astronomical Society},
	volume       = 320,
	number       = 4,
	pages        = {L55-L59},
	doi          = {10.1046/j.1365-8711.2001.04159.x},
	archiveprefix = {arXiv},
	eprint       = {astro-ph/0003421},
	primaryclass = {astro-ph},
	adsurl       = {https://ui.adsabs.harvard.edu/abs/2001MNRAS.320L..55M}
}

@article{Mastrobuono-Battisti2015,
	title        = {{A primordial origin for the compositional similarity between the Earth and the Moon}},
	author       = {{Mastrobuono-Battisti}, Alessandra and {Perets}, Hagai B. and {Raymond}, Sean N.},
	year         = 2015,
	month        = apr,
	journal      = {Nature},
	volume       = 520,
	number       = 7546,
	pages        = {212--215},
	doi          = {10.1038/nature14333},
	archiveprefix = {arXiv},
	eprint       = {1502.07346},
	primaryclass = {astro-ph.EP},
	adsurl       = {https://ui.adsabs.harvard.edu/abs/2015Natur.520..212M}
}

@article{mcdonough1995composition,
	title        = {The composition of the Earth},
	author       = {McDonough, William F and Sun, S-S},
	year         = 1995,
	month        = mar,
	journal      = {Chemical geology},
	publisher    = {Elsevier},
	volume       = 120,
	number       = {3-4},
	pages        = {223--253},
	doi          = {10.1016/0009-2541(94)00140-4},
	adsurl       = {https://ui.adsabs.harvard.edu/abs/1995ChGeo.120..223M}
}

@article{Meier2014,
	title        = {{On the origin and composition of Theia: Constraints from new models of the Giant Impact}},
	author       = {{Meier}, M.~M.~M. and {Reufer}, A. and {Wieler}, R.},
	year         = 2014,
	month        = nov,
	journal      = {Icarus},
	volume       = 242,
	pages        = {316--328},
	doi          = {10.1016/j.icarus.2014.08.003},
	archiveprefix = {arXiv},
	eprint       = {1410.3819},
	primaryclass = {astro-ph.EP},
	adsurl       = {https://ui.adsabs.harvard.edu/abs/2014Icar..242..316M}
}

@article{morbidelli2005chaotic,
	title        = {{Chaotic capture of Jupiter's Trojan asteroids in the early Solar System}},
	author       = {{Morbidelli}, A. and {Levison}, H.~F. and {Tsiganis}, K. and {Gomes}, R.},
	year         = 2005,
	month        = may,
	journal      = {Nature},
	volume       = 435,
	number       = 7041,
	pages        = {462--465},
	doi          = {10.1038/nature03540},
	adsurl       = {https://ui.adsabs.harvard.edu/abs/2005Natur.435..462M}
}

@article{morbidelli2012building,
	title        = {{Building Terrestrial Planets}},
	author       = {{Morbidelli}, A. and {Lunine}, J.~I. and {O'Brien}, D.~P. and {Raymond}, S.~N. and {Walsh}, K.~J.},
	year         = 2012,
	month        = may,
	journal      = {Annual Review of Earth and Planetary Sciences},
	volume       = 40,
	number       = 1,
	pages        = {251--275},
	doi          = {10.1146/annurev-earth-042711-105319},
	archiveprefix = {arXiv},
	eprint       = {1208.4694},
	primaryclass = {astro-ph.EP},
	adsurl       = {https://ui.adsabs.harvard.edu/abs/2012AREPS..40..251M}
}

@article{morbidelli2015great,
	title        = {{The great dichotomy of the Solar System: Small terrestrial embryos and massive giant planet cores}},
	author       = {{Morbidelli}, A. and {Lambrechts}, M. and {Jacobson}, S. and {Bitsch}, B.},
	year         = 2015,
	month        = sep,
	journal      = {Icarus},
	volume       = 258,
	pages        = {418--429},
	doi          = {10.1016/j.icarus.2015.06.003},
	archiveprefix = {arXiv},
	eprint       = {1506.01666},
	primaryclass = {astro-ph.EP},
	adsurl       = {https://ui.adsabs.harvard.edu/abs/2015Icar..258..418M}
}

@article{morbidelli2022contemporary,
	title        = {Contemporary formation of early Solar System planetesimals at two distinct radial locations},
	author       = {Morbidelli, Alessandro and Baillie, Kevin and Batygin, Konstantin and Charnoz, Sebastien and Guillot, Tristan and Rubie, David C and Kleine, Thorsten},
	year         = 2022,
	journal      = {Nature Astronomy},
	publisher    = {Nature Publishing Group},
	volume       = 6,
	number       = 1,
	pages        = {72--79}
}

@article{morbidelli2025did,
	title        = {{Did the terrestrial planets of the solar system form by pebble accretion?}},
	author       = {{Morbidelli}, A. and {Kleine}, T. and {Nimmo}, F.},
	year         = 2025,
	month        = jan,
	journal      = {Earth and Planetary Science Letters},
	volume       = 650,
	pages        = 119120,
	doi          = {10.1016/j.epsl.2024.119120},
	eid          = 119120,
	archiveprefix = {arXiv},
	eprint       = {2411.09271},
	primaryclass = {astro-ph.EP},
	adsurl       = {https://ui.adsabs.harvard.edu/abs/2025E\&PSL.65019120M}
}

@article{mothediniz2003distribution,
	title        = {{Distribution of taxonomic classes in the main belt of asteroids}},
	author       = {{Moth{\'e}-Diniz}, Thais and {Carvano}, Jorge M. {\'a}rcio and {Lazzaro}, Daniela},
	year         = 2003,
	month        = mar,
	journal      = {Icarus},
	volume       = 162,
	number       = 1,
	pages        = {10--21},
	doi          = {10.1016/S0019-1035(02)00066-0},
	adsurl       = {https://ui.adsabs.harvard.edu/abs/2003Icar..162...10M}
}

@article{Nakajima2021,
	title        = {{Scaling laws for the geometry of an impact-induced magma ocean}},
	author       = {{Nakajima}, Miki and {Golabek}, Gregor J. and {W{\"u}nnemann}, Kai and {Rubie}, David C. and {Burger}, Christoph and {Melosh}, Henry J. and {Jacobson}, Seth A. and {Manske}, Lukas and {Hull}, Scott D.},
	year         = 2021,
	month        = aug,
	journal      = {Earth and Planetary Science Letters},
	volume       = 568,
	pages        = 116983,
	doi          = {10.1016/j.epsl.2021.116983},
	eid          = 116983,
	archiveprefix = {arXiv},
	eprint       = {2004.04269},
	primaryclass = {astro-ph.EP},
	adsurl       = {https://ui.adsabs.harvard.edu/abs/2021E\&PSL.56816983N}
}

@article{nathan2023constraining,
	title        = {Constraining the origin of Mars with simulations of multi-stage core formation},
	author       = {Nathan, Gabriel and Rubie, David C and Jacobson, Seth A},
	year         = 2023,
	journal      = {Icarus},
	publisher    = {Elsevier},
	volume       = 401,
	pages        = 115596
}

@article{nesvorny2021role,
	title        = {{The Role of Early Giant-planet Instability in Terrestrial Planet Formation}},
	author       = {{Nesvorn{\'y}}, David and {Roig}, Fernando V. and {Deienno}, Rogerio},
	year         = 2021,
	month        = feb,
	journal      = {Astronomical Journal},
	volume       = 161,
	number       = 2,
	pages        = 50,
	doi          = {10.3847/1538-3881/abc8ef},
	eid          = 50,
	archiveprefix = {arXiv},
	eprint       = {2012.02323},
	primaryclass = {astro-ph.EP},
	adsurl       = {https://ui.adsabs.harvard.edu/abs/2021AJ....161...50N}
}

@article{Nesvorny2025terrestrial,
	title        = {{Terrestrial Planet Formation from Two Source Reservoirs}},
	author       = {{Nesvorn{\'y}}, David and {Morbidelli}, Alessandro and {Bottke}, William F. and {Deienno}, Rogerio and {Goldberg}, Max},
	year         = 2025,
	month        = sep,
	journal      = {The Astronomical Journal},
	volume       = 170,
	number       = 3,
	pages        = 180,
	doi          = {10.3847/1538-3881/adf20a},
	eid          = 180,
	archiveprefix = {arXiv},
	eprint       = {2507.14814},
	primaryclass = {astro-ph.EP},
	adsurl       = {https://ui.adsabs.harvard.edu/abs/2025AJ....170..180N}
}

@article{nimmo2007mars,
	title        = {How rapidly did mars accrete? Uncertainties in the hf--w timing of core formation},
	author       = {Nimmo, F. and Kleine, T.},
	year         = 2007,
	journal      = {Icarus},
	volume       = 191,
	number       = 2,
	pages        = {497--504}
}

@article{nimmo2010tungsten,
	title        = {Tungsten isotopic evolution during late-stage accretion: constraints on Earth--Moon equilibration},
	author       = {Nimmo, Francis and O'brien, DP and Kleine, Thorsten},
	year         = 2010,
	journal      = {Earth and Planetary Science Letters},
	publisher    = {Elsevier},
	volume       = 292,
	number       = {3-4},
	pages        = {363--370}
}

@article{nimmo2024mechanisms,
	title        = {Mechanisms and timing of carbonaceous chondrite delivery to the Earth},
	author       = {Nimmo, Francis and Kleine, Thorsten and Morbidelli, Alessandro and Nesvorny, David},
	year         = 2024,
	journal      = {Earth and Planetary Science Letters},
	publisher    = {Elsevier},
	volume       = 648,
	pages        = 119112
}

@article{Nittler2011,
	title        = {{The Major-Element Composition of Mercury{\textquoteright}s Surface from MESSENGER X-ray Spectrometry}},
	author       = {{Nittler}, Larry R. and {Starr}, Richard D. and {Weider}, Shoshana Z. and {McCoy}, Timothy J. and {Boynton}, William V. and {Ebel}, Denton S. and {Ernst}, Carolyn M. and {Evans}, Larry G. and {Goldsten}, John O. and {Hamara}, David K. and {Lawrence}, David J. and {McNutt}, Ralph L. and {Schlemm}, Charles E. and {Solomon}, Sean C. and {Sprague}, Ann L.},
	year         = 2011,
	month        = sep,
	journal      = {Science},
	volume       = 333,
	number       = 6051,
	pages        = 1847,
	doi          = {10.1126/science.1211567},
	adsurl       = {https://ui.adsabs.harvard.edu/abs/2011Sci...333.1847N}
}

@article{obrien2006terrestrial,
	title        = {{Terrestrial planet formation with strong dynamical friction}},
	author       = {{O'Brien}, David P. and {Morbidelli}, Alessandro and {Levison}, Harold F.},
	year         = 2006,
	month        = sep,
	journal      = {Icarus},
	volume       = 184,
	number       = 1,
	pages        = {39--58},
	doi          = {10.1016/j.icarus.2006.04.005},
	adsurl       = {https://ui.adsabs.harvard.edu/abs/2006Icar..184...39O}
}

@article{pahlevan2007equilibration,
	title        = {{Equilibration in the aftermath of the lunar-forming giant impact}},
	author       = {{Pahlevan}, Kaveh and {Stevenson}, David J.},
	year         = 2007,
	month        = oct,
	journal      = {Earth and Planetary Science Letters},
	volume       = 262,
	number       = {3-4},
	pages        = {438--449},
	doi          = {10.1016/j.epsl.2007.07.055},
	archiveprefix = {arXiv},
	eprint       = {1012.5323},
	primaryclass = {astro-ph.EP},
	adsurl       = {https://ui.adsabs.harvard.edu/abs/2007E\&PSL.262..438P}
}

@misc{palme2014cosmochemical,
	title        = {Cosmochemical Estimates of Mantle Composition. Planets, Asteriods, Comets and The Solar System, Volume 2 of Treatise on Geochemistry . Edited by Andrew M. Davis},
	author       = {Palme, H and O'Neill, H},
	year         = 2014,
	publisher    = {Elsevier}
}

@article{raymond2006high,
	title        = {High-resolution simulations of the final assembly of Earth-like planets I. Terrestrial accretion and dynamics},
	author       = {Raymond, Sean N and Quinn, Thomas and Lunine, Jonathan I},
	year         = 2006,
	journal      = {Icarus},
	publisher    = {Elsevier},
	volume       = 183,
	number       = 2,
	pages        = {265--282}
}

@article{raymond2009building,
	title        = {Building the terrestrial planets: Constrained accretion in the inner Solar System},
	author       = {Raymond, Sean N and O'Brien, David P and Morbidelli, Alessandro and Kaib, Nathan A},
	year         = 2009,
	journal      = {Icarus},
	publisher    = {Elsevier},
	volume       = 203,
	number       = 2,
	pages        = {644--662}
}

@article{Raymond2017,
	title        = {{Origin of water in the inner Solar System: Planetesimals scattered inward during Jupiter and Saturn's rapid gas accretion}},
	author       = {{Raymond}, Sean N. and {Izidoro}, Andre},
	year         = 2017,
	month        = nov,
	journal      = {Icarus},
	volume       = 297,
	pages        = {134--148},
	doi          = {10.1016/j.icarus.2017.06.030},
	archiveprefix = {arXiv},
	eprint       = {1707.01234},
	primaryclass = {astro-ph.EP},
	adsurl       = {https://ui.adsabs.harvard.edu/abs/2017Icar..297..134R}
}

@article{raymond2017empty,
	title        = {The empty primordial asteroid belt},
	author       = {Raymond, Sean N and Izidoro, Andre},
	year         = 2017,
	journal      = {Science advances},
	publisher    = {American Association for the Advancement of Science},
	volume       = 3,
	number       = 9,
	pages        = {e1701138}
}

@inproceedings{Raymond2022,
	title        = {{Planet Formation: Key Mechanisms and Global Models}},
	author       = {{Raymond}, Sean N. and {Morbidelli}, Alessandro},
	year         = 2022,
	month        = jan,
	booktitle    = {Demographics of Exoplanetary Systems, Lecture Notes of the 3rd Advanced School on Exoplanetary Science},
	series       = {Astrophysics and Space Science Library},
	volume       = 466,
	pages        = {3--82},
	doi          = {10.1007/978-3-030-88124-5_1},
	editor       = {{Biazzo}, Katia and {Bozza}, Valerio and {Mancini}, Luigi and {Sozzetti}, Alessandro},
	archiveprefix = {arXiv},
	eprint       = {2002.05756},
	primaryclass = {astro-ph.EP}
}

@article{reufer2012hit,
	title        = {A hit-and-run giant impact scenario},
	author       = {Reufer, Andreas and Meier, Matthias MM and Benz, Willy and Wieler, Rainer},
	year         = 2012,
	journal      = {Icarus},
	publisher    = {Elsevier},
	volume       = 221,
	number       = 1,
	pages        = {296--299}
}

@article{righter2003partition,
	title        = {Partition coefficients at high pressure and temperature},
	author       = {Righter, K and Drake, MJ},
	year         = 2003,
	journal      = {Treatise on geochemistry},
	volume       = 2,
	pages        = 568
}

@article{righter2016redox,
	title        = {Redox variations in the inner solar system with new constraints from vanadium XANES in spinels},
	author       = {Righter, Kevin and Sutton, Steve R and Danielson, Lisa and Pando, Kellye and Newville, Matt},
	year         = 2016,
	journal      = {American Mineralogist},
	publisher    = {Mineralogical Society of America},
	volume       = 101,
	number       = 9,
	pages        = {1928--1942}
}

@article{righter2020redox,
	title        = {Redox processes in early Earth accretion and in terrestrial bodies},
	author       = {Righter, Kevin and Herd, Christopher DK and Boujibar, Asmaa},
	year         = 2020,
	journal      = {Elements: An International Magazine of Mineralogy, Geochemistry, and Petrology},
	publisher    = {Mineralogical Society of America},
	volume       = 16,
	number       = 3,
	pages        = {161--166}
}

@article{rubie2011heterogeneous,
	title        = {Heterogeneous accretion, composition and core--mantle differentiation of the Earth},
	author       = {Rubie, David C and Frost, Daniel J and Mann, Ute and Asahara, Yuki and Nimmo, Francis and Tsuno, Kyusei and Kegler, Philip and Holzheid, Astrid and Palme, Herbert},
	year         = 2011,
	journal      = {Earth and Planetary Science Letters},
	publisher    = {Elsevier},
	volume       = 301,
	number       = {1-2},
	pages        = {31--42}
}

@incollection{RUBIE201543,
	title        = {9.03 - Formation of the Earth's Core},
	author       = {D.C. Rubie and F. Nimmo and H.J. Melosh},
	year         = 2015,
	booktitle    = {Treatise on Geophysics (Second Edition)},
	publisher    = {Elsevier},
	address      = {Oxford},
	pages        = {43--79},
	doi          = {https://doi.org/10.1016/B978-0-444-53802-4.00154-8},
	isbn         = {978-0-444-53803-1},
	url          = {https://www.sciencedirect.com/science/article/pii/B9780444538024001548},
	editor       = {Gerald Schubert},
	edition      = {Second Edition}
}

@article{rubie2015accretion,
	title        = {Accretion and differentiation of the terrestrial planets with implications for the compositions of early-formed Solar System bodies and accretion of water},
	author       = {Rubie, David C and Jacobson, Seth A and Morbidelli, Alessandro and O'Brien, David P and Young, Edward D and de Vries, Jellie and Nimmo, Francis and Palme, Herbert and Frost, Daniel J},
	year         = 2015,
	journal      = {Icarus},
	publisher    = {Elsevier},
	volume       = 248,
	pages        = {89--108}
}

@article{rubie2016highly,
	title        = {Highly siderophile elements were stripped from Earth's mantle by iron sulfide segregation},
	author       = {Rubie, David C and Laurenz, Vera and Jacobson, Seth A and Morbidelli, Alessandro and Palme, Herbert and Vogel, Antje K and Frost, Daniel J},
	year         = 2016,
	journal      = {Science},
	publisher    = {American Association for the Advancement of Science},
	volume       = 353,
	number       = 6304,
	pages        = {1141--1144}
}

@article{rufu2017multiple,
	title        = {{A multiple-impact origin for the Moon}},
	author       = {{Rufu}, Raluca and {Aharonson}, Oded and {Perets}, Hagai B.},
	year         = 2017,
	month        = jan,
	journal      = {Nature Geoscience},
	volume       = 10,
	number       = 2,
	pages        = {89--94},
	doi          = {10.1038/ngeo2866},
	archiveprefix = {arXiv},
	eprint       = {1903.02525},
	primaryclass = {astro-ph.EP},
	adsurl       = {https://ui.adsabs.harvard.edu/abs/2017NatGe..10...89R}
}

@article{schonbachler2005nb,
	title        = {{Nb/Zr fractionation on the Moon and the search for extinct $^{92}$Nb}},
	author       = {{Sch{\"o}nb{\"a}chler}, Maria and {Lee}, Der-Chuen and {Rehk{\"a}mper}, Mark and {Halliday}, Alex N. and {Hattendorf}, Bodo and {G{\"u}nther}, Detlef},
	year         = 2005,
	month        = feb,
	journal      = {Geochimica et Cosmochimica Acta},
	volume       = 69,
	number       = 3,
	pages        = {775--785},
	doi          = {10.1016/j.gca.2004.08.003},
	adsurl       = {https://ui.adsabs.harvard.edu/abs/2005GeCoA..69..775S}
}

@article{siebert2013terrestrial,
	title        = {Terrestrial accretion under oxidizing conditions},
	author       = {Siebert, Julien and Badro, James and Antonangeli, Daniele and Ryerson, Frederick J},
	year         = 2013,
	journal      = {Science},
	publisher    = {American Association for the Advancement of Science},
	volume       = 339,
	number       = 6124,
	pages        = {1194--1197}
}

@article{sossi2025physicochemical,
	title        = {Physicochemical Controls on the Compositions of the Earth and Planets},
	author       = {Sossi, Paolo A and Hin, Remco C and Kleine, Thorsten and Morbidelli, Alessandro and Nimmo, Francis},
	year         = 2025,
	journal      = {Space Science Reviews},
	publisher    = {Springer},
	volume       = 221,
	number       = 8,
	pages        = 118
}

@article{sossi2026homogeneous,
	title        = {Homogeneous accretion of the Earth in the inner Solar System},
	author       = {Sossi, Paolo A and Bower, Dan J},
	year         = 2026,
	journal      = {Nature Astronomy},
	publisher    = {Nature Publishing Group UK London},
	pages        = {1--8}
}

@article{spicuzza2007oxygen,
	title        = {{Oxygen isotope constraints on the origin and differentiation of the Moon}},
	author       = {{Spicuzza}, Michael J. and {Day}, James M.~D. and {Taylor}, Lawrence A. and {Valley}, John W.},
	year         = 2007,
	month        = jan,
	journal      = {Earth and Planetary Science Letters},
	volume       = 253,
	number       = {1-2},
	pages        = {254--265},
	doi          = {10.1016/j.epsl.2006.10.030},
	adsurl       = {https://ui.adsabs.harvard.edu/abs/2007E&PSL.253..254S}
}

@article{tian2017coupled,
	title        = {{Coupled orbital-thermal evolution of the early Earth-Moon system with a fast-spinning Earth}},
	author       = {{Tian}, ZhenLiang and {Wisdom}, Jack and {Elkins-Tanton}, Linda},
	year         = 2017,
	month        = jan,
	journal      = {Icarus},
	volume       = 281,
	pages        = {90--102},
	doi          = {10.1016/j.icarus.2016.08.030},
	adsurl       = {https://ui.adsabs.harvard.edu/abs/2017Icar..281...90T}
}

@article{tian2020vertical,
	title        = {{Vertical angular momentum constraint on lunar formation and orbital history}},
	author       = {{Tian}, ZhenLiang and {Wisdom}, Jack},
	year         = 2020,
	month        = jul,
	journal      = {Proceedings of the National Academy of Science},
	volume       = 117,
	number       = 27,
	pages        = {15460--15464},
	doi          = {10.1073/pnas.2003496117},
	adsurl       = {https://ui.adsabs.harvard.edu/abs/2020PNAS..11715460T}
}

@article{Tonks1992,
	title        = {Core formation by giant impacts},
	author       = {Tonks, W Brian and Melosh, H Jay},
	year         = 1992,
	journal      = {Icarus},
	volume       = 100,
	number       = 2,
	pages        = {326--346}
}

@article{Tonks1993,
	title        = {{Magma ocean formation due to giant impacts}},
	author       = {{Tonks}, W. Brian and {Melosh}, H. Jay},
	year         = 1993,
	month        = mar,
	journal      = {Journal of Geophysical Research},
	volume       = 98,
	number       = {E3},
	pages        = {5319--5333},
	doi          = {10.1029/92JE02726},
	adsurl       = {https://ui.adsabs.harvard.edu/abs/1993JGR....98.5319T}
}

@article{touboul2007late,
	title        = {Late formation and prolonged differentiation of the Moon inferred from W isotopes in lunar metals},
	author       = {Touboul, Mathieu and Kleine, Thorsten and Bourdon, Bernard and Palme, Herbert and Wieler, Rainer},
	year         = 2007,
	journal      = {Nature},
	publisher    = {Nature Publishing Group UK London},
	volume       = 450,
	number       = 7173,
	pages        = {1206--1209}
}

@article{touboul2015tungsten,
	title        = {Tungsten isotopic evidence for disproportional late accretion to the Earth and Moon},
	author       = {Touboul, Mathieu and Puchtel, Igor S and Walker, Richard J},
	year         = 2015,
	journal      = {Nature},
	publisher    = {Nature Publishing Group UK London},
	volume       = 520,
	number       = 7548,
	pages        = {530--533}
}

@article{Tronnes2002,
	title        = {{Peridotite melting and mineral-melt partitioning of major and minor elements at 22-24.5 GPa}},
	author       = {{Tr{\o}nnes}, Reidar G. and {Frost}, Daniel J.},
	year         = 2002,
	month        = mar,
	journal      = {Earth and Planetary Science Letters},
	volume       = 197,
	number       = {1-2},
	pages        = {117--131},
	doi          = {10.1016/S0012-821X(02)00466-1},
	adsurl       = {https://ui.adsabs.harvard.edu/abs/2002E&PSL.197..117T}
}

@article{tsiganis2005origin,
	title        = {{Origin of the orbital architecture of the giant planets of the Solar System}},
	author       = {{Tsiganis}, K. and {Gomes}, R. and {Morbidelli}, A. and {Levison}, H.~F.},
	year         = 2005,
	month        = may,
	journal      = {Nature},
	volume       = 435,
	number       = 7041,
	pages        = {459--461},
	doi          = {10.1038/nature03539},
	adsurl       = {https://ui.adsabs.harvard.edu/abs/2005Natur.435..459T}
}

@article{Urey1953,
	title        = {{The composition of the stone meteorites and the origin of the meteorites}},
	author       = {{Urey}, Harold C. and {Craig}, Harmon},
	year         = 1953,
	month        = aug,
	journal      = {Geochimica et Cosmochimica Acta},
	volume       = 4,
	number       = 1,
	pages        = {36--82},
	doi          = {10.1016/0016-7037(53)90064-7},
	adsurl       = {https://ui.adsabs.harvard.edu/abs/1953GeCoA...4...36U}
}

@article{wade2005core,
	title        = {Core formation and the oxidation state of the Earth},
	author       = {Wade, Jon and Wood, BJ},
	year         = 2005,
	journal      = {Earth and Planetary Science Letters},
	publisher    = {Elsevier},
	volume       = 236,
	number       = {1-2},
	pages        = {78--95}
}

@article{Wade2016,
	title        = {{The oxidation state and mass of the Moon-forming impactor}},
	author       = {{Wade}, Jon and {Wood}, Bernard J.},
	year         = 2016,
	month        = may,
	journal      = {Earth and Planetary Science Letters},
	volume       = 442,
	pages        = {186--193},
	doi          = {10.1016/j.epsl.2016.02.053},
	adsurl       = {https://ui.adsabs.harvard.edu/abs/2016E\&PSL.442..186W}
}

@article{walker2009highly,
	title        = {Highly siderophile elements in the Earth, Moon and Mars: update and implications for planetary accretion and differentiation},
	author       = {Walker, Richard J},
	year         = 2009,
	journal      = {Geochemistry},
	publisher    = {Elsevier},
	volume       = 69,
	number       = 2,
	pages        = {101--125}
}

@article{walsh2011effect,
	title        = {The effect of an early planetesimal-driven migration of the giant planets on terrestrial planet formation},
	author       = {Walsh, Kevin J and Morbidelli, Alessandro},
	year         = 2011,
	journal      = {Astronomy \& Astrophysics},
	publisher    = {EDP Sciences},
	volume       = 526,
	pages        = {A126}
}

@article{walsh2011low,
	title        = {A low mass for Mars from Jupiter's early gas-driven migration},
	author       = {Walsh, Kevin J and Morbidelli, Alessandro and Raymond, Sean N and O'Brien, David P and Mandell, Avi M},
	year         = 2011,
	journal      = {Nature},
	publisher    = {Nature Publishing Group},
	volume       = 475,
	number       = 7355,
	pages        = {206--209}
}

@article{walsh2012populating,
	title        = {{Populating the asteroid belt from two parent source regions due to the migration of giant planets{\textemdash}``The Grand Tack''}},
	author       = {{Walsh}, Kevin J. and {Morbidelli}, A. and {Raymond}, S.~N. and {O'Brien}, D.~P. and {Mandell}, A.~M.},
	year         = 2012,
	month        = dec,
	journal      = {Meteoritics and Planetary Science},
	volume       = 47,
	number       = 12,
	pages        = {1941--1947},
	doi          = {10.1111/j.1945-5100.2012.01418.x},
	adsurl       = {https://ui.adsabs.harvard.edu/abs/2012M&PS...47.1941W}
}

@article{warren2005new,
	title        = {“New” lunar meteorites: Implications for composition of the global lunar surface, lunar crust, and the bulk Moon},
	author       = {Warren, Paul H},
	year         = 2005,
	journal      = {Meteoritics \& Planetary Science},
	publisher    = {Wiley Online Library},
	volume       = 40,
	number       = 3,
	pages        = {477--506}
}

@article{warren2011stable,
	title        = {Stable-isotopic anomalies and the accretionary assemblage of the Earth and Mars: A subordinate role for carbonaceous chondrites},
	author       = {Warren, Paul H},
	year         = 2011,
	month        = nov,
	journal      = {Earth and Planetary Science Letters},
	publisher    = {Elsevier},
	volume       = 311,
	number       = {1-2},
	pages        = {93--100},
	doi          = {10.1016/j.epsl.2011.08.047},
	adsurl       = {https://ui.adsabs.harvard.edu/abs/2011E\&PSL.311...93W}
}

@article{weidenschilling1977aerodynamics,
	title        = {Aerodynamics of solid bodies in the solar nebula},
	author       = {Weidenschilling, SJ},
	year         = 1977,
	journal      = {Monthly Notices of the Royal Astronomical Society},
	publisher    = {The Royal Astronomical Society},
	volume       = 180,
	number       = 2,
	pages        = {57--70}
}

@incollection{weisberg2006systematics,
	title        = {{Systematics and Evaluation of Meteorite Classification}},
	author       = {{Weisberg}, M.~K. and {McCoy}, T.~J. and {Krot}, A.~N.},
	year         = 2006,
	booktitle    = {Meteorites and the Early Solar System II},
	publisher    = {Univ. of Arizona Press},
	pages        = 19,
	editor       = {{Lauretta}, Dante S. and {McSween}, Harry Y.},
	adsurl       = {https://ui.adsabs.harvard.edu/abs/2006mess.book...19W}
}

@inproceedings{Wetherill1978,
	title        = {{Accumulation of the Terrestrial Planets}},
	author       = {{Wetherill}, G.~W.},
	year         = 1978,
	month        = jan,
	booktitle    = {IAU Colloq. 52: Protostars and Planets},
	pages        = 565,
	editor       = {{Gehrels}, Tom and {Matthews}, Mildred Shapley},
	adsurl       = {https://ui.adsabs.harvard.edu/abs/1978prpl.conf..565W}
}

@article{Wetherill1989,
	title        = {{Accumulation of a swarm of small planetesimals}},
	author       = {{Wetherill}, G.~W. and {Stewart}, G.~R.},
	year         = 1989,
	month        = feb,
	journal      = {Icarus},
	volume       = 77,
	number       = 2,
	pages        = {330--357},
	doi          = {10.1016/0019-1035(89)90093-6},
	adsurl       = {https://ui.adsabs.harvard.edu/abs/1989Icar...77..330W}
}

@article{wetherill1989accumulation,
	title        = {Accumulation of a swarm of small planetesimals},
	author       = {Wetherill, GW and Stewart, Glen R},
	year         = 1989,
	journal      = {Icarus},
	publisher    = {Elsevier},
	volume       = 77,
	number       = 2,
	pages        = {330--357}
}

@inproceedings{wetherill1991isn,
	title        = {Why isn't Mars as big as Earth?},
	author       = {Wetherill, GW},
	year         = 1991,
	booktitle    = {Abstracts of the Lunar and Planetary Science Conference, volume 22, page 1495,(1991)},
	volume       = 22,
	pages        = 1495
}

@article{wiechert2001oxygen,
	title        = {Oxygen isotopes and the Moon-forming giant impact},
	author       = {Wiechert, Uwe and Halliday, Alex N and Lee, D-C and Snyder, Gittel A and Taylor, Larry A and Rumble, Douglas},
	year         = 2001,
	journal      = {Science},
	publisher    = {American Association for the Advancement of Science},
	volume       = 294,
	number       = 5541,
	pages        = {345--348}
}

@article{wisdom2015early,
	title        = {{Early evolution of the Earth-Moon system with a fast-spinning Earth}},
	author       = {{Wisdom}, Jack and {Tian}, ZhenLiang},
	year         = 2015,
	month        = aug,
	journal      = {Icarus},
	volume       = 256,
	pages        = {138--146},
	doi          = {10.1016/j.icarus.2015.02.025},
	adsurl       = {https://ui.adsabs.harvard.edu/abs/2015Icar..256..138W}
}

@article{woo2024terrestrial,
	title        = {Terrestrial planet formation from a ring: long-term simulations accounting for the giant planet instability},
	author       = {Woo, JMY and Nesvorn{\`y}, D and Scora, J and Morbidelli, A},
	year         = 2024,
	journal      = {Icarus},
	publisher    = {Elsevier},
	volume       = 417,
	pages        = 116109
}

@article{wood2006accretion,
	title        = {Accretion of the Earth and segregation of its core},
	author       = {Wood, Bernard J and Walter, Michael J and Wade, Jonathan},
	year         = 2006,
	journal      = {Nature},
	publisher    = {Nature Publishing Group},
	volume       = 441,
	number       = 7095,
	pages        = {825--833}
}

@article{wordsworth2018redox,
	title        = {Redox evolution via gravitational differentiation on low-mass planets: implications for abiotic oxygen, water loss, and habitability},
	author       = {Wordsworth, RD and Schaefer, LK and Fischer, RA},
	year         = 2018,
	journal      = {The Astronomical Journal},
	publisher    = {IOP Publishing},
	volume       = 155,
	number       = 5,
	pages        = 195
}

@article{Yap2025,
	title        = {{Early Solar System Turbulence Constrained by High Oxidation States in the Oldest Noncarbonaceous Planetesimals}},
	author       = {{Yap}, Teng Ee and {Batygin}, Konstantin and {Tissot}, Fran{\c{c}}ois L.~H.},
	year         = 2025,
	month        = jan,
	journal      = {Planetary Science Journal},
	volume       = 6,
	number       = 1,
	pages        = 2,
	doi          = {10.3847/PSJ/ad92fa},
	eid          = 2,
	archiveprefix = {arXiv},
	eprint       = {2412.07211},
	primaryclass = {astro-ph.EP},
	adsurl       = {https://ui.adsabs.harvard.edu/abs/2025PSJ.....6....2Y}
}

@article{young2016oxygen,
	title        = {{Oxygen isotopic evidence for vigorous mixing during the Moon-forming giant impact}},
	author       = {{Young}, Edward D. and {Kohl}, Issaku E. and {Warren}, Paul H. and {Rubie}, David C. and {Jacobson}, Seth A. and {Morbidelli}, Alessandro},
	year         = 2016,
	month        = jan,
	journal      = {Science},
	volume       = 351,
	number       = 6272,
	pages        = {493--496},
	doi          = {10.1126/science.aad0525},
	archiveprefix = {arXiv},
	eprint       = {1603.04536},
	primaryclass = {astro-ph.EP},
	adsurl       = {https://ui.adsabs.harvard.edu/abs/2016Sci...351..493Y}
}

@article{zhang2012proto,
	title        = {The proto-Earth as a significant source of lunar material},
	author       = {Zhang, Junjun and Dauphas, Nicolas and Davis, Andrew M and Leya, Ingo and Fedkin, Alexei},
	year         = 2012,
	journal      = {Nature Geoscience},
	publisher    = {Nature Publishing Group UK London},
	volume       = 5,
	number       = 4,
	pages        = {251--255}
}

@article{zhang2024compositions,
	title        = {Compositions of iron-meteorite parent bodies constrain the structure of the protoplanetary disk},
	author       = {Zhang, Bidong and Chabot, Nancy L and Rubin, Alan E},
	year         = 2024,
	journal      = {Proceedings of the National Academy of Sciences},
	publisher    = {National Acad Sciences},
	volume       = 121,
	number       = 23,
	pages        = {e2306995121}
}

@article{zube2019constraints,
	title        = {Constraints on terrestrial planet formation timescales and equilibration processes in the Grand Tack scenario from Hf-W isotopic evolution},
	author       = {Zube, Nicholas G and Nimmo, Francis and Fischer, Rebecca A and Jacobson, Seth A},
	year         = 2019,
	journal      = {Earth and planetary science letters},
	publisher    = {Elsevier},
	volume       = 522,
	pages        = {210--218}
}

\clearpage
\appendix
\onecolumn
\section{Additional Figures}
\label{app:additional_figures}

\renewcommand{\thefigure}{A.\arabic{figure}}
\setcounter{figure}{0}

\begin{figure}[!htbp]
    \centering
    \includegraphics[width=0.97\textwidth,height=0.72\textheight,keepaspectratio]{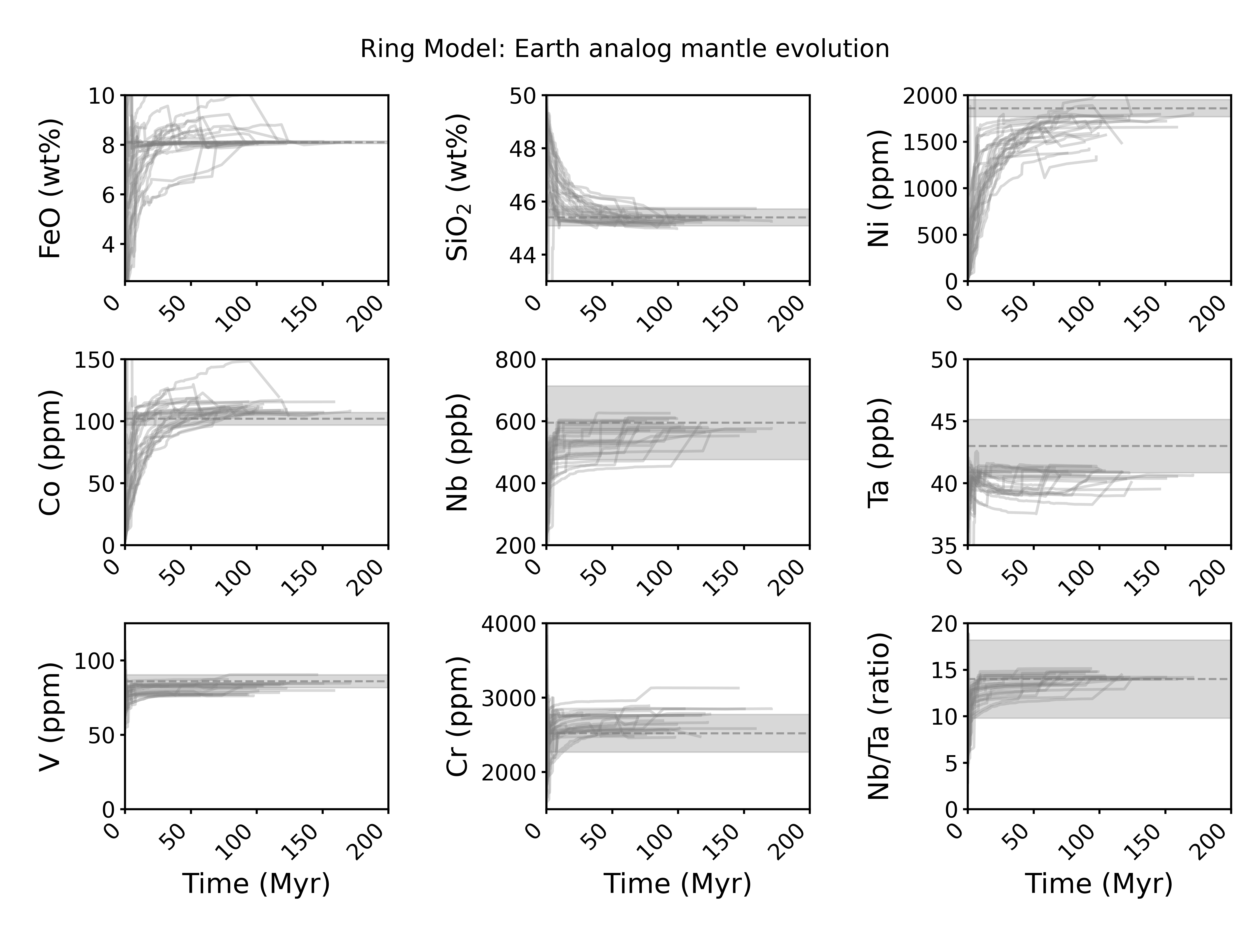}
    \caption{Mantle compositional evolution is shown for Earth analogs in Ring Model scenario simulations \citep{nesvorny2021role}. Subplots show the mantle abundance of fitted constraints, including FeO, SiO$_2$, Ni, Co, Nb, Ta, V, Cr, and the mantle Nb/Ta ratio on the y-axis. Time in millions of years is shown on the x-axis. Grey dashed lines and shaded region show the observed Earth mantle value and uncertainty, respectively, as estimated by \citet{palme2014cosmochemical}.}
    \label{fig:ring_elements}
\end{figure}

Figures \ref{fig:CJS_elements}--\ref{fig:ring_elements} show the evolving mantle composition for Earth analogs across the various terrestrial planet formation scenarios.

\begin{figure}[!htbp]
    \centering
    \includegraphics[width=0.97\textwidth,height=0.72\textheight,keepaspectratio]{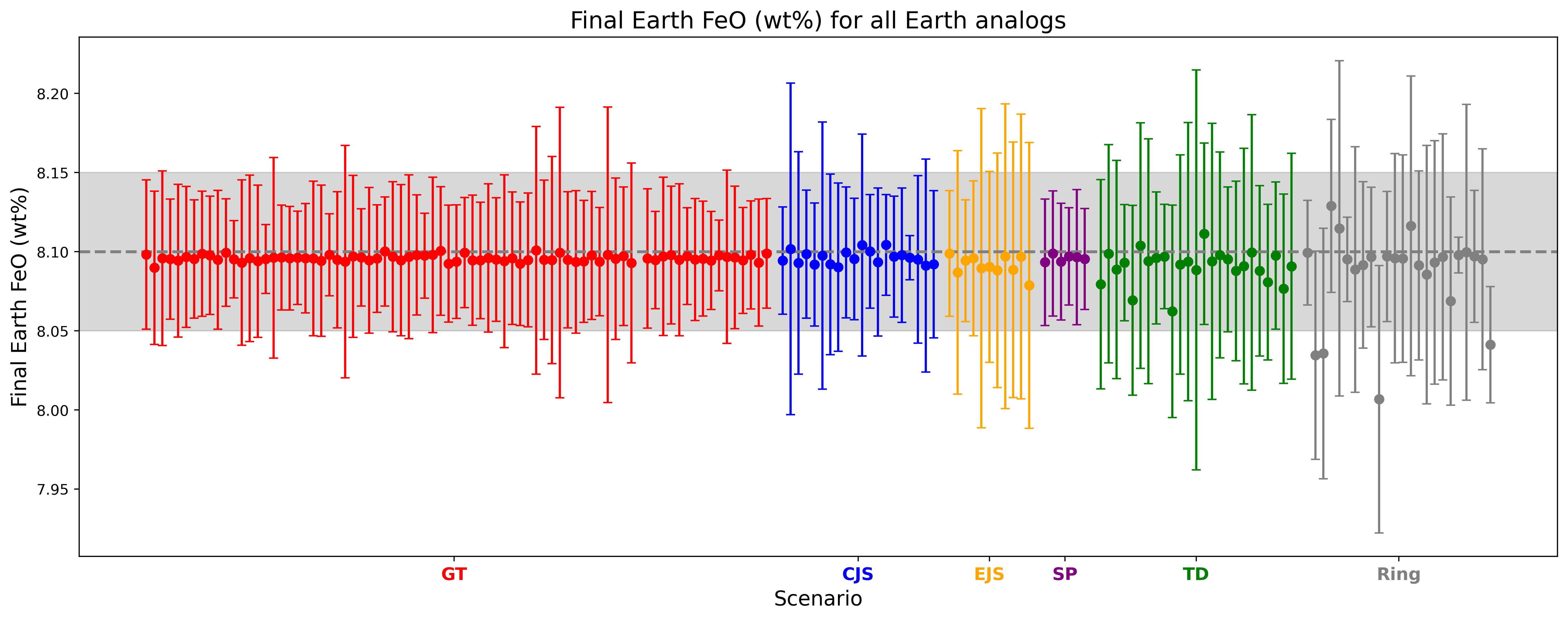}
    \caption{Final FeO wt\% of simulated Earth analog mantles. Error bars represent uncertainties in laboratory-measured elemental partitioning coefficients propagated through the core–mantle differentiation model. Colors and labels indicate the formation scenario of each simulated Earth. Abundance plus uncertainty in the Earth mantle \citep{palme2014cosmochemical} is shown in the grey band.}
    \label{fig:plot_earth_information_Final_Earth_FeO}
\end{figure}

\begin{figure}[!htbp]
    \centering
    \includegraphics[width=0.97\textwidth,height=0.72\textheight,keepaspectratio]{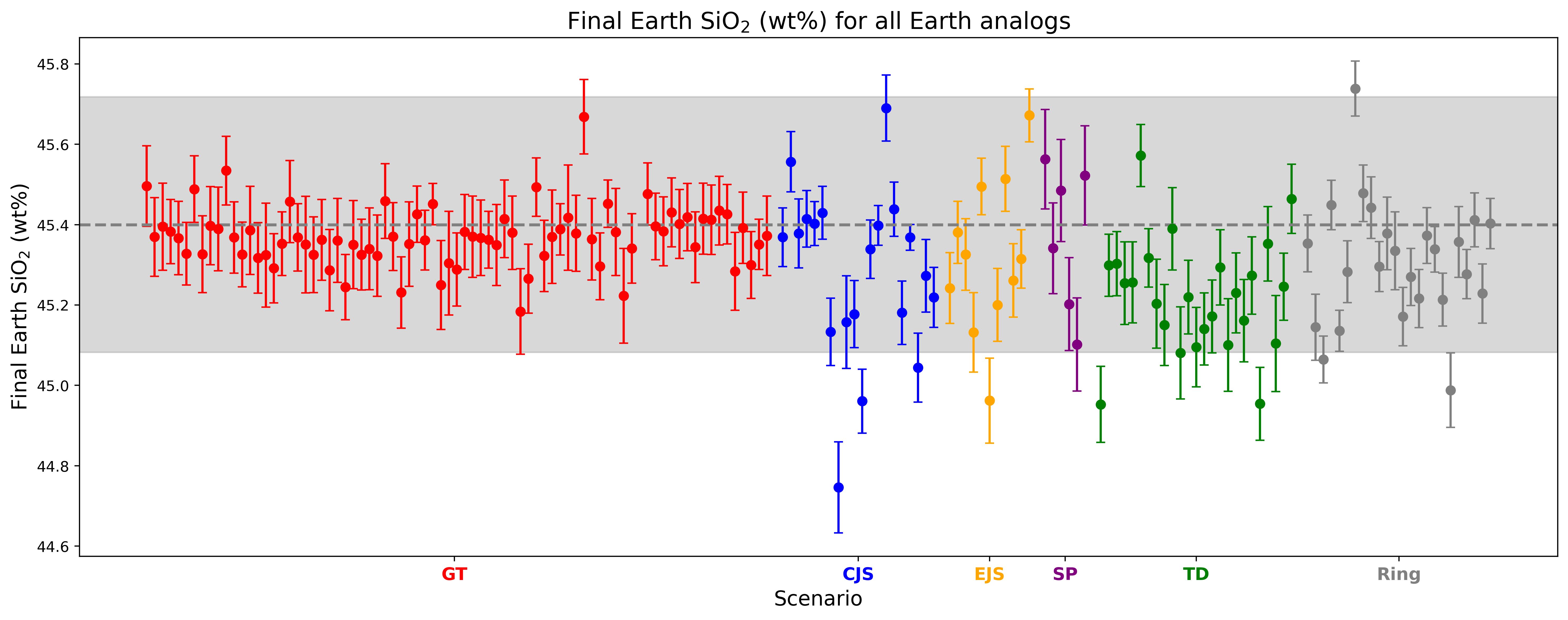}
    \caption{Final SiO$_2$ wt\% of simulated Earth analog mantles. Error bars represent uncertainties in laboratory-measured elemental partitioning coefficients propagated through the core–mantle differentiation model. Colors and labels indicate the formation scenario of each simulated Earth. Abundance plus uncertainty in the Earth mantle \citep{palme2014cosmochemical} is shown in the grey band.}
    \label{fig:plot_earth_information_Final_Earth_SiO2}
\end{figure}

\begin{figure}[!htbp]
    \centering
    \includegraphics[width=0.97\textwidth,height=0.72\textheight,keepaspectratio]{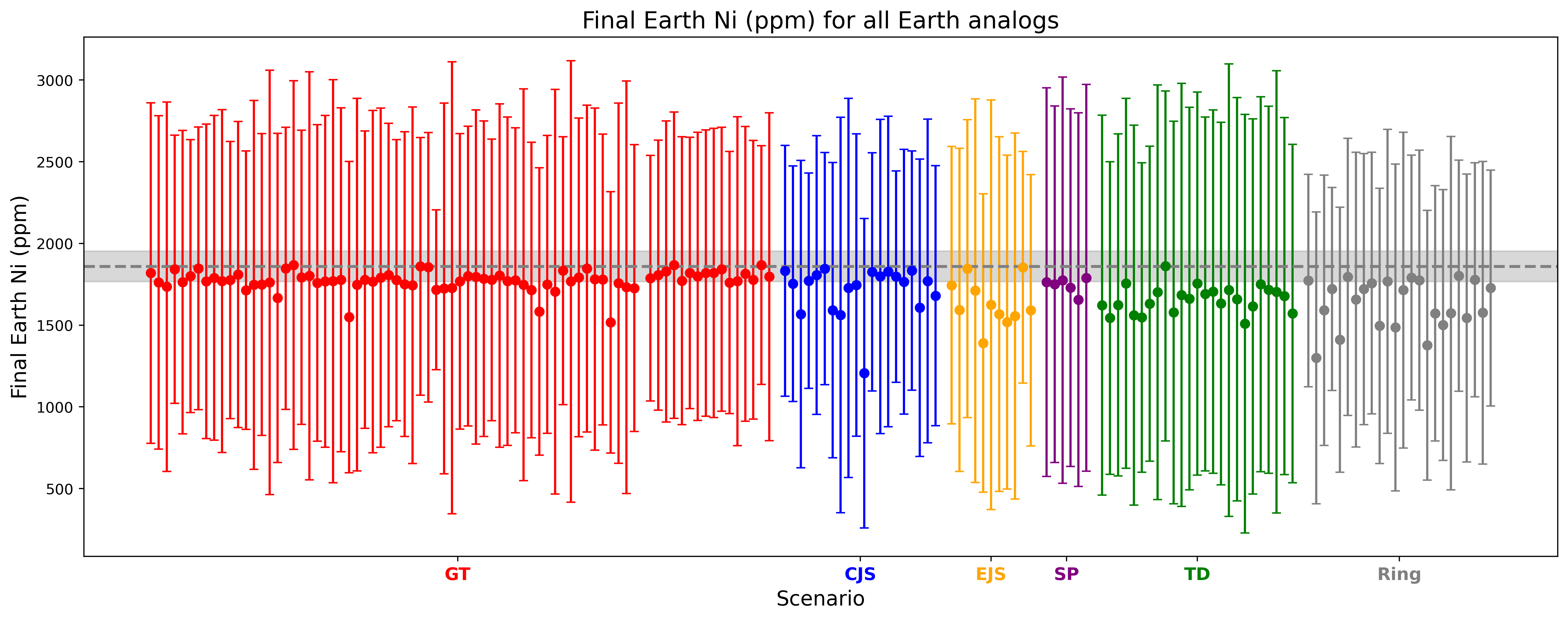}
    \caption{Final Ni (ppm) of simulated Earth analog mantles. Error bars represent uncertainties in laboratory-measured elemental partitioning coefficients propagated through the core–mantle differentiation model. Colors and labels indicate the formation scenario of each simulated Earth. Abundance plus uncertainty in the Earth mantle \citep{palme2014cosmochemical} is shown in the grey band.}
    \label{fig:plot_earth_information_Final_Earth_Ni}
\end{figure}

\begin{figure}[!htbp]
    \centering
    \includegraphics[width=0.97\textwidth,height=0.72\textheight,keepaspectratio]{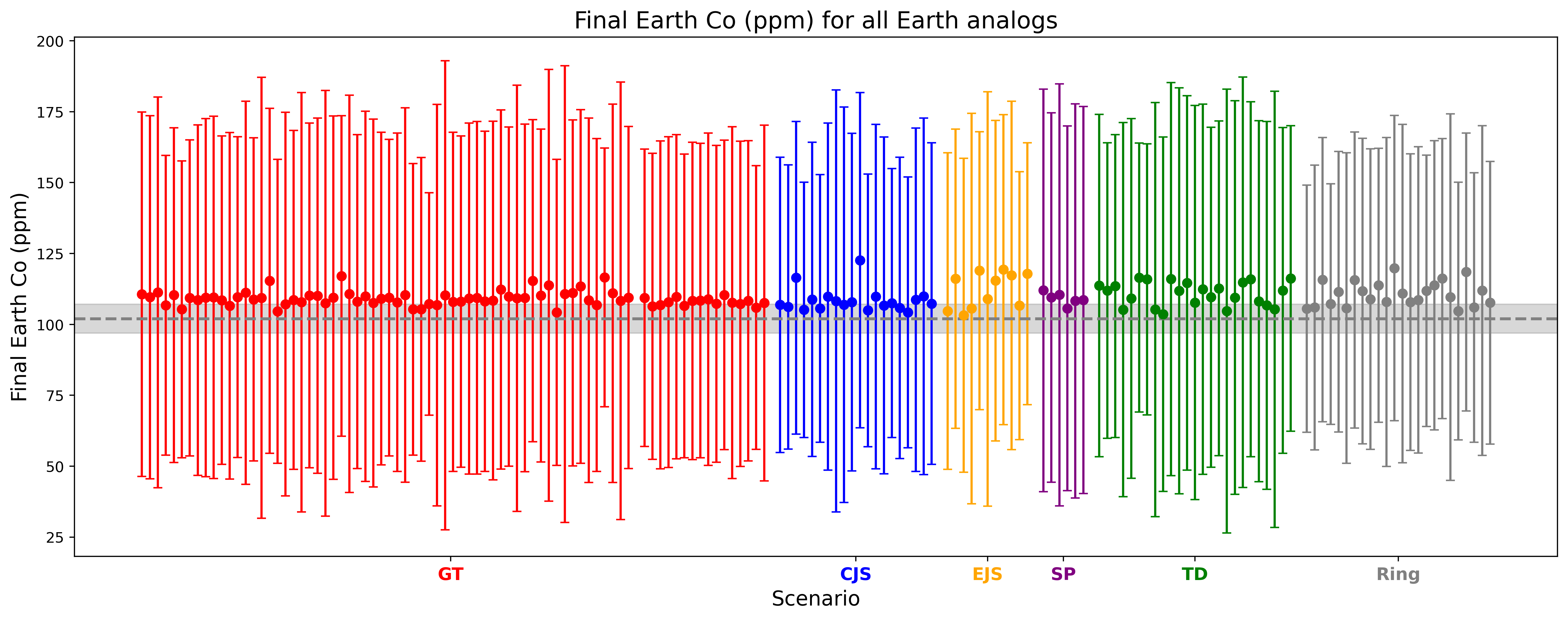}
    \caption{Final Co (ppm) of simulated Earth analog mantles. Error bars represent uncertainties in laboratory-measured elemental partitioning coefficients propagated through the core–mantle differentiation model. Colors and labels indicate the formation scenario of each simulated Earth. Abundance plus uncertainty in the Earth mantle \citep{palme2014cosmochemical} is shown in the grey band.}
    \label{fig:plot_earth_information_Final_Earth_Co}
\end{figure}

\begin{figure}[!htbp]
    \centering
    \includegraphics[width=0.97\textwidth,height=0.72\textheight,keepaspectratio]{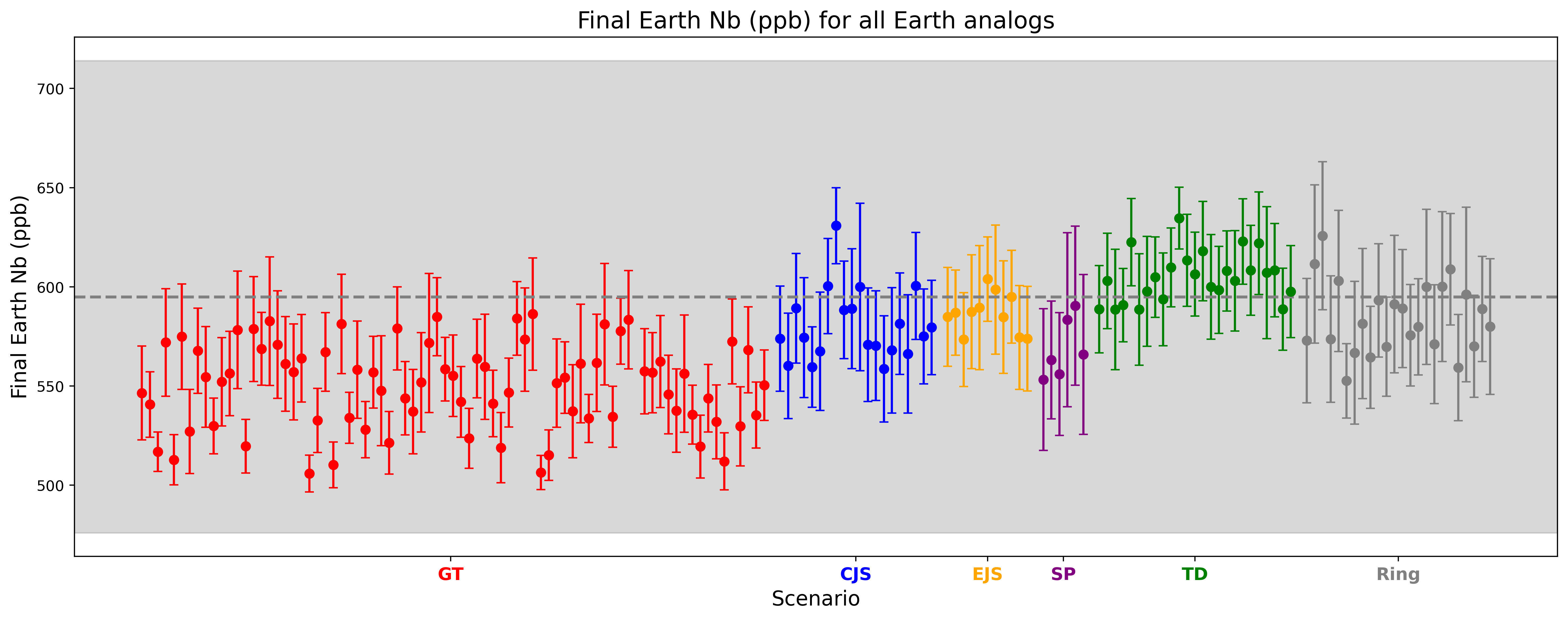}
    \caption{Final Nb (ppb) of simulated Earth analog mantles. Error bars represent uncertainties in laboratory-measured elemental partitioning coefficients propagated through the core–mantle differentiation model. Colors and labels indicate the formation scenario of each simulated Earth. Abundance plus uncertainty in the Earth mantle \citep{palme2014cosmochemical} is shown in the grey band.}
    \label{fig:plot_earth_information_Final_Earth_Nb}
\end{figure}

\begin{figure}[!htbp]
    \centering
    \includegraphics[width=0.97\textwidth,height=0.72\textheight,keepaspectratio]{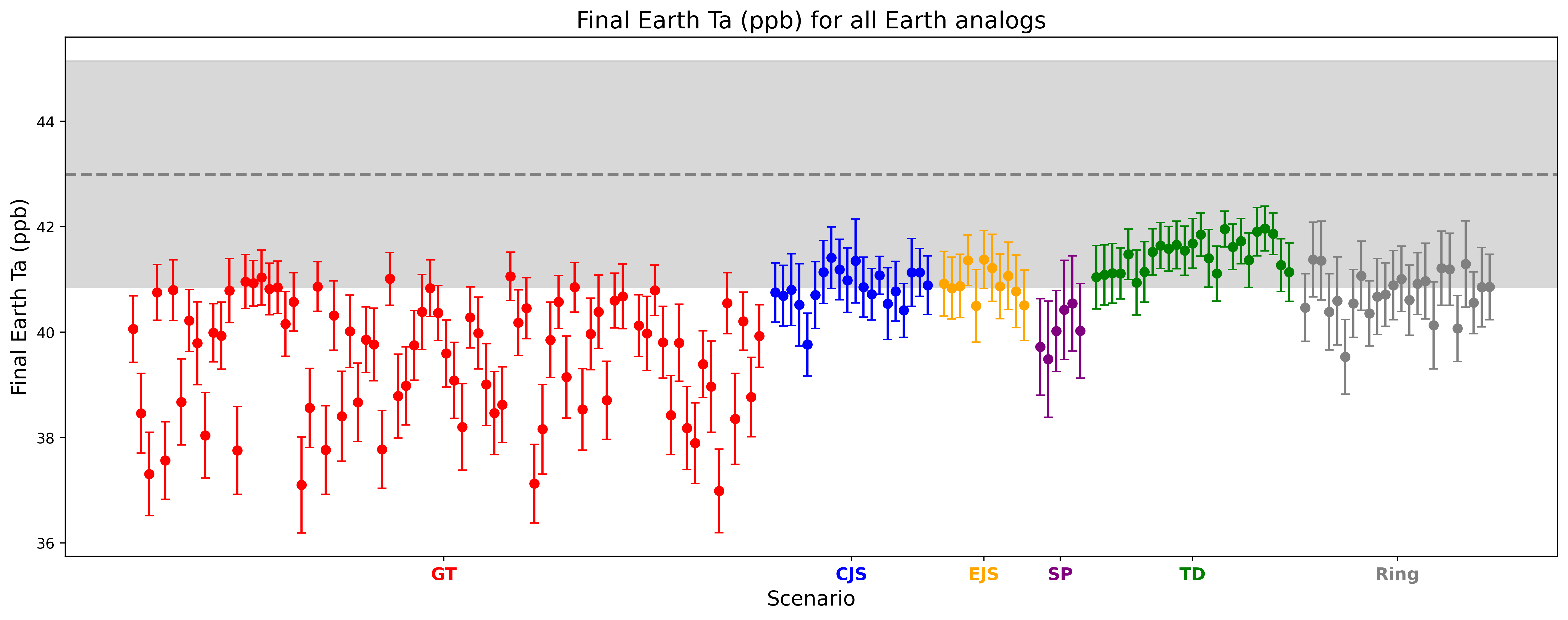}
    \caption{Final Ta (ppb) of simulated Earth analog mantles. Error bars represent uncertainties in laboratory-measured elemental partitioning coefficients propagated through the core–mantle differentiation model. Colors and labels indicate the formation scenario of each simulated Earth. Abundance plus uncertainty in the Earth mantle \citep{palme2014cosmochemical} is shown in the grey band.}
    \label{fig:plot_earth_information_Final_Earth_Ta}
\end{figure}

\begin{figure}[!htbp]
    \centering
    \includegraphics[width=0.97\textwidth,height=0.72\textheight,keepaspectratio]{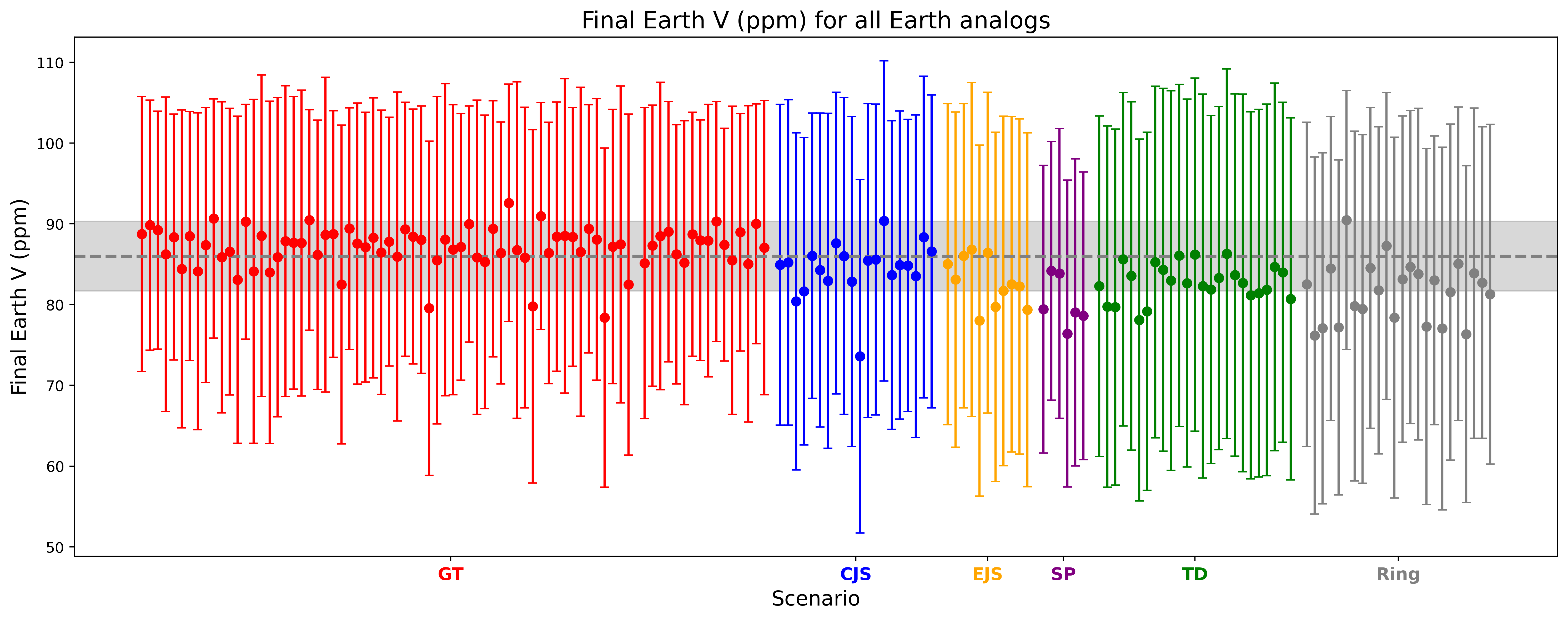}
    \caption{Final V (ppm) of simulated Earth analog mantles. Error bars represent uncertainties in laboratory-measured elemental partitioning coefficients propagated through the core–mantle differentiation model. Colors and labels indicate the formation scenario of each simulated Earth. Abundance plus uncertainty in the Earth mantle \citep{palme2014cosmochemical} is shown in the grey band.}
    \label{fig:plot_earth_information_Final_Earth_V}
\end{figure}

\begin{figure}[!htbp]
    \centering
    \includegraphics[width=0.97\textwidth,height=0.72\textheight,keepaspectratio]{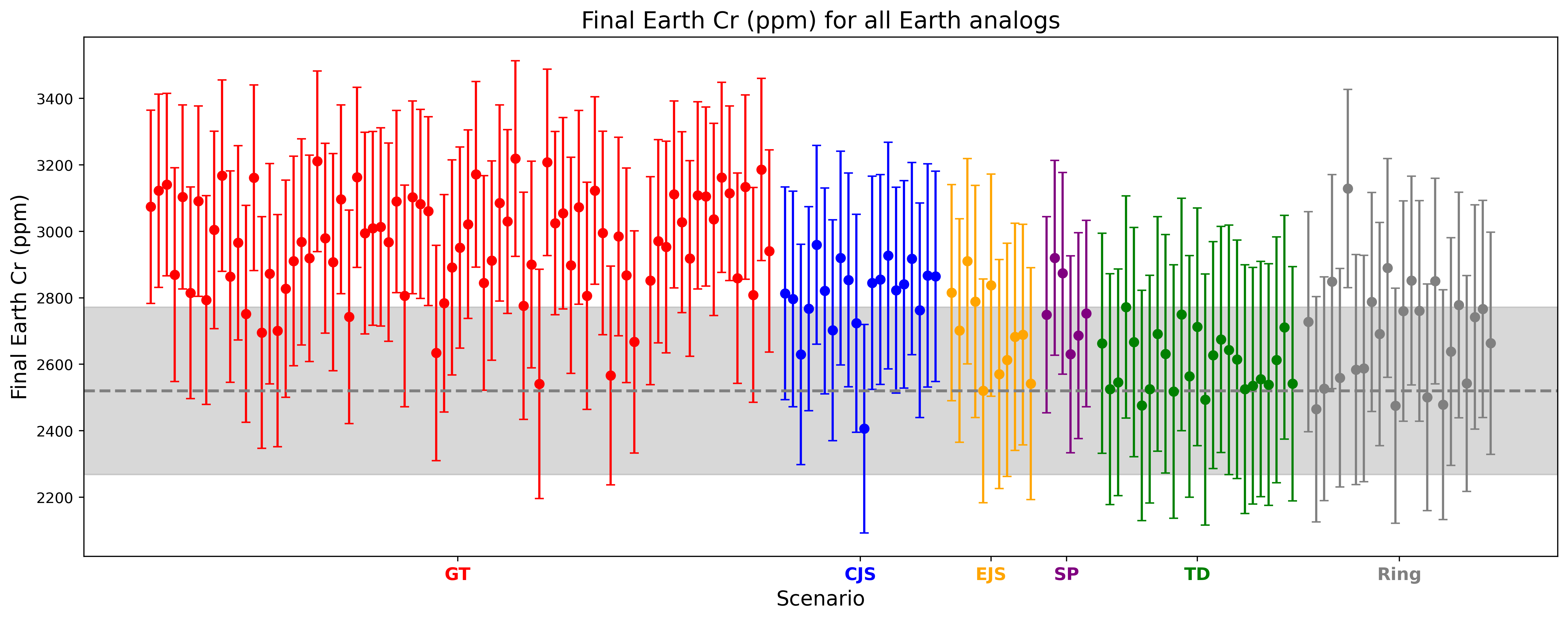}
    \caption{Final Cr (ppm) of simulated Earth analog mantles. Error bars represent uncertainties in laboratory-measured elemental partitioning coefficients propagated through the core–mantle differentiation model. Colors and labels indicate the formation scenario of each simulated Earth. Abundance plus uncertainty in the Earth mantle \citep{palme2014cosmochemical} is shown in the grey band.}
    \label{fig:plot_earth_information_Final_Earth_Cr}
\end{figure}

\begin{figure}[!htbp]
    \centering
    \includegraphics[width=0.97\textwidth,height=0.72\textheight,keepaspectratio]{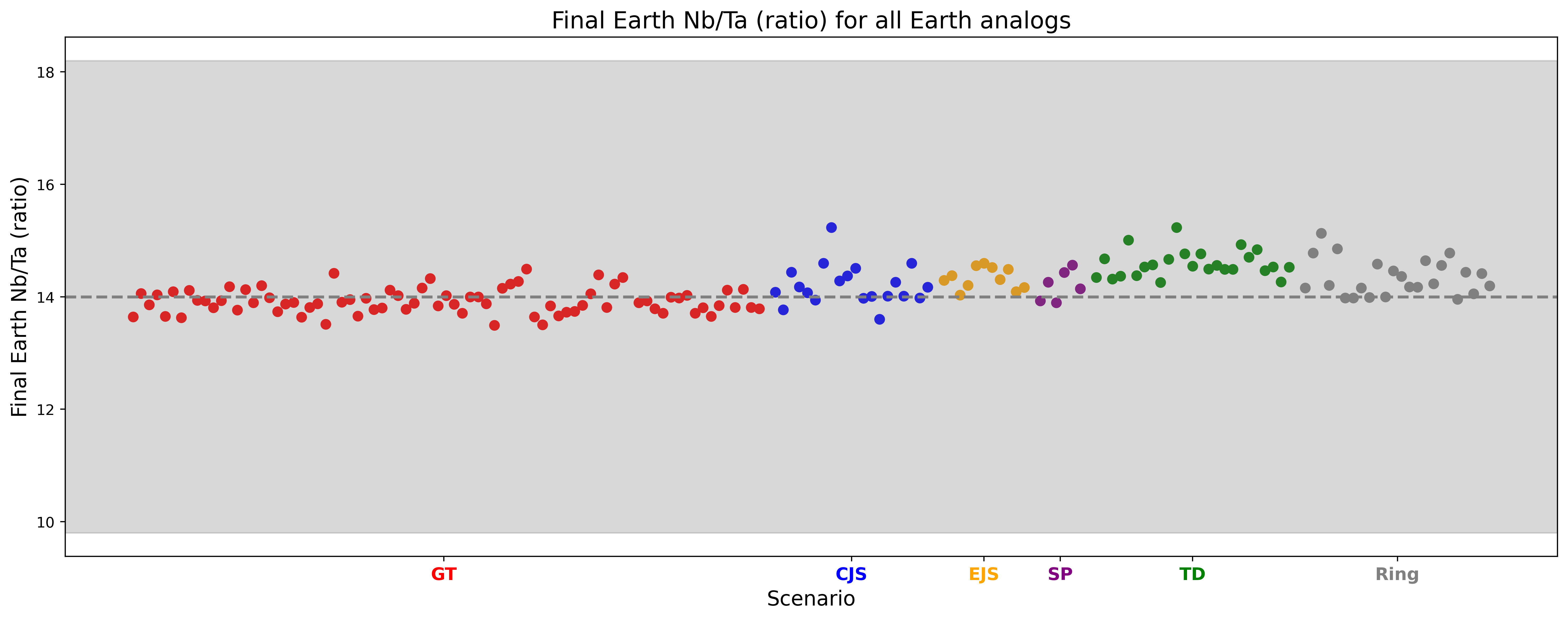}
    \caption{Final Nb/Ta ratio of simulated Earth analog mantles. Error bars represent uncertainties in laboratory-measured elemental partitioning coefficients propagated through the core–mantle differentiation model. Colors and labels indicate the formation scenario of each simulated Earth. Ratio plus uncertainty in the Earth mantle \citep{palme2014cosmochemical} is shown in the grey band.}
    \label{fig:plot_earth_information_Final_Earth_NbTa}
\end{figure}

Figures \ref{fig:plot_earth_information_Final_Earth_FeO}--\ref{fig:plot_earth_information_Final_Earth_NbTa} show the final mantle composition of specific element abundance constraints for Earth analogs across the various terrestrial planet formation scenarios.

\begin{figure}[!htbp]
    \centering
    \includegraphics[width=0.97\textwidth,height=0.72\textheight,keepaspectratio]{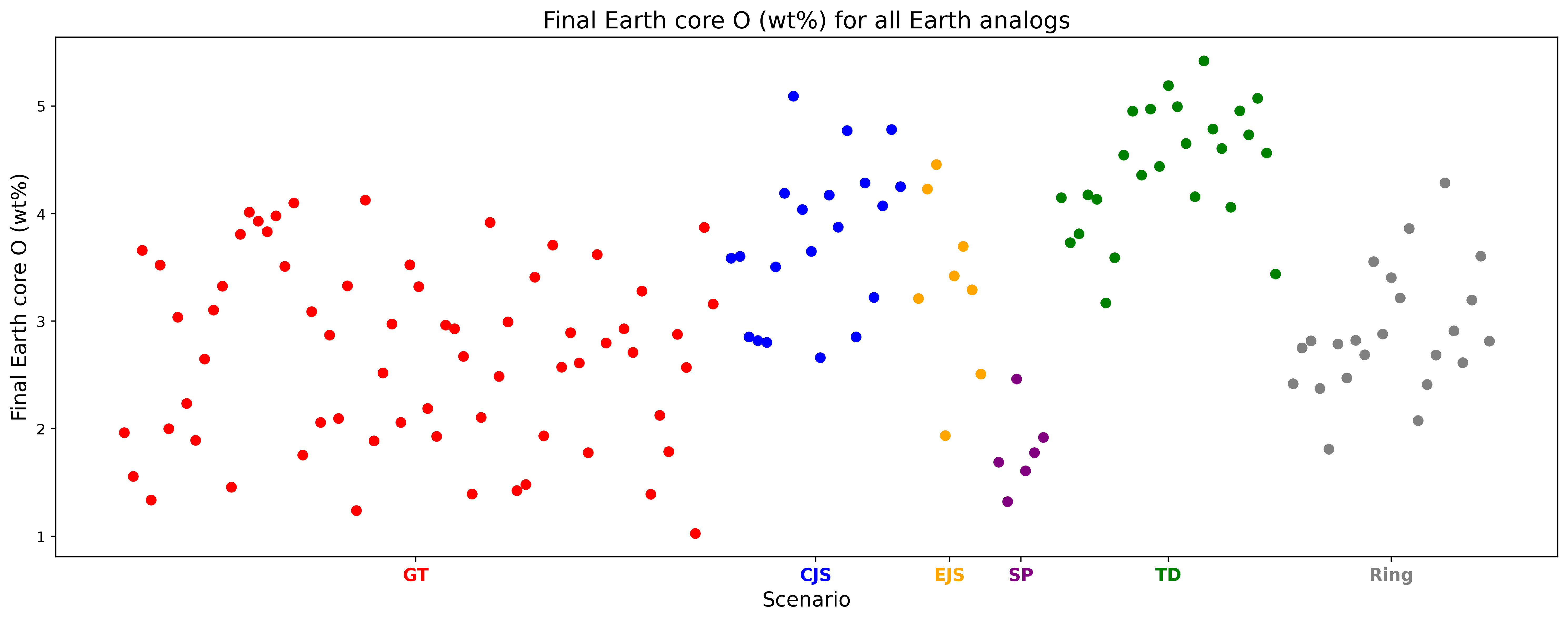}
    \caption{\GIN{Final oxygen concentration in the cores of simulated Earth analogs. Colors and labels indicate the formation scenario of each simulated Earth.}}
    \label{fig:final_earth_core_oxygen}
\end{figure}

\begin{figure}[!htbp]
    \centering
    \includegraphics[width=0.97\textwidth,height=0.72\textheight,keepaspectratio]{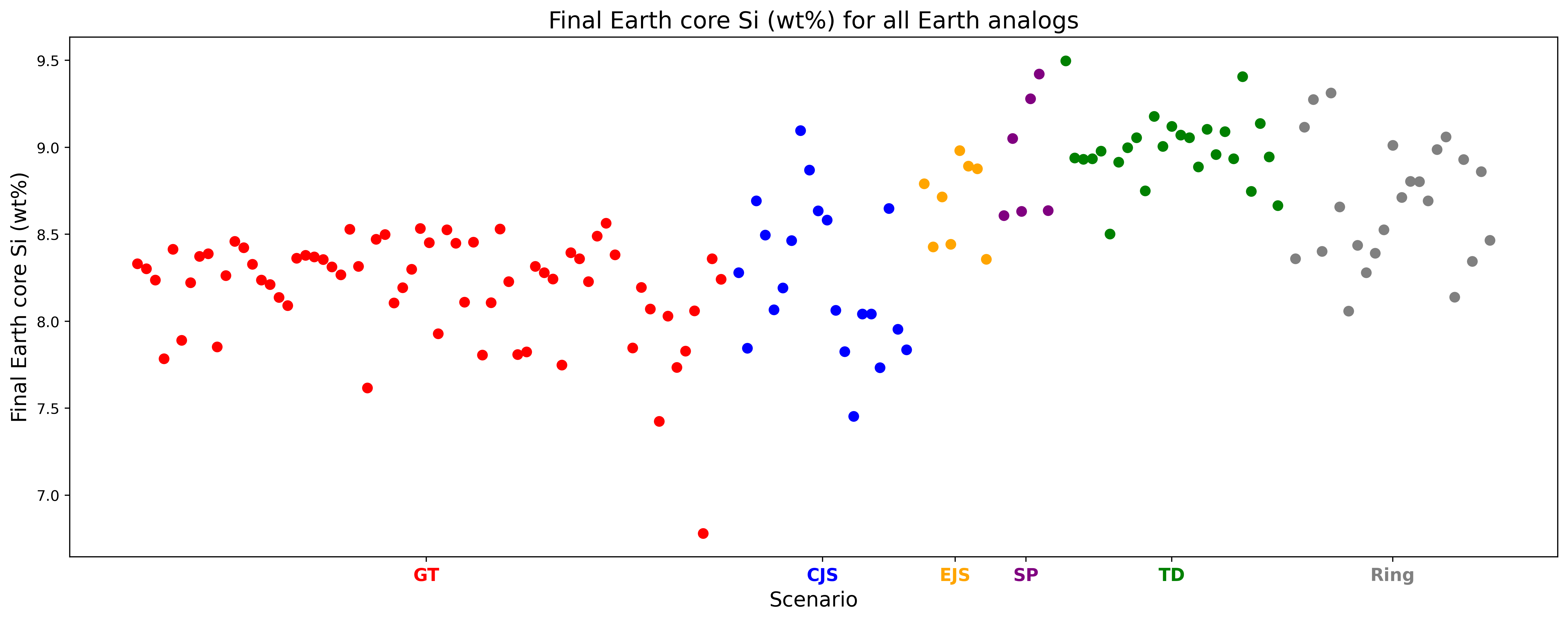}
    \caption{\GIN{Final silicon concentration in the cores of simulated Earth analogs. Colors and labels indicate the formation scenario of each simulated Earth.}}
    \label{fig:final_earth_core_silicon}
\end{figure}

\GIN{Figures \ref{fig:final_earth_core_oxygen} and \ref{fig:final_earth_core_silicon} show the oxygen and silicon compositions of Earth analog cores across the various terrestrial planet formation scenarios.}

\begin{center}
    \centering
    \includegraphics[width=0.95\textwidth,height=0.72\textheight,keepaspectratio]{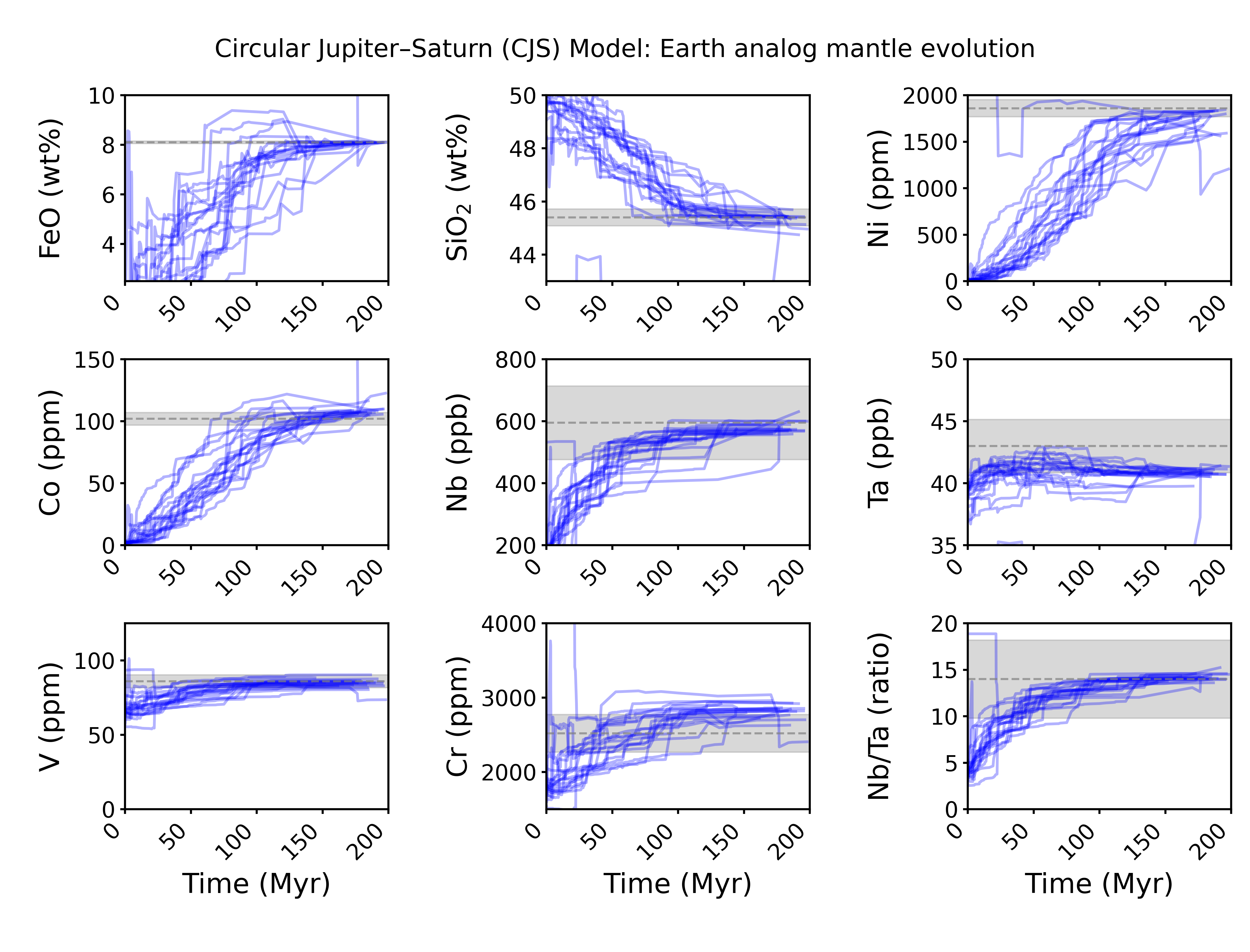}
    \captionof{figure}{Mantle compositional evolution is shown for Earth analogs in Circular Jupiter-Saturn (CJS) scenario simulations \citep{raymond2009building}. Subplots show the mantle abundance of fitted constraints, including FeO, SiO$_2$, Ni, Co, Nb, Ta, V, Cr, and the mantle Nb/Ta ratio on the y-axis. Time in millions of years is shown on the x-axis. Grey dashed lines and shaded region show the observed Earth mantle value and uncertainty, respectively, as estimated by \citet{palme2014cosmochemical}.}
    \label{fig:CJS_elements}
\end{center}

\begin{figure}[!htbp]
    \centering
    \includegraphics[width=0.97\textwidth,height=0.72\textheight,keepaspectratio]{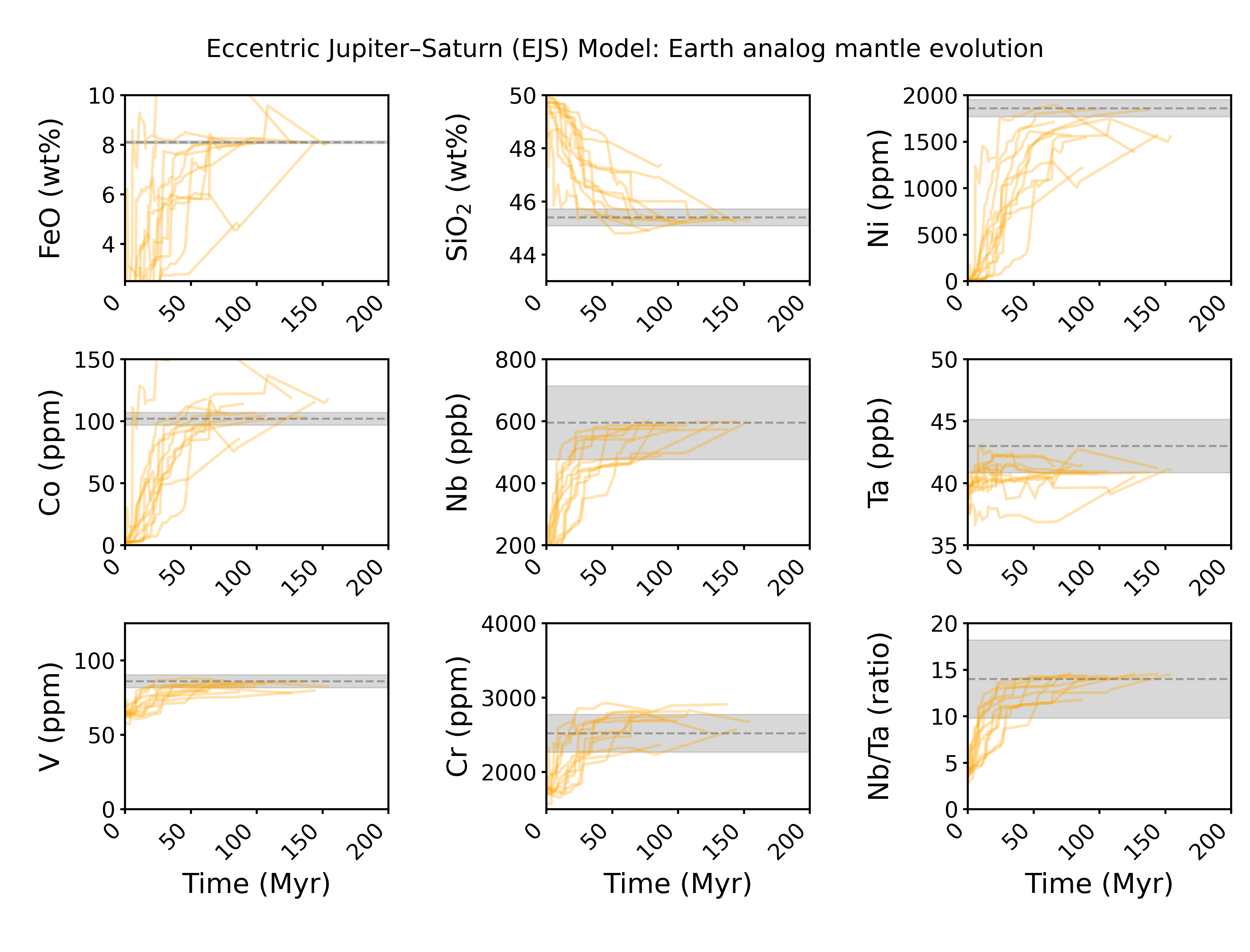}
    \caption{Mantle compositional evolution is shown for Earth analogs in Eccentric Jupiter-Saturn (EJS) scenario simulations \citep{raymond2009building}. Subplots show the mantle abundance of fitted constraints, including FeO, SiO$_2$, Ni, Co, Nb, Ta, V, Cr, and the mantle Nb/Ta ratio on the y-axis. Time in millions of years is shown on the x-axis. Grey dashed lines and shaded region show the observed Earth mantle value and uncertainty, respectively, as estimated by \citet{palme2014cosmochemical}.}
    \label{fig:EJS_elements}
\end{figure}

\begin{figure}[!htbp]
    \centering
    \includegraphics[width=0.97\textwidth,height=0.72\textheight,keepaspectratio]{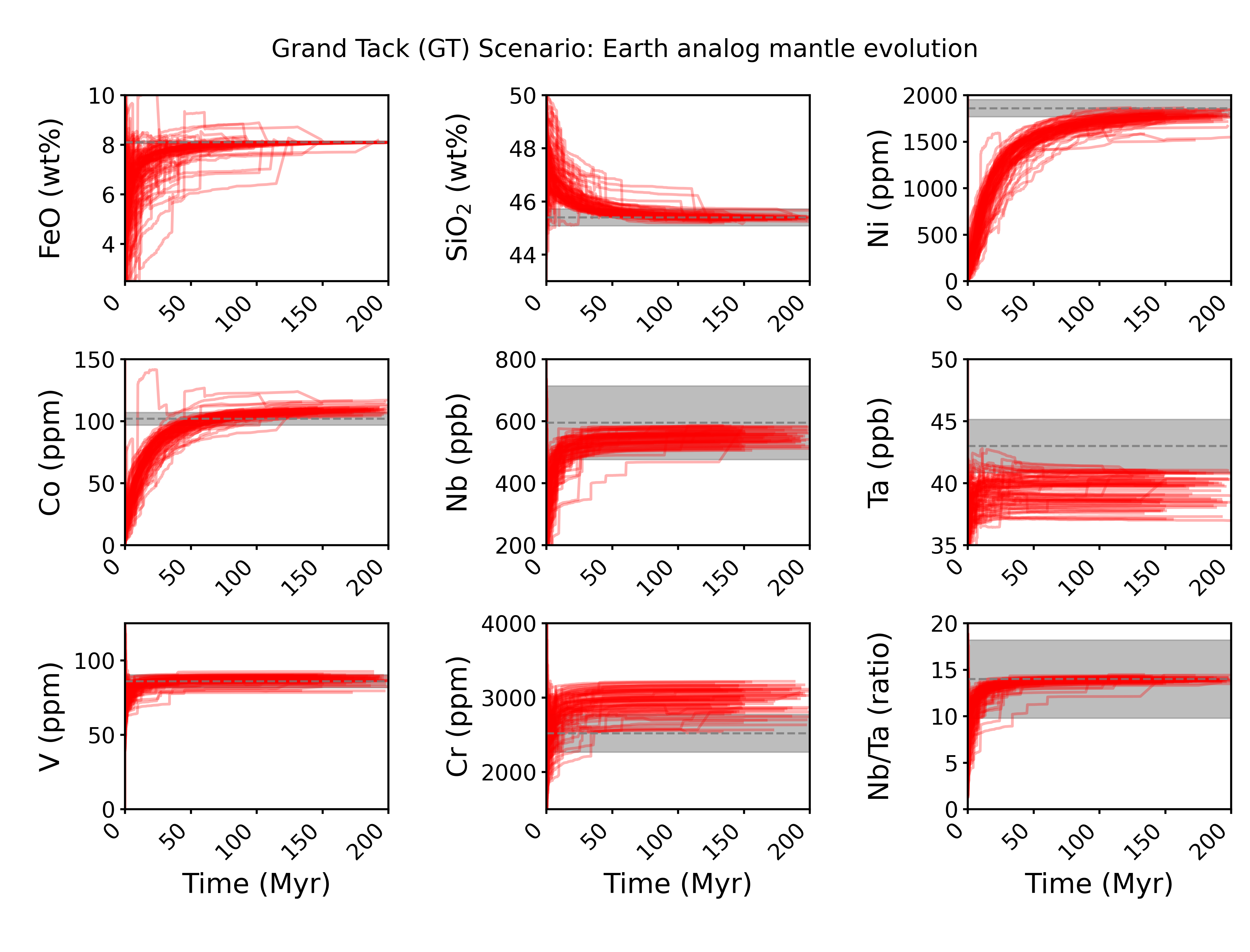}
    \caption{Mantle compositional evolution is shown for Earth analogs in Grand Tack (GT) scenario simulations \citep{walsh2011effect, O'Brien2014, jacobson2014lunar}. Subplots show the mantle abundance of fitted constraints, including FeO, SiO$_2$, Ni, Co, Nb, Ta, V, Cr, and the mantle Nb/Ta ratio on the y-axis. Time in millions of years is shown on the x-axis. Grey dashed lines and shaded region show the observed Earth mantle value and uncertainty, respectively, as estimated by \citet{palme2014cosmochemical}.}
    \label{fig:GT_elements}
\end{figure}

% \begin{figure}
%     \centering
%     \includegraphics[width=0.97\textwidth,height=0.72\textheight,keepaspectratio]{pt_1_figures_arxiv_upload/GTv_elements_wrt_time.png}
%     \caption{Element accretion histories over time for the Grand Tack - Venus (GTv) scenario.}
%     \label{fig:GTv_elements}
% \end{figure}

\begin{figure}[!htbp]
    \centering
    \includegraphics[width=0.97\textwidth,height=0.72\textheight,keepaspectratio]{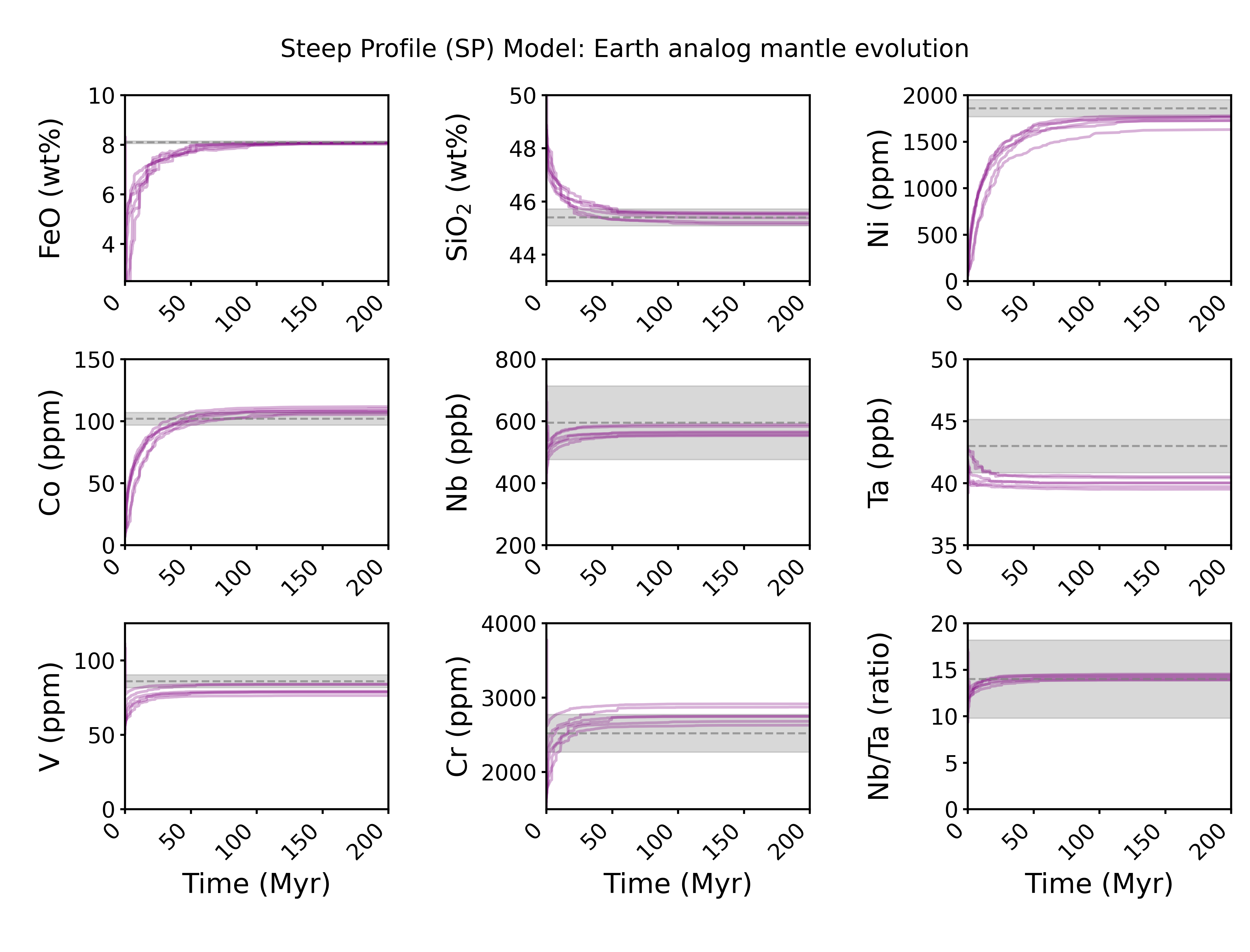}
    \caption{Mantle compositional evolution is shown for Earth analogs in Steep Profile (SP) scenario simulations \citep{izidoro2015terrestrial}. Subplots show the mantle abundance of fitted constraints, including FeO, SiO$_2$, Ni, Co, Nb, Ta, V, Cr, and the mantle Nb/Ta ratio on the y-axis. Time in millions of years is shown on the x-axis. Grey dashed lines and shaded region show the observed Earth mantle value and uncertainty, respectively, as estimated by \citet{palme2014cosmochemical}.}
    \label{fig:SP_elements}
\end{figure}

\begin{figure}[!htbp]
    \centering
    \includegraphics[width=0.97\textwidth,height=0.72\textheight,keepaspectratio]{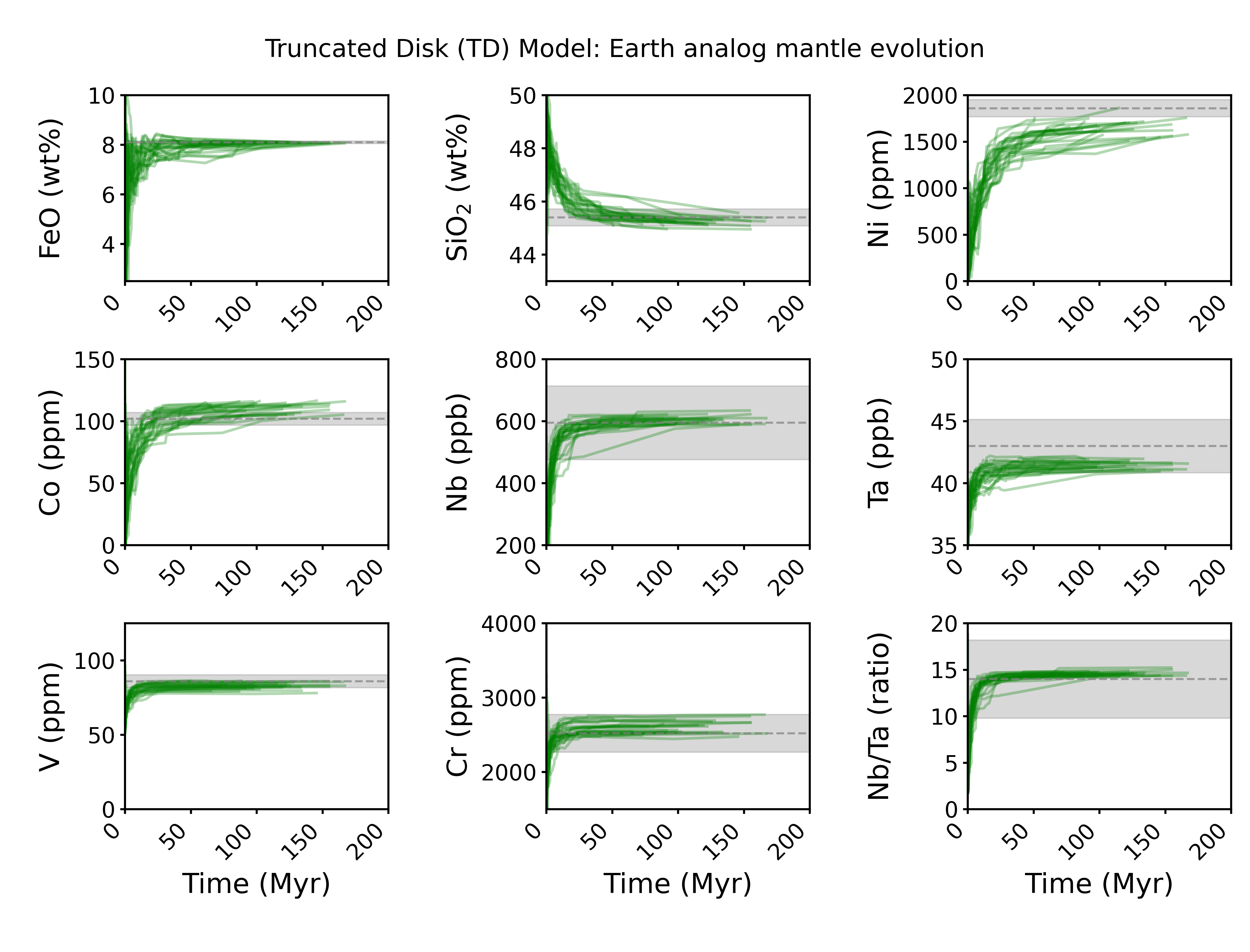}
    \caption{Mantle compositional evolution is shown for Earth analogs in Truncated Disk (TD) scenario simulations \citep{hansen2009formation}. Subplots show the mantle abundance of fitted constraints, including FeO, SiO$_2$, Ni, Co, Nb, Ta, V, Cr, and the mantle Nb/Ta ratio on the y-axis. Time in millions of years is shown on the x-axis. Grey dashed lines and shaded region show the observed Earth mantle value and uncertainty, respectively, as estimated by \citet{palme2014cosmochemical}.}
    \label{fig:TD_elements}
\end{figure}

% FeO, 

% SiO$_2$,

 % Ni, 

 % Co, 

 % Nb,

 % Ta, 

 % V, 

 % Cr, 

 % and the mantle Nb/Ta 

% Oxygen in the core

% Silicon in the core

% CAS writes the page total before deferred appendix figures are flushed.
% Override that early value with the number of actually shipped manuscript
% pages; the highlights page precedes page 1 and is not counted.
% After the final clearpage, the page counter points to the next unused page.
\makeatletter
\AtEndDocument{%
  \clearpage
  \immediate\write\@auxout{\string\csxdef{lastpage}{\number\numexpr\value{page}-1\relax}}%
}
\makeatother
\end{document}